\documentclass{aa}  

\usepackage{amsmath, amssymb, amsfonts, amscd}
\usepackage{natbib}
\usepackage{graphicx}
\usepackage{txfonts}
\usepackage[final]{hyperref}
\usepackage{mathrsfs}
\usepackage{color}
\usepackage{mathabx}
\usepackage{supertabular}

\hypersetup{
   colorlinks=true,
   citecolor=blue,
   linkcolor=blue,
   urlcolor=black
   }

\definecolor{emerald}{rgb}{0.0,0.5,0.0}
\definecolor{smcolor}{rgb}{0.7,0.3,0.0}
\definecolor{blue-violet}{rgb}{0.54, 0.17, 0.89}

\newcommand{\smc}[1]{#1}
\newcommand{\jlc}[1]{#1}
\newcommand{\padc}[1]{#1}
\newcommand{\ebc}[1]{#1}
\newcommand{\rec}[1]{#1}

\def\kro{\delta}					
\def\inumber{i}					
\def\GammaF{\Gamma}			
\def\vprod{\wedge}				
\def\Zset{\mathbb{Z}}			
\def\Rset{\mathbb{R}}			
\def\define{\equiv}				
\def\sign{{\rm sign}}				
\def\scale{\, \propto \,}			
\def\logdix{{\log_{10}}}

\newcommand{\notation}[4]{#1_{#2 ; #3}^{#4}}		
\newcommand{\ddroit}{{\rm d}}				
\newcommand{\vect}[1]{\boldsymbol{#1}}		
\newcommand{\evect}[1]{\vect{{\rm e}}_{#1}}	
\newcommand{\Cvar}[1]{\tilde{#1}}			
\newcommand{\Rvar}[1]{\hat{#1}}			
\newcommand{\dd}[2]{\partial_{#2} #1}		
\newcommand{\DD}[2]{\dfrac{\ddroit #1}{\ddroit #2}}			
\newcommand{\DDn}[3]{\dfrac{\ddroit^{#3} #1}{\ddroit #2^{#3}}} 
\newcommand{\Dt}[1]{\dot{#1}}				
\newcommand{\Dtt}[1]{\ddot{#1}}			
\newcommand{\real}[1]{\Re \left\{ #1 \right\}}	
\newcommand{\imag}[1]{\Im \left\{ #1 \right\}}	
\newcommand{\abs}[1]{\left| #1 \right|}		
\newcommand{\indvect}[2]{\vect{#1}_{#2}}		
\newcommand{\tensor}[1]{\underline{\underline{\boldsymbol{#1}}}} 
\newcommand{\transp}[1]{#1^{\rm T}}		
\newcommand{\expdec}[1]{ \times 10^{#1}}		

\def\ihoriz{{\rm h}}
\def\nab{\nabla}					
\def\grad{\nab}					
\def\gradh{\grad_{\ihoriz}}			
\def\divh{\gradh \cdot}			
\def\XX{X}	

\def\llat{l}						
\def\mm{m}					
\def\kk{k}						
\def\ss{s}						
\def\Legendre{P}				

\def\nn{n}						
\def\Hough{\Theta}				
\def\spinpar{\Cvar{\nu}}			
\def\mnu{{\mm,\spinpar}}			
\def\Laplace{\mathcal{L}^{\mnu}}	

\def\Ahoughnot{A}				
\def\Bhoughnot{B}				
\def\Choughnot{C}				

\newcommand{\Ahough}[4]{\Ahoughnot_{#1,#2}^{#3,#4}}
\newcommand{\Bhough}[4]{\Bhoughnot_{#1,#2}^{#3,#4}}
\newcommand{\Chough}[5]{\Choughnot_{#1,#2,#3}^{#4,#5}}

\def\Anl{\Ahough{\nn}{\llat}{\mm}{\spinpar}}
\def\Bkn{\Bhough{\kk}{\nn}{\mm}{\spinpar}}
\def\Akl{\Ahough{\kk}{\llat}{\mm}{\spinpar}}
\def\Bln{\Bhough{\llat}{\nn}{\mm}{\spinpar}}
\def\Clnk{\Chough{\llat}{\nn}{\kk}{\mm}{\spinpar}}
\def\Cnquad{\Chough{2}{\nn}{2}{2}{\spinpar}}
\def\Coquadsemid{\Chough{2}{0}{2}{2}{\spinparquadsemid}}
\def\Coquads{\Chough{2}{0}{2}{2}{\spinparquads}}
\def\Nhough{N}
\def\gamCn{\gamma_{\ss}}

\newcommand{\LegP}[1]{\bar{\Legendre}_{#1}}			
\newcommand{\LegFnorm}[2]{\Legendre_{#1}^{#2}}		
\newcommand{\LegF}[2]{\bar{\Legendre}_{#1}^{#2}}		
\newcommand{\Hansen}[3]{X_{#1}^{#2,#3}}			
\newcommand{\gHansen}[2]{g_{#1}^{#2}}				
\newcommand{\expo}[1]{{\rm e}^{#1}}				
\newcommand{\integ}[4]{\int_{#3}^{#4} #1 {\rm d} #2 }	
\newcommand{\kroind}[2]{\kro_{#1,#2}}				
\newcommand{\kroinf}[1]{\kro_{#1<0}}				
\newcommand{\HoughF}[3]{\Hough_{#1}^{#2,#3}}		
\newcommand{\Houghval}[3]{\Lambda_{#1}^{#2,#3}}		
\newcommand{\maxi}[2]{\max \left\{ #1 \right\}_{#2} }		

\def\Ggrav{G}					

\def\rr{r}						
\def\col{\theta}					
\def\lon{\varphi}					
\def\time{t}					

\def\er{\evect{\rr}}				
\def\etheta{\evect{\col}}			
\def\ephi{\evect{\lon}}			

\def\ipla{{\rm p}}				
\def\istar{\star}					
\def\ecc{e}					
\def\meana{M}					
\def\truea{v}					
\def\spinangle{\vartheta}			
\def\smaxis{a}					
\def\norb{n_\istar}				
\def\spinrate{\Omega}			
\def\spinvect{\vect{\spinrate}}		
\def\norbvect{\vect{\norb}}			
\def\eccasync{\ecc_{\rm AR}}		
\def\ecctrans{\ecc_{\rm trans}}		

\def\Mbody{M}					
\def\Rbody{R}					
\def\Mpla{\Mbody_\ipla}			
\def\Mstar{\Mbody_\istar}			
\def\rstar{\rr_\istar}				
\def\Rpla{\Rbody_\ipla}			
\def\centerpla{O}				

\def\iearth{\Earth}				
\def\Mearth{\Mbody_{\iearth}}		
\def\Rearth{\Rbody_{\iearth}}		

\def\isun{\sun}					
\def\Msun{\Mbody_{\isun}}			

\def\iocean{{\rm oc}}				
\def\isol{{\rm sol}}				
\def\idrag{{\rm R}}				
\def\icore{{\rm c}}				
\def\Hlayer{H}					
\def\densite{\rho}				
\def\chartime{\tau}				
\def\freq{\sigma}				
\def\ggravi{g}					
\def\Hoc{\Hlayer_\iocean}			
\def\Moc{\Mbody_\iocean}			
\def\fdrag{\freq_\idrag}			
\def\taudrag{\chartime_\idrag}
\def\rhooc{\densite_\iocean}		
\def\Rcore{\Rbody_\icore}			
\def\Mcore{\Mbody_\icore}			
\def\straintens{\tensor{\varepsilon}}	
\def\stresstens{\tensor{\zeta}}		
\def\rhoc{\densite_\icore}			
\def\rhopla{\densite_{\ipla}}		

\def\Cinertie{C}
\def\Binertie{B}
\def\Ainertie{A}
\def\Cinertpla{\Cinertie}
\def\Ainertpla{\Ainertie}
\def\Binertpla{\Binertie}
\def\inertiepla{\Cinertpla}		
\def\inertieplanorm{\check{\Cinertpla}}

\def\mupla{\mu}					
\def\tauA{\chartime_{\rm A}} 		
\def\tauM{\chartime_{\rm M}}		
\def\alphaA{\alpha}				
\def\betaA{\beta}				
\def\rhocore{\densite_\icore}		
\def\Cmu{\Cvar{\mu}}			
\def\viscosity{\eta}				
\def\ualpha{u_{\alphaA}}			
\def\Csol{\mathcal{B}_{\icore}}

\def\framesymb{\mathcal{R}}			
\def\igeo{{\rm G}}				
\def\irot{\ipla}
\def\framegeo{\framesymb_{\igeo}}		
\def\framerot{\framesymb_{\irot}}		
\def\Xvect{X}					
\def\Yvect{Y}					
\def\Zvect{Z}					
\def\Xgeo{\indvect{\Xvect}{\igeo}}	
\def\Ygeo{\indvect{\Yvect}{\igeo}}	
\def\Zgeo{\indvect{\Zvect}{\igeo}}	
\def\Xrot{\indvect{\Xvect}{\irot}}		
\def\Yrot{\indvect{\Yvect}{\irot}}		
\def\Zrot{\indvect{\Zvect}{\irot}}		

\newcommand{\refframe}[2]{\framesymb_{#1} {:} \left( #2 , \indvect{\Xvect}{#1},\indvect{\Yvect}{#1},\indvect{\Zvect}{#1} \right)}

\def\tauevol{\chartime_{\rm ev}}		

\def\ftide{\freq}					
\def\period{P}					
\def\Porb{\period_{\istar}}			
\def\Prot{\period_{\rm rot}}
\def\ftidedrag{\Cvar{\freq}}			
\def\msigma{\mm,\ftide}			
\def\ftidequads{\ftide_{2,\ss}}		
\def\ftidequadsemid{\ftide_{2,2}}	
\newcommand{\ftidequad}[1]{\ftide_{2,#1}}
\def\spinparquads{\spinpar_{2,\ss}}
\def\spinparquadsemid{\spinpar_{2,2}}
\def\Hocs{\Hlayer_{\iocean ; \ss}}			
\def\Cmul{\Cmu_\llat^{\ftide}}		
\def\coeffAmu{A}
\def\Al{\coeffAmu_{\llat}}
\def\Aquad{\coeffAmu_{2}}		
\def\gravpot{U}					
\def\vel{V}						
\def\dep{\xi}					
\def\ibottom{\icore}				
\def\isurf{\iocean}				
\def\iforcing{{\rm T}}				
\def\tautide{\period_{\iforcing}}	
\def\Ftide{\Psi}					
\def\Utide{\gravpot_\iforcing}		
\def\UstarR{\Rvar{\gravpot}_\istar}	
\def\xibot{\dep_\ibottom}			
\def\xisurf{\dep_\isurf}			
\def\Vvect{\vect{\vel}}			
\def\Vtheta{\vel_\col}				
\def\Vphi{\vel_\lon}				
\def\xivect{\vect{\dep}}			
\def\xilayer{\eta}				
\def\kwave{k}				
\def\khori{\kwave_{\ihoriz ; \nn}} 		
\def\Ulms{\notation{\gravpot}{\iforcing}{\llat}{\mm,\ss}} 	
\def\Ulmsig{\notation{\gravpot}{\iforcing}{\llat}{\mm,\ftide}}	
\def\Xslm{\Hansen{\ss}{\llat}{\mm}}		
\def\Alms{A_{\llat,\mm,\ss}}			
\def\UtideR{\Rvar{\gravpot}_\iforcing}	
\def\Utidequad{\notation{\gravpot}{\iforcing}{2}{2,\ss}} 
\def\quanti{q}					
\def\quantisig{\quanti^{\ftide}}		
\def\quantiR{\Rvar{\quanti}}		

\def\Vvectms{\Vvect^{\msigma}}	
\def\Vthetams{\Vtheta^{\msigma}}	
\def\Vphims{\Vphi^{\msigma}}	
\def\Ftidems{\Ftide^{\msigma}}	
\def\xilayerms{\xilayer^{\msigma}}	

\def\Ftidelms{\Ftide_{\llat}^{\msigma}} 
\def\xilayerlms{\xilayer_{\llat}^{\msigma}}
\def\Vvectlms{\Vvect_{\llat}^{\msigma}}

\def\Ftiden{\Ftide_{\nn}^{\msigma}}				
\def\xilayern{\xilayer_{\nn}^{\msigma}}			
\def\xilayerk{\xilayer_{\kk}^{\msigma}}			
\def\Fforce{F}
\newcommand{\Fforceind}[1]{\Fforce_{#1}^{\msigma}}
\newcommand{\xilayerind}[1]{\xilayer_{#1}^{\msigma}}
\def\Fforcen{\Fforceind{\nn}}

\def\heq{h}
\def\heqn{\heq_{\nn}^{\msigma}}	

\newcommand{\Utidelms}[3]{\notation{\gravpot}{\iforcing}{#1}{#2,#3}}

\newcommand{\sig}[2]{\ftide_{#1,#2}} 

\def\ipeak{{\rm max}}
\def\focnmoins{\Cvar{\ftide}_{\nn}^{-}}
\def\focnplus{\Cvar{\ftide}_{\nn}^{+}}
\def\focnpm{\Cvar{\ftide}_{\nn}^{\pm}}
\def\focn{\ftide_{\nn}}
\def\foco{\ftide_{0}}
\def\fpeakn{\ftide_{\ipeak ; \nn}} 

\def\ieq{{\rm eq}}
\def\spinrateeq{\spinrate_{\ieq}}		
\def\spinrateeqs{\spinrate_{\ieq ; \ss}}	
\newcommand{\spinrateeqsnum}[1]{\spinrate_{\ieq ;#1}}

\def\lovegrav{k}					
\def\lovedep{h}					
\def\torque{\mathcal{T}}			

\def\kl{\lovegrav_\llat^\ftide}			
\def\hl{\lovedep_\llat^\ftide}			
\def\kloadl{\mathfrak{\lovegrav}_{\llat}^\ftide}	
\def\hloadl{\mathfrak{\lovedep}_{\llat}^\ftide}	

\def\kquad{\lovegrav_{2}^{\ftide}}
\def\kquadsemid{\lovegrav_2^{2,\ftidequadsemid}}
\def\kp{\notation{\lovegrav}{\ipla}{\llat}{\msigma}}	
\def\kpquad{\notation{\lovegrav}{\ipla}{2}{2,\ftide}}
\def\kpquadsemid{\notation{\lovegrav}{\ipla}{2}{2,\ftidequadsemid}}

\def\kocquad{\notation{\lovegrav}{\iocean}{2}{2,\ftide}}	
\def\kocquads{\notation{\lovegrav}{\iocean}{2}{2,\ftidequads}}

\def\torquepla{\torque_\ipla}		
\def\torquesol{\torque_\isol}		
\def\torqueoc{\torque_\iocean}		
\def\tiltfactor{\gamma}			
\def\tiltxi{\notation{\tiltfactor}{\dep}{\llat}{\ftide}}		
\def\tiltU{\notation{\tiltfactor}{\gravpot}{\llat}{\ftide}}	

\def\Xfreq{X}	
\def\Aconst{\mathcal{A}}	

\def\torquetri{\torque_{\ipla}^{\rm tri}}	
\def\torquetriC{\Cvar{\torque}_{\ipla}^{\rm tri}}
\def\angletriax{\gamma}
\def\anglequads{\angletriax_{2,\ss}}	
\def\Deltares{\Delta}
\def\DeltaProt{\Deltares_{\period}}
\def\Psol{\period_{\rm sol}}

\def\Protres{\Prot^\Deltares}
\def\Psolres{\Psol^\Deltares}
\def\itide{{\rm \smc{tide}}}
\def\ieven{{\rm even}}
\def\iodd{{\rm odd}}
\def\torqueplatid{\torquepla^\itide}
\def\torqueeven{\torque_\ieven^\itide}
\def\torqueodd{\torque_\iodd^\itide}
\def\probability{\mathcal{P}}
\def\captureprob{\probability_{{\rm cap ;} \ss}}

\def\exptphi{\expo{\inumber \left( \ftide \time + \mm \lon  \right)}} 
\def\Pl{\LegP{\llat}}						
\def\Plmnorm{\LegFnorm{\llat}{\mm}}			
\def\Plm{\LegF{\llat}{\mm}}				
\def\Thetan{\HoughF{\nn}{\mm}{\spinpar}}		
\def\Lambdan{\Houghval{\nn}{\mm}{\spinpar}}	
\def\Lambdanquads{\Houghval{\nn}{2}{\spinparquads}}
\def\Lambdaoquads{\Houghval{0}{2}{\spinparquads}}
\def\Plmnormquad{\LegFnorm{2}{2}}

\newcommand{\eq}[1]{Eq.~\ref{#1}}
\newcommand{\eqs}[2]{Eqs.~\ref{#1} and~\ref{#2}}
\newcommand{\eqsto}[2]{Eqs.~\ref{#1}-\ref{#2}}

\newcommand{\append}[1]{Appendix~\ref{#1}}
\newcommand{\units}[1]{~${\rm #1}$}
\newcommand{\fig}[1]{Fig.~\ref{#1}}

\newcommand{\sect}[1]{Sect.~\ref{#1}}
\newcommand{\sects}[2]{Sects.~\ref{#1} and~\ref{#2}}
\newcommand{\sectsto}[2]{Sects.~\ref{#1} to~\ref{#2}}
\newcommand{\tab}[1]{Table~\ref{#1}}
\newcommand{\comments}[1]{}
\def\days{\units{days}}

\begin{document} 
  \title{\padc{Final spin states of eccentric ocean planets}}

  \author{P. Auclair-Desrotour\inst{1} \fnmsep \inst{2}
     \and  J. Leconte\inst{1}  
     \and E. Bolmont\inst{3}
     \and S. Mathis\inst{4} 
          }

  \institute{Laboratoire d'Astrophysique de Bordeaux, Univ. Bordeaux, CNRS, B18N, allée Geoffroy Saint-Hilaire, 33615 Pessac, France
  \and University of Bern, Center for Space and. Habitability, Gesellschaftsstrasse 6, CH-3012, Bern, Switzerland \\
   \email{pierre.auclair-desrotour@csh.unibe.ch} 
  \and \ebc{Observatoire de Genève, Université de Genève, 51 Chemin des Maillettes, CH-1290 Sauvergny, Switzerland}
  \and \smc{AIM, CEA, CNRS, Université Paris-Saclay, Université Paris Diderot, Sorbonne Paris Cité, F-91191 Gif-sur-Yvette Cedex, France}
  }
 

  \date{Received ...; accepted ...}

  \abstract
   {Eccentricity tides generate a torque that can drive an ocean planet towards asynchronous rotation states of equilibrium when enhanced by resonances associated with the oceanic tidal modes.}
   {We investigate the impact of eccentricity tides on the rotation of rocky planets hosting a thin uniform ocean and orbiting cool dwarf stars such as TRAPPIST-1, with orbital periods $\sim 1-10$\units{days}.  }
   {Combining the linear theory of oceanic tides in the shallow water approximation with the Andrade model for the solid part of the planet, we develop a global model including the coupling effects of ocean loading, self-attraction, and deformation of the solid regions. We derive from this model analytic solutions for the tidal Love numbers and torque exerted on the planet. These solutions are used with realistic values of parameters provided by advanced models of the internal structure \smc{and tidal oscillations of solid bodies} to explore the parameter space both analytically and numerically. }
   {Our model allows us to fully characterise the frequency-resonant tidal response of the planet, and particularly the features of resonances associated with the oceanic tidal modes (eigenfrequencies, resulting maxima of the tidal torque and Love numbers) as functions of the planet parameters (mass, radius, Andrade parameters, ocean depth and Rayleigh drag frequency). Resonances associated with the oceanic tide decrease the critical eccentricity beyond which asynchronous rotation states distinct from the usual spin-orbit resonances can exist. We provide an estimation and scaling laws for this critical eccentricity, which is found to be lowered by roughly one order of magnitude, switching from $\sim 0.3$ to $\sim 0.06$ in typical cases and to $\sim 0.01$ in extremal ones.}
   {}


\maketitle


\section{Introduction}
\label{sec:intro}
One of the most essential questions raised by the discovery of rocky exoplanets is the nature of their climate and surface conditions. Particularly, this question motivated the major part of studies dealing with the TRAPPIST-1 system \citep[][]{Gillon2017}, where an ultra-cool M-dwarf star harbours seven Earth-sized planets among which four -- namely planets d, e, f, and  g -- may be potentially habitable \citep[e.g.][]{Bolmont2017,Bourrier2017,Grimm2018,Papaloizou2018,Unterborn2018b,Barr2018,Turbet2018,Dobos2019}. To constrain the climate of such planets, it is crucial to preliminary constrain their rotation. As tides drive the long term evolution of planetary systems, this requires to characterise the possible asynchronous states of equilibrium where rocky planets may be tidally locked. 

Rocky planets orbiting cool dwarf stars in tightly packed systems combine temperate surfaces conditions with small orbital periods, which makes them privileged objects of study. The TRAPPIST-1 system illustrates well this configuration since its seven rocky planets orbit within a disk of radius 0.065\units{au}, and two of them -- planets d and e -- are located in the habitable zone of the host star with orbital periods of 4.05 and 6.10~days, respectively \citep[][]{Gillon2017,Grimm2018}. Under these conditions, the presence of an important amount of liquid water on the planet makes the existence of global oceans likely \citep[e.g.][]{Bolmont2017,Bourrier2017,Turbet2018}. 

The tidal response of a free-surface oceanic layer strongly differs from that of solid bodies. Dry rocky planets undergoing the tidal gravitational \smc{potential} of their host star, or perturber, are subject to small distortions due to their elasticity \citep[e.g.][]{Henning2009,Efroimsky2012,Remus2012}. The resulting tidal elongation corresponds to the hydrostatic adjustment between \smc{gravity and elasticity}, which is the so-called 'equilibrium tide' \citep[][]{Zahn1966a,OL2004}. Because of the lag induced by internal dissipative processes, the tidal bulge generates a torque acting on the rotation of the body. In the case of a circular and coplanar planet, this torque is created by the semidiurnal tide solely and drives the planet towards spin-orbit synchronous rotation. Unlike solid tides, which are not strongly dependent on the tidal frequency, oceanic tides exhibit a frequency-resonant behaviour that enables large variations of the tidal torque \citep[e.g.][]{Webb1980,Tyler2011,Chen2014,Matsuyama2014,ADLML2018}. 

In thin fluid shells, such a behaviour results from the propagation of gravito-inertial surface modes  \smc{(or barotropic modes)} forced by the tidal gravitational potential, which corresponds to the solutions of the Laplace's tidal equation \citep[e.g.][]{LH1968,Webb1980}. In deeper oceans, stable stratification leads in addition to the propagation of internal gravito-inertial waves, which are restored by the Archimedean force and provide a supplementary resonant contribution to the oceanic tidal torque \citep[e.g.][]{Tyler2011,ADLML2018}. \smc{These modes form the baroclinic tide.} The component of the tidal response associated with the propagation of waves is named 'dynamical tide' \smc{\citep[e.g.][]{Zahn1975,OL2004}} in opposition with the equilibrium tide.

The tidal response of the ocean is tightly coupled with that of the solid part by self-attraction and loading, meaning that the contributions of the two layers cannot be separated but merge into an effective tidal response of the planet. Although these couplings are often ignored for the sake of simplicity \citep[e.g.][]{Tyler2011,Chen2014,ADLML2018}, they significantly modify the global tidal response of the planet \citep[e.g.][]{Matsuyama2014,Matsuyama2018} and should be taken into account in the calculation of the effective tidal torque and Love numbers \citep[][]{Love1911}, which quantify the impact of tidal dissipation on the evolution of the planet-perturber system \citep[see e.g.][for solid bodies]{MM1960}. 

The specific action of tides on the planetary spin can be characterised in the idealised coplanar and circular configuration, that is in the absence of obliquity and eccentricity \citep[e.g.][]{ADLML2018}. In this configuration, the semidiurnal tidal component predominates and drives the planet towards the spin-orbit synchronous rotation, except in the case of thermal atmospheric tides \citep[e.g.][]{LC1969}, where the body is torqued away from synchronization \citep[e.g.][]{GS1969,ID1978,DI1980,CL2001,CL2003,Leconte2015,ADLM2017a,ADLM2017b,ADL2019}. 

Eccentricity affects this evolution by introducing supplementary forcing terms, weighted by the Hansen coefficients \citep[e.g.][]{Hughes1981}, which can torque the body away from synchronization \citep[e.g.][]{Greenberg2009,ME2013,Correia2014}. A dry rocky planet with an eccentric orbit may thus be tidally locked into one of the spin-orbit resonances induced by eccentricity tides \citep[$3{:}2$, $2{:}1$, $5{:}2$, $3{:}1$, etc.; see][]{Makarov2012,ME2013,Correia2014}, which correspond to asynchronous rotation rates. \rec{This question has been investigated mainly for the rocky bodies of the Solar system, such as Mercury, by means of two-layer tidal models composed of a molten core and a solid crust \citep[e.g.][]{PB1977,CL2009,Henning2014,Noyelles2014}.} The \rec{spin-orbit} rotation equilibria were also retrieved in the case of giant planets by studies using a visco-elastic model of the tidally dissipated energy \citep[e.g.][]{SL2014}. 

As shown by many studies examining the case of icy satellites in the Solar system \citep[e.g.][]{Tyler2008,Tyler2009,Tyler2011,Chen2014,Beuthe2016,Matsuyama2014,Matsuyama2018}, the resonances of oceanic modes may enhance eccentricity tidal components by several orders of magnitude and increase as well the resulting tidal heating. As a consequence, they are also likely to generate asynchronous rotation states of equilibrium distinct from the spin-orbit resonances identified by early works for solid bodies, including for low eccentricities. This means that the combination of eccentricity tides with resonances enables the existence of such states for rocky planets exhibiting quasi-circular orbits. 

In the present work, we investigate this mechanism by considering the case of an idealized ocean planet with an eccentric orbit and no obliquity, the central object of the system being assumed to be a TRAPPIST-1-like dwarf star. Treating the ocean as a spherical thin shell in the shallow water approximation, we follow \cite{Matsuyama2014} and take the effects of ocean loading, self-attraction, and deformation of solid regions into account self-consistently. \rec{We shall emphasize here that we do not include the coupling between the internal tidal heating of the planet and its structure and surface conditions, these laters being fixed in the model. Therefore, the planetary system is essentially parametrized by the mass of the host star, which is set to the TRAPPIST-1 value. The implications of tidal heating on the possible existence of water oceans are discussed in conclusions.} 

To describe the tidal response of the solid part, we use the Andrade model \citep[][]{Andrade1910,CR2011,Efroimsky2012} with values of parameters provided by \smc{a spectral code that computes the visco-elastic tidal oscillations of the body expanded in spherical harmonics by taking as inputs its density, rigidity, and viscosity profiles \citep[][]{TS1972,Tobie2005}.} \ebc{For a given mass and composition for the planet, the internal structure used for these calculations is derived from the model detailed in \cite{Sotin2007}. For the derived internal structure of the planet, we then use the model of \cite{Tobie2005}, which solves an equation of state self-consistently across the radial direction.}

\padc{The choice of the Andrade model} is mainly motivated by the abundance of experimental data supporting it \smc{for telluric planets}, \padc{the model having} been reported to match over a wide range of experimental conditions \citep[e.g.][]{Andrade1910,Andrade1914,CA1947,Duval1978,Jackson1993}. \rec{Particularly, the Andrade model is believed to better describe the behaviour of terrestrial bodies than the Maxwell rheology in the high-frequency range \citep[e.g.][]{EL2007,Efroimsky2012}}. However, the methodology applied in this article is general and can thus be easily \smc{applied to} any rheology, \rec{including more complex ones such as the Sundberg-Cooper rheology \citep[e.g.][]{RH2018}}.

The ability of eccentricity tides to drive the planet away from the spin-orbit synchronous rotation is quantified by the minimal eccentricity for which asynchronous states may exist. All along the article, this eccentricity is called the 'critical eccentricity' and denoted by $\eccasync $ ('AR' refers to 'Asynchronous Rotation'). As we focus on the frequency interval bounded by the $1{:}1$ (synchronization) and $3{:}2$ spin-orbit resonances, $\ecc < \eccasync$ means that the planet is driven towards the spin-orbit \ebc{synchronous} rotation, while $\ecc > \eccasync$ corresponds to asynchronous final rotation states of equilibrium. 

In \sect{sec:setup} we introduce the physical setup, the main parameters, and the reference frames of the study. In \sect{sec:tidal_dynamics}, we expand the perturbing tidal gravitational potential in eccentricity series, and establish the equations and quantities governing the tidal response of the solid part and ocean. Particularly, we detail the features of oceanic tidal modes derived from the Laplace's tidal equation. In \sect{sec:tidal_torque}, we establish the analytic expressions of the tidal torque and Love numbers in the general case, and in the asymptotic cases corresponding to pure solid and oceanic tidal responses. This allows us to analyse the effect of resonances on eccentricity terms, and to characterise the frequency behaviour of the tidal torque. In \sect{sec:critical_eccentricity}, we derive analytic estimations of the critical eccentricity in  the quasi-adiabatic asymptotic regime. 

In \sect{sec:final_rotation}, the final rotation states of Earth and super-Earth-sized planets are calculated numerically as functions of the eccentricity and ocean depth for various orbital periods and Rayleigh drag timescales, which highlights that the critical eccentricity may be decreased by one order of magnitude owing to the action of resonances associated with oceanic tidal modes. \padc{The existence of these final spin states is discussed in term of possible capture in the $1{:}1$ spin-orbit resonance using the theory of \cite{GP1966}.} In \sect{sec:critical_ecc}, the critical eccentricity is calculated as a function of the orbital period and ocean depth in order to unravel the regions of the parameter space where asynchronous spin equilibria are compatible with low eccentricities. In \sect{sec:evolution_timescale}, we integrate the evolution of the planet spin over time for various eccentricities, and thus quantify the evolution timescale associated with the tidal torque. Finally, we give our conclusions and discuss the limitations of the model in \sect{sec:conclusions}.

\section{Physical setup}
\label{sec:setup}

We examine the simplified case of a single terrestrial planet orbiting its host star with an eccentric orbit. The planet, of mass $\Mpla$ and radius $\Rpla$, is treated as a bi-layered body basically composed of an internal solid part and an external thin incompressible ocean of uniform thickness $\Hoc \ll \Rpla$ and density $\rhooc$. Its orbital motion is described by its mean motion $\norb$ and eccentricity~$\ecc$. The orbital \smc{angular} momentum vector is denoted $\norbvect$. We introduce the \smc{planeto-centric} referential $\refframe{\igeo}{\centerpla}$, where $\Zgeo \define \norbvect / \norb$ (the symbol $\define$ meaning 'defined by'), $\Xgeo$ and $\Ygeo$ designate the directions of two distant stars defining the orbital plane, and $\centerpla$ the planet gravity center. 

We assume the absence of obliquity, which allows us to define the reference frame co-rotating with the planet as $\refframe{\irot}{\centerpla}$ with $\Zrot = \Zgeo$. The vectors $\Xrot$ and $\Yrot$ define the equatorial plane of the planet, which is also its orbital plane in the present case. The rotating motion of $\framerot$ with respect to $\framegeo$ is defined by the spin vector $\spinvect = \spinrate \Zrot$, \smc{$\spinrate$ being} the rotation rate of the planet. Hence, denoting $\time$ the time, $\Xrot = \cos \left( \spinrate \time \right) \Xgeo + \sin \left( \spinrate \time \right) \Ygeo$ and $\Yrot = - \sin \left( \spinrate \time \right) \Xgeo + \cos \left( \spinrate \time \right) \Ygeo $. We finally need to introduce the spherical coordinates, namely the radius $\rr$, colatitude $\col$, and longitude $\lon$, and the associated \smc{unit-vector} basis $\left( \er , \etheta, \ephi \right)$. 

The surface planet gravity is denoted $\ggravi$. We introduce the planet mean density \smc{$\rhopla \define \Mpla  / \left( \frac{4 \pi}{3} \Rpla^3 \right)$} and the mass of the ocean $\Moc  \define 4 \pi \Rpla^2 \Hoc \rhooc$. Besides, the mass and radius of the solid part are denoted $\Mcore \define \Mpla - \Moc  \approx \Mpla$ and $\Rcore \define \Rpla - \Hoc \approx \Rpla$, respectively. To simplify calculations, we assume in the following that this layer can be treated as a homogeneous body of density \smc{$\rhocore \define  \Mcore / \left( \frac{4 \pi}{3} \Rcore^3 \right) \approx \rhopla$}. 

\rec{The ocean is considered as perfectly coupled with the solid part of the planet by frictional forces, meaning that the whole planet rotates as a solid body. This assumption holds provided that the ocean is thin with respect to the planet radius. In the case of a deep ocean, the solid part of the planet and the upper layers of the ocean decouple, which allow zonal flows to develop and potentially give rise to differential rotation depending on the strength of friction. Owing to this decoupling, one may envision that a sufficiently deep ocean and the solid part could be rotating in different spin states, similarly as the core and the mantle in rocky planets \citep[e.g.][]{CL2009}.} 

\rec{In the framework of Rayleigh drag approximation chosen for this study \citep[e.g.][]{Vallis2006}, the typical timescale of the core-ocean coupling by viscous friction is $\taudrag \define \fdrag^{-1}$, where $\fdrag$ is the effective Rayleigh drag frequency parametrising the dissipative term in the momentum equation (see \eq{momentum_eq}). In the case of the Earth, with $\fdrag = 10^{-5}$\units{s^{-1}} \citep[][]{Webb1980}, $\taudrag \approx 30$~hr, which is far smaller than the evolution timescale of the planet spin. }

\begin{figure}[t]
   \centering
   \includegraphics[width=0.48\textwidth,trim = 0.5cm 1.0cm 3.6cm 2.1cm,clip]{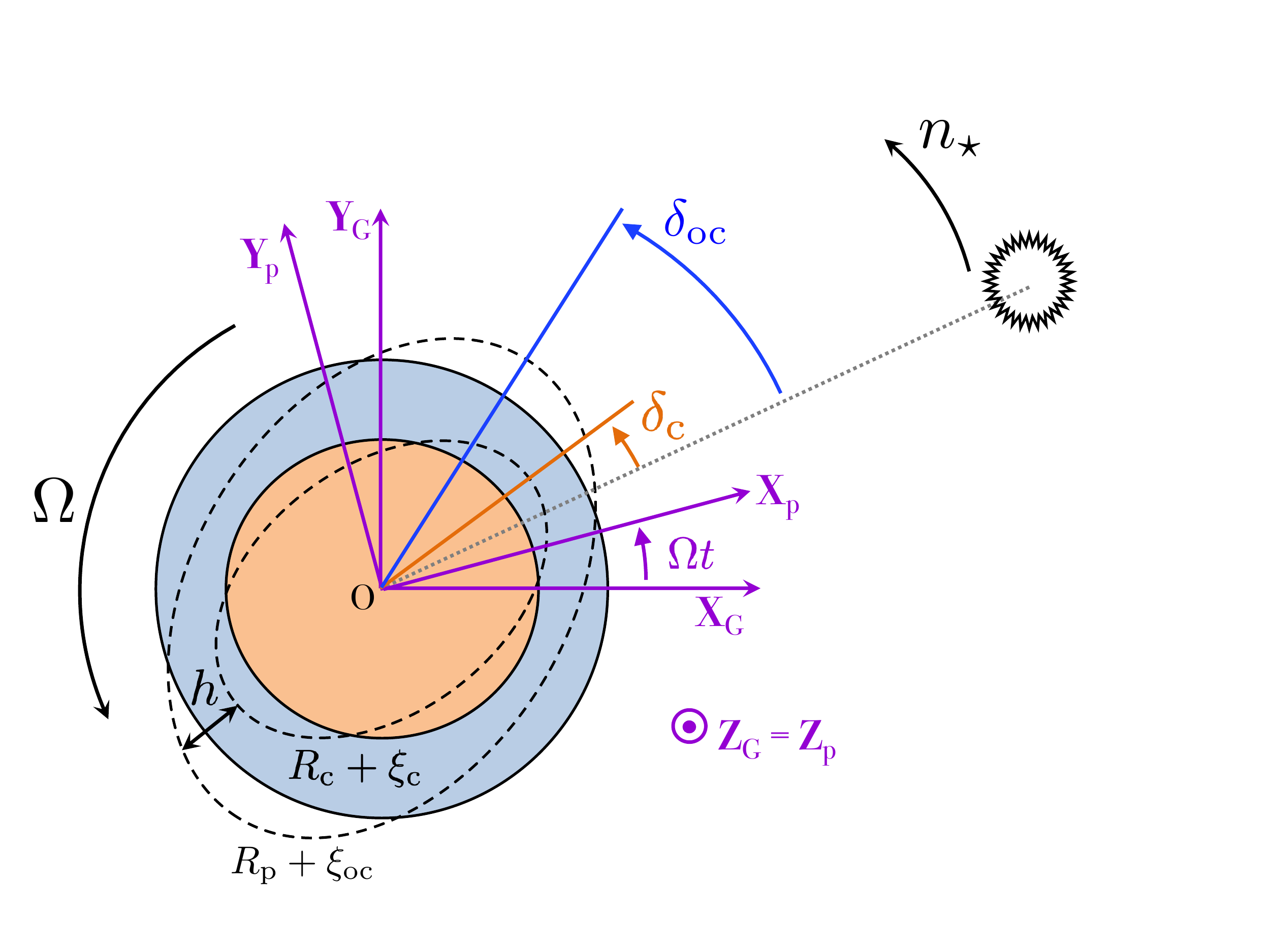}
   \caption{Tidal distortion of the terrestrial planet resulting from the gravitational forcing of its host star. The planet structure is composed of a solid part \smc{of radius $\Rcore$} and a thin uniform ocean \smc{of external radius $\Rpla$} rotating as a solid body at the spin rotation rate $\spinrate$. The tidal responses of the two layers are coupled \ebc{together} and characterised by different frequency behaviours. The corresponding angular lags are designated by $\delta_{\icore}$ and $\delta_{\iocean}$, \smc{and the surface displacements by $\xibot$ and $\xisurf$}. \padc{The ocean depth modified by solid and oceanic tides is denoted by $h$.} References frames introduced in \sect{sec:setup} are drawn \padc{in violet}. }
       \label{fig:setup}%
\end{figure}


\section{Tidal dynamics}
\label{sec:tidal_dynamics}

Owing to its eccentric orbit, the planet is subject to eccentricity tides, which act on its spin rotation in a different way from the standard semidiurnal tide. While the semidiurnal tide drives the body towards the spin-orbit synchronous rotation, eccentricity tides tend to desynchronise it, and enable thereby the existence of non-synchronised states of equilibrium. In this section, we establish the components of the tidal potential and discuss the desynchronising mechanism of eccentricity tides. We then use the classical tidal theory to describe analytically the dynamics of the planet tidal response, by treating \smc{successively the solid part and the oceanic layer}. 

\subsection{Perturbing tidal gravitational potential}
\label{ssec:perturbing_potential}
The whole planet is tidally forced by the gravitational potential of the host star. For large star-planet distances $\rstar$, that is $\rstar \gg \Rpla$ typically, this gravitational potential is expressed in the accelerated frame of the planet as

\begin{equation}
\UstarR \left( \rr , \rstar \right) = \frac{\Ggrav \Mstar}{\abs{\rr - \rstar}} - \frac{\Ggrav \Mstar}{\rstar^2} \rr \cos \col ,
\label{stellar_potential}
\end{equation}

\noindent the first term of the right-hand side of the equation corresponding to the attraction by the host star, \rec{of mass $\Mstar$}, and the second term to the centrifugal force due to the orbital motion. We note that the symbol $\Rvar{~}$ is employed here and all along the article to highlight real quantities with respect to complex ones, the laters being preferentially used in analytical developments in the general case. Conversely, a few specific complex quantities that are usually real will be highlighted by the symbol $\Cvar{~}$ in order to avoid confusion.

In the thin layer approximation, the tidal gravitational potential is approximated by its value at the planet surface ($\rr = \Rpla$),

\begin{equation}
\UtideR \left( \col , \lon , \rstar \right) \define \UstarR \left( \Rpla , \col , \varphi, \rstar \right) - \frac{\Ggrav \Mstar}{\rstar}, 
\label{UtideR}
\end{equation}

\noindent where we have removed the constant component as it does not contribute to the tidal force. The associated complex gravitational potential $\Utide$, such that $\UtideR = \real{\Utide} $, is expanded in Fourier series of the time and spherical harmonics, following Kaula's theory \citep[e.g.][]{Kaula1966}. In the absence of obliquity, it is thus written

\begin{equation}
\Utide = \sum_{\llat=2}^{+ \infty} \sum_{\mm=0}^{\llat} \sum_{\ss = - \infty}^{+ \infty} \Ulms \left( \Rpla \right) \Plmnorm \left( \cos \col \right) \expo{\inumber \left[ \ftide_{\mm,\ss} t + \mm \lon \right] },
\label{Utide}
\end{equation}

\noindent where $\llat$ and $\mm$ designate the latitudinal and longitudinal degrees, $\ss $ an integer, $\ftide_{\mm,\ss} = \mm \spinrate - \ss \norb $ the forcing \smc{tidal} frequency of the mode associated with the doublet $\left( \mm , \ss \right)$, $\Plmnorm$ the normalised associated Legendre function associated with the doublet $\left( \llat , \mm \right)$ (see \append{app:legendre_functions}), and $\Ulms$ the surface tidal gravitational potential associated with the triplet $\left( \llat , \mm , \ss \right)$. This later is given by 

\begin{equation}
\Ulms \left( \Rpla \right)  = \frac{\Ggrav \Mstar}{\smaxis} \left(\frac{\Rpla}{\smaxis} \right)^\llat \Alms \left( \ecc \right),
\label{Utidelms}
\end{equation}

\noindent where we have introduced \rec{the semi-major axis $\smaxis$, and} the dimensionless coefficients $\Alms$, following the notation by \cite{Ogilvie2014} (Eq.~(3) of the review; we note that we use here the normalised associated Legendre functions, which explains we get different numerical factors). Let us introduce the Kronecker symbol $\kroind{\llat}{\kk}$, such that $\kroind{\llat}{\kk} = 1$ if $\llat = \kk$ and $\kroind{\llat}{\kk} = 0$ otherwise. By analogy with the Kronecker symbol, we define by $\kroinf{\ss}$ the coefficient such that $\kroinf{\ss}= 1$ if $\ss < 0$ and $0$ otherwise. The $\Alms$ coefficients are thus expressed as 

\begin{align}
\label{Alms_ecc}
\Alms \left( \ecc \right) = & \left( 2 - \kroind{\mm}{0} \kroind{\ss}{0} \right) \left( 1 - \kroind{\mm}{0} \kroinf{\ss} \right) \\
	& \times \sqrt{\frac{2 \left( \llat - \mm \right) ! }{\left( 2 \llat + 1 \right) \left( \llat + \mm \right) !}} \LegF{\llat}{\mm} \left( 0 \right) \Hansen{\ss}{-\left( \llat + 1 \right)}{\mm} \left( \ecc \right).\nonumber
\end{align}

\noindent In the above expression, \rec{the $\Plm$ are the unnormalised associated Legendre functions, and} the eccentricity functions $\Xslm$ are the so-called Hansen coefficients \citep[][]{Hughes1981,PS1990,Laskar2005}, \ebc{which are calculated numerically in the study using a fast Fourier transform (see \append{app:Hansen} for details).} 

\begin{figure}[t]
   \centering
   \includegraphics[width=0.48\textwidth,trim = 0cm 0cm 3.5cm 0cm,clip]{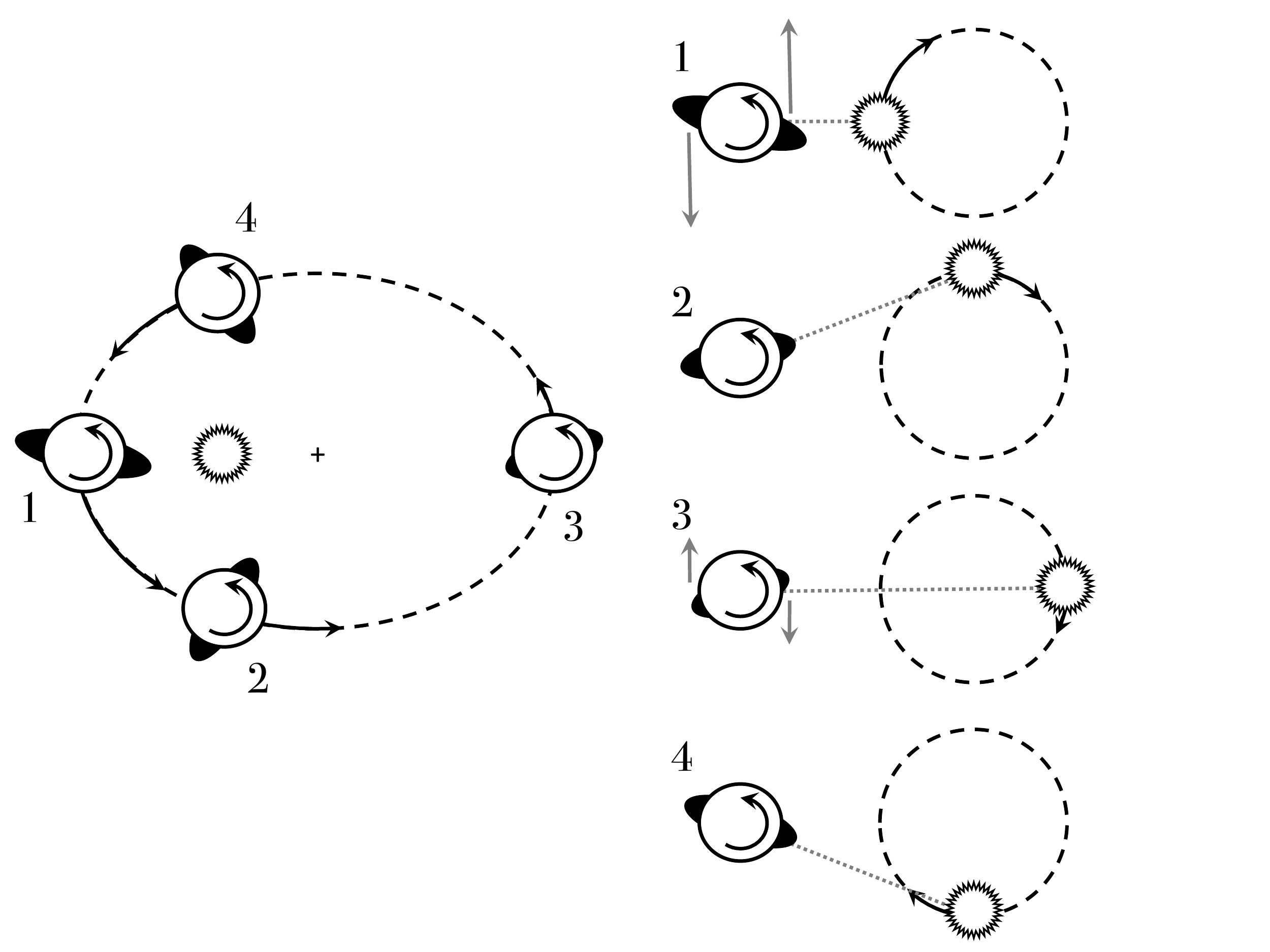}
   \caption{Desynchronising mechanism of eccentricity tides. Tidal bulge raised on a planet by the host star in the heliocentric reference frame (left) and in a \smc{planeto-centric} reference frame rotating with the planet mean motion (right). The tidal gravitational forcing is assumed to be dominated by its eccentricity components, given by \eqs{U221}{U223}. Because of the angular lag generated by dissipative processes, the rotation of the planet is accelerated by the component of degree $\ss = 3$ (eastward propagating potential) in the vicinity of the periastron. Conversely, it is decelerated by the component of degree $\ss = 1$ (westward propagating potential) to a lesser extent in the vicinity of the apoapsis. }
       \label{fig:eccentricity_tides}%
\end{figure}

Owing to the \smc{radial behaviour} $\Ulms \propto \left( \rr / \smaxis \right)^\llat$, terms of degrees $\llat > 2$ are negligible with respect to second order components when the star-planet distance far exceeds the radius of the planet \smc{\citep[][]{MLP2009}}. They can thus be ignored in the framework of this study. Quadrupolar terms ($\llat = \mm =2$) are associated with the eccentricity frequencies

\begin{equation}
\ftidequads = 2 \spinrate - \ss \norb, 
\label{ftide2s}
\end{equation}

\noindent and expressed as 

\begin{equation}
\Utidequad = \sqrt{\frac{3}{5}} \frac{\Ggrav \Mstar}{\smaxis} \left( \frac{\Rpla}{\smaxis} \right)^2 \Hansen{\ss}{-3}{2} \left( \ecc \right),
\label{U22s}
\end{equation}

\noindent with $\ss \in \Zset $. In the case of a circular orbit ($\ecc = 0$), the tidal potential reduces to the semidiurnal component 

\begin{equation}
\Utidelms{2}{2}{2} = \sqrt{\frac{3}{5}} \frac{\Ggrav \Mstar}{a} \left( \frac{\Rpla}{a} \right)^2 .
\label{U222}
\end{equation}

At small eccentricities ($\ecc \ll 1$), eccentricity tides are predominantly forced by terms associated with $\ss = 1$ (westward propagating potential) and $\ss = 3$ (eastward propagating potential), given by \padc{\citep[e.g.][Table~1]{Ogilvie2014}}

\begin{align}
\label{U221}
& \Utidelms{2}{2}{1} = - \frac{\ecc}{2} \sqrt{\frac{3}{5}} \frac{\Ggrav \Mstar}{a} \left( \frac{\Rpla}{a} \right)^2 , \\
\label{U223}
& \Utidelms{2}{2}{3} = \frac{7 \ecc}{2} \sqrt{\frac{3}{5}} \frac{\Ggrav \Mstar}{a} \left( \frac{\Rpla}{a} \right)^2,
\end{align}

\noindent which are associated with the tidal frequencies $\sigma_{2,1} = 2 \spinrate - \norb $ and $\sigma_{2,3} = 2 \spinrate - 3 \norb$, respectively. At high eccentricities ($\ecc \lesssim 1$), the order $\ss$ of the predominant component increases with $\ecc$, while the spectrum of forcing terms widens \smc{\citep[see][Fig.~3]{Ogilvie2014}}. Besides, we note that terms associated with negative $\ss$ are always negligible compared to those associated with positive~$\ss$.

The transition between low- and high-eccentricity regimes occurs for $\Utidelms{2}{2}{3} \approx \Utidelms{2}{2}{2}$, that is at the \smc{transition} eccentricity \padc{$\ecctrans \approx 0.243$ in the second order approximation in $\ecc$ \citep[see Table~C.1 in][]{Correia2014}}, independently from the system parameters. In the absence of resonances, the semidiurnal tide remains predominant as long as $\ecc \lesssim \ecctrans$, meaning that the critical eccentricity beyond which asynchronous final rotation states of equilibrium can exist is $\eccasync = \ecctrans$ in this case. As shown in the following, $\eccasync$ can be strongly decreased by the resonances associated with the oceanic tidal response.

Figure~\ref{fig:eccentricity_tides} illustrates the desynchronising mechanism of eccentricity tides, through the action on rotation of the components of the tidal gravitational potential given by \eqsto{U222}{U223}. The semidiurnal component, given by \eq{U222}, is locked on the average direction of the satellite. The two other components, given by \eqs{U221}{U223} and induced by eccentricity, are traveling clockwise and counterclockwise, respectively \citep[see][Fig.~3]{Greenberg2009}. Dissipative processes, such as viscous friction, induce an angular lag between the tidal bulge and the direction of the star. 

The configuration shown by \fig{fig:eccentricity_tides} \smc{is super-synchronous and} corresponds to $\norb \leq \spinrate < \left( 3/2 \right) \norb$. In this case, the planet is torqued towards the spin-orbit synchronous rotation by the semidiurnal component and the eccentricity component of degree $\ss = 1$, which results from the westward displacement of the star in the reference frame of the planet in the vicinity of the \smc{apoapsis}. Conversely, its rotation is accelerated by the component of degree $\ss = 3$ resulting from the westward displacement of the star in the vicinity of the periastron. In the figure, this component is assumed to be predominant. Hence the spin rotation of the planet is accelerated in the vicinity of the periastron. 

In the case of sub-synchronous rotation ($\spinrate \leq \norb$), the eccentricity component of degree $\ss = 3$ torques the planet towards synchronisation, as the semidiurnal one does. The desynchronising component is that of degree $\ss = 1$ in this case, from the moment that $\spinrate > \left( 1/2 \right) \norb$. The role played the two predominating eccentricity components in the approximation of small eccentricities may be generalised to other ones ($\ss \in \Zset$). The degree-$\ss$ eccentricity component tends to drive the planet away from synchronisation provided that $\spinrate$ satisfies the conditions

\begin{equation}
\begin{array}{ll}
	\displaystyle \frac{\ss}{2} \norb < \spinrate \leq  \norb & \mbox{if} \ \ss \leq 1, \\
	\displaystyle \norb \leq  \spinrate < \frac{\ss}{2} \norb & \mbox{if} \ \ss \geq 3. 
\end{array}
\label{condition_desynch}
\end{equation}

\noindent Otherwise, the component pushes the planet towards synchronisation. The global effect resulting from the combinations of tidal components depends on the behaviour of \smc{the} solid part and \smc{the} oceanic layer coupled together. Particularly, it is sensitive to resonances proper to the oceanic tidal response \citep[][]{Webb1980,ADLML2018}. \smc{These resonances induce a strong dependence of the torque on the tidal frequency, which is not the case of the behaviour of the solid part \citep[e.g.][]{EL2007}}. We thus characterise the global tidal response of the planet in the following.


\subsection{Tidal response of the solid part}
\label{ssec:solid_part}

As a first step, let us focus on the tidal response of the solid part of the planet. In the general case, the distortion associated with the gravitational tidal forcing of the host star can be considered as a small perturbation, meaning that variations of physical quantities characterising the planet are small with respect to \smc{the hydrostatic background}. This is the framework where the linear tidal theory used in this work can be applied. 

Computing the response of the solid part requires to preliminary assume a rheology for the material that composes it. Hence, following \cite{Efroimsky2012}, we consider that the solid part behaves as an isotropic visco-elastic body of unrelaxed shear modulus $\mupla$. This body undergoes a time-periodic tidal force of frequency $\ftide$. Thus, in the linear approximation, any quantity $\quantiR$ describing the distortion can be written as $\quantiR = \real{\quantisig \expo{\inumber \ftide \time }}$, where $\quantisig$ stands for the associated complex quantity, which depends on the forcing frequency. \ebc{We} introduce the stress tensor $\stresstens$ and the strain tensor in the linear approximation

\begin{equation}
\straintens = \frac{1}{2} \left[ \grad \xivect + \transp{\left( \grad \xivect \right)} \right],
\label{strain_stress}
\end{equation}

\noindent the notation $\xivect$ designating the displacement vector. The rheology of the solid is determined by the relationship between $\stresstens$ and $\straintens$. 

To model this relationship, we make use of \ebc{a generalised version of} Hooke's law, which is a linear constitutive law that governs the distortion of isotropic elastic materials as long as this distortion does not exceed the elastic limit of the material. Neglecting compressibility, we only take the deviatoric stresses and strain into consideration. Therefore, Hooke's law reduces in the present case to 

\begin{equation}
\stresstens = 2 \Cmu \straintens,
\label{Hooke}
\end{equation}
 
\noindent where $\Cmu$ is the complex shear modulus accounting for the material rheology. We note that the imaginary part of $\Cmu$ characterises the anelasticity of the material due to the action of internal dissipative processes, such as friction. In the case of a purely elastic body, $\Cmu$ would be real. 

In this study, we opt for an Andrade rheology \citep[see][for details]{Andrade1910,CR2011,Efroimsky2012}, which allows us to write $\Cmu$ as an explicit function of the forcing frequency,

\begin{equation}
\Cmu = \frac{\mupla}{1 + \left( \inumber \ftide \tauA \right)^{- \alphaA} \GammaF \left( 1 + \alphaA \right) + \left( \inumber \ftide \tauM \right)^{-1}},
\label{Andrade_mu}
\end{equation}

\noindent where we have introduced the gamma function $\GammaF$ \citep[][]{AS1972}, the dimensionless rheological exponent $\alphaA$ \citep[the values of $\alphaA$ are determined experimentally and typically fall within the interval $0.2 - 0.4$ for olivine-rich rocks, see][]{Castelnau2008}, the Maxwell relaxation time $\tauM$ associated with the viscous response of the material \citep[$\tauM$ is defined by the ratio of viscosity $\viscosity$ to unrelaxed rigidity, $\tauM \define \viscosity / \mupla$, see e.g.][]{CR2011}, and the Andrade -- or anelastic -- time $\tauA$ related to anelasticity. This latter parameter is the characteristic timescale associated with the material creep.

\begin{figure}[t]
   \centering
   \includegraphics[width=0.48\textwidth,trim = 2.4cm 2.5cm 1.5cm 2cm,clip]{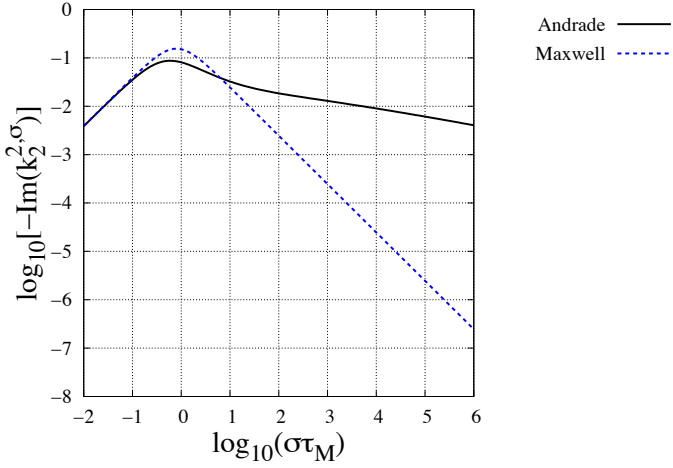}
   \caption{Andrade (solid black line) and Maxwell (blue dotted line) models in the frequency domain. The negative imaginary part of the complex quadrupole Love number $\kquad$ (see \eq{love_solid}) characterising a solid planet is plotted as a function of the normalised tidal frequency $\ftide \tauM$ in logarithmic scales. We use the values of the Andrade parameters given by \tab{tab:andrade_para} for the Earth.}
       \label{fig:andrade_maxwell}%
\end{figure}

By setting $\tauA = + \infty$, one retrieves the standard Maxwell viscoelastic rheology \citep[][]{Greenberg2009,Efroimsky2012,Correia2014}. In this case, the behaviour of the solid part follows two asymptotic regimes defined by the hierarchy between the tidal period $\tautide$ and the Maxwell time. In the zero-frequency limit ($\tautide \gg \tauM$), the response is dominated by viscous friction, leading to the purely imaginary complex shear modulus $\Cmu \sim  \inumber \mupla \tauM \ftide$ \rec{(we remind here that $\ftide$ designates the forcing frequency of the tidal mode)}. In the high-frequency regime ($\tautide \ll \tauM$), the behaviour of the solid part tends to be purely elastic, with $\Cmu \sim \mupla$. \smc{In the case of telluric planets}, the order of magnitude of the Maxwell time is of several centuries \citep[typically $\tauM \sim 10^{4}-10^{5}$ \days; for the Earth $\tauM$ is about 500~years, see e.g.][]{Efroimsky2012}, that is far greater than usual tidal periods ($\tautide \sim 10^0 - 10^2$ \days). As a consequence, the imaginary part of $\Cmu$ drops rapidly as $\ftide$ increases in the high-frequency regime, which leads to underestimate \ebc{the energy tidally dissipated in the planet interior (this effect has repercussions on the frequency-dependence of the imaginary part of Love numbers, as shown by \fig{fig:andrade_maxwell} in the following)}. 

The main interest of the Andrade model is to improve this behaviour by providing more realistic orders of magnitude of the energy tidally dissipated within planetary interiors with a minimal number of additional parameters, namely $\alphaA$ and $\tauA$. In the Andrade model, the decay of the anelastic component with the forcing frequency is attenuated with respect to that described by the Maxwell model, the slope of the decay being determined by the rheological parameter $\alphaA$ (this is illustrated by \fig{fig:andrade_maxwell}, discussed further, where the imaginary part of the quadrupole Love number is plotted as a function of the tidal frequency). \rec{This description of the tidal behaviour may be refined using more sophisticated rheological models like the Sundberg-Cooper rheology, which implies a larger tidal dissipation around a critical temperature and frequency \citep[][]{RH2018}.}

As discussed by early studies \citep[][]{CR2011,Efroimsky2012}, estimating the Andrade time is far from being trivial. Particularly, there is no reason for $\tauA$ and $\tauM$ to be comparable in the general case since they are related to two different physical mechanisms. The order of magnitude of $\tauA$ can be derived from microphysics through the study of the propagation of seismic waves, as done for instance in the case of the Earth mantle \citep[e.g.][]{KS1990,Tan2001,Jackson2002}. For more details, we refer the reader to \cite{CR2011}, where different sets of data are compared in order to determine the Andrade parameter $\betaA = \mupla^{-1} \tauA^{- \alphaA}$.

For the purpose of this work, we will use in calculations the effective values of parameters derived from the \smc{numerical} integration of the equations of the elasto-gravitational theory \citep[][]{Love1911,TS1972} with realistic radial profiles of background quantities (see \tab{tab:andrade_para}). The tidal model used to obtain these values is described in \cite{Tobie2005}, and treats the deformation of a spherically symmetric, non-rotating, and elastic body, for which the momentum, Poisson and mass conservation equations are solved by means of the classical propagator matrix method \citep[e.g.][]{SV2004} over a sampling of the tidal frequency.

 \smc{The radial profiles of background quantities used as inputs of the tidal model are computed with a code -- detailed in \cite{Sotin2007} -- that solves numerically the hydrostatic balance across the radial direction assuming an equation of state and a composition for the planet. In these calculations of the internal structure, the planet is considered as a dry rocky body owing to the negligible impact of the thin oceanic shell on the parameters of the solid visco-elastic tidal response. In the case of a thick oceanic layer, these parameters should be derived by taking the ocean into account as this later can no longer be neglected \citep[][]{Dermott1979,Remus2012}. }

\begin{table}[h]
\centering
 \textsf{\caption{\label{tab:andrade_para} Parameters of the Andrade model derived from the elasto-gravitational theory \citep[][]{Tobie2005,Breton2018}.  }}
\begin{small}
    \begin{tabular}{ l l l}
      \hline
      \hline
      \textsc{Parameter} & \textsc{Earth}  & \textsc{Super-Earth}  \\ 
      \hline 
      Planet mass $\Mpla$ ($\Mearth$) & 1.0 & 10  \\
      Planet radius $\Rpla$ ($\Rearth$) & 1.0 & 1.86 \\
      Shear modulus $\mupla$ (GPa) & 14.6 & 75.2 \\
      Rheological exponent $\alphaA$ & 0.25 & 0.25 \\
      Maxwell time $\tauM$ (yr) & $741$ & $1.87\expdec{5}$ \\
      Andrade time $\tauA$ (yr) & $2.19 \expdec{4}$ & $1.66\expdec{6}$ \\
      \hline 
    \end{tabular}
\begin{flushleft}
\textbf{Notes.}  The subscript $\iearth$ refers to the Earth.
\end{flushleft}
\end{small}
 \end{table}

Values of Andrade parameters are obtained by fitting the Andrade model to the frequency-spectra \smc{derived from the calculations of tidal oscillations} \ebc{using the method of \cite{Tobie2005}}. These values may be regarded as the effective parameters of an \smc{equivalent body homogeneous in density presenting} the same frequency behaviour as the studied body of realistic internal structure. An application of this method to the TRAPPIST-1 system may be found in \cite{Breton2018}, and will be detailed in a forthcoming article \ebc{(Bolmont et al. 2019, in preparation)}. 


The variation of mass distribution associated with the tidal distortion is described by Love numbers. In the linear approach, each degree-$\llat$ mode of the expansion in spherical harmonics (see \eq{Utide}) has its associated Love numbers, denoted by $\kl$, $\hl$, $\kloadl$, and $\hloadl$ in the present work. The first two parameters, $\kl$ and $\hl$, stand for the tidal gravitational and displacement Love numbers, respectively. They characterise the variations of the self-gravitational potential and vertical displacement induced by the internal gravitational tidal forcing at the surface of the solid body. 

The notations $\kloadl$ and $\hloadl$ designate the load Love numbers for the gravitational potential and vertical displacement of the surface, respectively. These numbers characterise the response of the planet to the gravitational force and surface pressure induced by the variation of mass distribution of a thin external fluid layer. In the case of a uniform interior, the degree-$\llat$ Love numbers are given by \citep[e.g.][]{MM1960}

\begin{equation}
\left\{ \kl , \hl , \kloadl, \hloadl \right\} = \frac{1}{1+\Cmul} \left\{ \frac{3}{2 \left( \llat - 1 \right)} , \frac{2 \llat +1}{2 \left( \llat - 1 \right)} , - 1 , -  \frac{2 \llat +1}{3} \right\},
\label{love_solid}
\end{equation}

\noindent where we have introduced the dimensionless effective rigidity 

\begin{equation}
\Cmul  \define \Al \frac{\Cmu}{\mupla}, 
\label{Cmul}
\end{equation}

\noindent with 

\begin{equation}
\Al \define \frac{4 \left( 2 \llat^2 + 4 \llat + 3 \right) \pi \Rcore^4 \mupla}{3 \llat \Ggrav \Mcore^2}.
\label{Al_Cmu}
\end{equation}

The contribution of the degree-$\llat$ component to the tidal torque exerted on the solid part is proportional to $\imag{\kl}$ \citep[e.g.][]{Makarov2012}. For comparison, the imaginary part of the quadrupole Love number $\kquad$ is plotted in \fig{fig:andrade_maxwell} for the Andrade and Maxwell rheology \citep[e.g.][]{ME2013,Correia2014} as a function of the normalised tidal frequency $\ftide \tauM$. The behaviour of the two models does not differ in the low-frequency regime, where $\imag{\kquad}$ increases with $\ftide$ linearly. A maximum is reached for $\tautide \sim \tauM$. In the high-frequency regime, the torque decays as $\ftide $ increases, but the Andrade model is characterised by a weaker slope than the Maxwell model, as discussed above. \smc{Basically, the slope of the imaginary part of the complex degree-2 Love number is $-0.25$ per decade for the Andrade model ($\alphaA = 0.25$), and $-1$ per decade for the Maxwell model (\fig{fig:andrade_maxwell}).}

\subsection{Tidal response of the ocean}
\label{sec:ocean_response}

We now consider the tidal response of the ocean. Similarly as the solid part, the oceanic layer undergoes the gravitational tidal forcing of the perturber. It is also affected by the distortion of the solid part, which induces both a variation of the planet self-gravitational potential and a displacement of the oceanic floor, denoted by $\xibot$. We take this solid-ocean coupling into account by following along the line by \cite{Matsuyama2014}. In this \ebc{earlier} work, the equations of dynamics are written in the shallow water approximation, the fluid being supposed incompressible. Mean flows are ignored (the whole planet rotates as a solid, ocean included), and the fluctuations of quantities associated with the perturbation are assumed to be small with respect to background quantities, which is the postulate required by the linear approach. 

The perturbation is thus described by \smc{an} horizontal velocity field $\Vvect = \Vtheta \etheta + \Vphi \ephi$ and the variation of the oceanic depth $\xilayer = \xisurf - \xibot$, the notation $\xisurf$ designating the oceanic surface displacement \ebc{(see \fig{fig:setup})}. In the \smc{frame} rotating with the planet ($\framerot$), the dynamics of the oceanic tidal response is governed by the momentum equation \citep[][]{Matsuyama2014},

\begin{equation}
\dd{\Vvect}{\time} + 2 \spinvect \vprod \Vvect + \fdrag \Vvect = \gradh \Ftide,
\label{momentum_eq}
\end{equation}

\noindent and the equation of mass conservation,

\begin{equation}
\dd{\xilayer}{\time} + \Hoc \divh  \Vvect  = 0,
\label{mass_eq}
\end{equation}

\noindent where we have introduced the horizontal gradient operator in spherical coordinates,

\begin{equation}
\gradh \define \Rpla^{-1} \left[ \etheta  \dd{}{\col} + \ephi \left( \sin \col \right)^{-1} \dd{}{\lon} \right],
\label{gradh}
\end{equation}

\noindent the horizontal component of the velocity divergence, 

\begin{equation}
\divh \Vvect \define \left( \Rpla \sin \col \right)^{-1} \left[ \dd{\left( \sin \col \Vtheta \right)}{\col} + \dd{\Vphi}{\lon}  \right],
\label{divh}
\end{equation}

\noindent the Rayleigh drag frequency $\fdrag$ characterising the  \smc{friction with the oceanic floor and the conversion of barotropic tidal flows into internal gravity waves \citep[][]{Wunsch1975}}, and a perturbed potential, denoted by $\Ftide = \Ftide \left( \time , \col , \lon \right) $ and defined further, which encompasses both the tidal gravitational forcing and coupling with the solid part. We note that the radial component of the Coriolis acceleration $- 2 \spinrate \Vphi \er$ is neglected in the shallow water approximation since fluid motions are supposed to be dominated by horizontal flows. As a consequence, the Coriolis acceleration in \eq{momentum_eq} is simply given by  $2 \spinvect \vprod \Vvect = 2 \spinrate \cos \col \left( \Vtheta \ephi - \Vphi \etheta \right) $.

In the linear approximation, the response of a given mode is proportional to the corresponding component of the forcing. It follows that $\Ftide$, $\Vvect$, and $\xilayer$ can be expanded in Fourier series of $ \time$ and $\lon$ similarly as the tidal gravitational potential (\eq{Utide}),

\begin{equation}
\begin{array}{ll}
 \displaystyle   \Ftide  = \sum_{\msigma}  \Ftidems \left( \col \right) \exptphi , &
 \displaystyle   \Vvect = \sum_{\msigma}  \Vvectms \left( \col \right) \exptphi , \\
 \displaystyle   \xilayer = \sum_{\msigma}  \xilayerms \left( \col \right) \exptphi . &
\end{array}
\label{Fourier_var}
\end{equation}

The Fourier coefficients $\Ftidems$, $\Vvectms$, and $\xilayerms$ may themselves be expanded as series of the normalised associated Legendre functions $\Plmnorm$, 

\begin{equation}
\begin{array}{ll}
 \displaystyle   \Ftidems  = \sum_{\llat \geq \mm}  \Ftidelms \Plmnorm \left( \cos \col \right) , &
 \displaystyle   \Vvectms  = \sum_{\llat \geq \mm}  \Vvectlms \Plmnorm \left( \cos \col \right) , \\
 \displaystyle   \xilayerms  = \sum_{\llat \geq \mm}  \xilayerlms \Plmnorm \left( \cos \col \right) , &
\end{array}
\label{legendre_series}
\end{equation}

\noindent where $\Ftidelms$, $\Vvectlms$ and $\xilayerlms$ designate the degree-$\llat$ components of the expansion. 

By introducing the complex tidal frequency $\ftidedrag$ and spin parameter $\spinpar$, defined by

\begin{equation}
\begin{array}{lrl}
 \ftidedrag \define \ftide - \inumber \fdrag, & \mbox{and} & \displaystyle \spinpar \define \frac{2 \spinrate}{\ftidedrag} ,
\end{array}
\label{freq_spinpar}
\end{equation}

\noindent and substituting perturbed quantities by their Fourier expansions (\eq{Fourier_var}) in \eq{momentum_eq}, we may express the components of the horizontal velocity field as functions of $\Ftide$,

\begin{align}
& \Vthetams = - \frac{\inumber}{\Rpla \ftidedrag \left( 1 - \spinpar^2 \cos^2 \col \right)} \left( \DD{}{\col} + \mm \spinpar \cot \col \right) \Ftidems ,  \\
& \Vphims = \frac{1}{\Rpla \ftidedrag \left( 1 - \spinpar^2 \cos^2 \col \right)} \left( \spinpar \cos \col \DD{}{\col} + \frac{\mm}{\sin \col} \right) \Ftidems. 
\end{align}

Then, substituting $\Vthetams$ and $\Vphims$ by the above expressions in \eq{mass_eq}, we end up with

\begin{equation}
\Laplace \left( \Ftidems \right) = \frac{\Rpla^2 \ftide \ftidedrag}{\Hoc} \xilayerms, 
\label{Laplace_eq}
\end{equation}

\noindent the notation $\Laplace$ referring to the Laplace's tidal operator, defined by \citep[e.g.][]{LS1997}

\begin{align}
\Laplace \define & \frac{1}{\sin \col} \DD{}{\col} \left( \frac{\sin \col}{1 - \spinpar^2 \cos^2 \col} \DD{}{\col} \right) \\
  & - \frac{1}{1 - \spinpar^2 \cos^2 \col} \left( \mm \spinpar \frac{1 + \spinpar^2 \cos^2 \col}{1 - \spinpar^2 \cos^2 \col} + \frac{\mm^2}{\sin^2 \col} \right). \nonumber
\end{align}

We recognise in \eq{Laplace_eq} the Laplace's Tidal Equation, which determines the horizontal structure of the fluid tidal response \citep[e.g.][]{LC1969}. If we assume that the Fourier components $\Ftidems$ and $\xilayerms$ may be written as 

\begin{equation}
\begin{array}{ll}
 \displaystyle  \Ftidems  = \sum_{\nn} \Ftiden  \Thetan \left( \col \right), & 
 \displaystyle  \xilayerms  = \sum_{\nn} \xilayern  \Thetan \left( \col \right),
\end{array}
\label{Hough_var}
\end{equation}

\noindent the parameter $\nn$ being an integer, and consider that the \rec{latitudinal} functions $\Thetan$ are bounded at the poles, \eq{Laplace_eq} defines an eigenfunction-eigenvalue problem. Hence, the set of eigenfunctions $\left\{ \Thetan \right\}$, called Hough functions after Hough's work \citep[][]{Hough1898}, is associated with a set of eigenvalues $\left\{ \Lambdan \right\}$ through the relationship 

\begin{equation}
\Laplace \Thetan = - \Lambdan \Thetan. 
\label{Laplace}
\end{equation}

One shall also introduce here the corresponding complex equivalent depths $\heqn$, which are defined by analogy with real equivalent depths \citep[e.g.][]{Taylor1936,LC1969}, by

\begin{equation}
\heqn \define \frac{\Rpla^2 \ftide \ftidedrag}{\Lambdan \ggravi}. 
\label{heqn}
\end{equation}

The Hough function $\Thetan$ may be written as a combination of the normalised associated Legendre functions,

\begin{equation}
\Thetan \left( \col \right) = \sum_{\llat \geq \mm} \Anl \Plmnorm \left( \cos \col \right).  
\label{Hough_Plm}
\end{equation}

\noindent Conversely, the normalised associated Legendre function $\Plmnorm$ may be written as

\begin{equation}
\Plmnorm \left( \cos \theta \right) = \sum_{\nn} \Bln \Thetan \left( \col \right). 
\label{Plm_Hough}
\end{equation}

\noindent In the preceding expressions, the $\Anl$ and $\Bln$ are complex overlap coefficients. They are computed in the meantime as the eigenvalues $\Lambdan$ using the standard method detailed by \cite{Wang2016}, which is based on series of the normalised associated Legendre functions. Besides, we introduce here the overlap coefficients 

\begin{equation}
\Clnk \define \Bkn \Anl,
\label{Clnk}
\end{equation}

\noindent which will be used further to weight the degree-$\nn$ components of the oceanic tidal torque.


In the absence of friction ($\fdrag = 0$), $\heqn = \Hoc$ corresponds to the resonant configuration where the phase velocity of the forced degree-$\nn$ mode equalises the characteristic propagation velocity of large-wavelength surface gravity waves. The non-frictional case has been thoroughly discussed in early studies \citep[e.g.][]{LH1968,LC1969,LS1997} so that we do not need to enter here into details. We just recall the main aspects of this regime. 

When $\fdrag = 0$, the complex quantities defined by \eq{freq_spinpar} reduce to $\ftidedrag = \ftide$ and $\spinpar = 2 \spinrate / \ftide$, which are the real forcing frequency and spin parameter usually met in literature. The set of Hough functions divides into two families. A first family, refereed to as gravity modes or g-modes \citep[see e.g.][]{LS1997}, is defined for $\spinpar \in \Rset$. This family corresponds to the ordinary spherical modes modulated by rotation and naturally reduce to the set of the associated Legendre functions in the \smc{non-rotating} case. For $\abs{\spinpar} > 1 $, an other family of functions appear. These functions are generally called rotational modes or r-modes, and develop outside of the equatorial band where g-modes are confined by Coriolis effects. 

When a Rayleigh drag is introduced, the sets of Hough functions and associated eigenvalues become complex in the general case \citep[][]{Volland1974a,Volland1974b,ADLML2018}. The real case discussed above corresponds to the asymptotic regime where $\abs{\ftide} \gg \fdrag$. For $\abs{\ftide} \lesssim \fdrag $, the friction affects the behaviour of tidal modes. While the ratios $\abs{\ftide} / \fdrag$ and $\abs{2 \spinrate} / \fdrag$ decay, g-modes and r-modes tend to merge together and converge towards the functions of the \smc{non-rotating} case, namely the associated Legendre functions. Hence, in the asymptotic limit ($\abs{\ftide} / \fdrag \rightarrow 0$), the friction is strong enough to annihilate the distortion caused by \smc{the} Coriolis effects. \padc{From a mathematical point of view,} the introduction of friction regularises the solution in the zero-frequency limit. In the absence of dissipation, the number of Hough modes necessary to approximate the solution given by \eq{Hough_var} diverges as $ \abs{\spinpar} \rightarrow 0 $ \smc{(i.e. $\spinrate \rightarrow 0$ or $\ftide \rightarrow + \infty$)}. This is no longer the case when a drag is taken into account.

We now come back to $\Ftide$, which still has to be defined. Following \cite{Matsuyama2014}, we introduce the tilt factors associated with the tidal gravitational forcing of the perturber and the distortion of the oceanic layer, denoted by $\tiltU$ and $\tiltxi$, respectively. \smc{In the framework of the thin shell approximation\footnote{\smc{In the case of a thick oceanic layer, the ocean depth intervenes in the expressions of tilt factors as well as in those of the solid Love numbers \citep[][]{Dermott1979,Remus2012}.}},} these factors are defined by 

\begin{align}
\label{tiltU}
& \tiltU \define 1 + \kl - \hl,  \\
\label{tiltxi}
& \tiltxi \define 1 - \left( 1 + \kloadl - \hloadl \right) \frac{3 \rhooc}{\left( 2 \llat + 1 \right) \rhocore}.
\end{align}

They characterise the effective forcing of the oceanic layer including the effects of the tidal distortion of the solid part, which undergoes both the tidal gravitational forcing and the ocean loading. The degree-$\llat$ component of the potential $\Ftidems$ is thus defined as 

\begin{equation}
\Ftidems = \sum_{\llat \geq \mm} \left( - \ggravi \tiltxi \xilayerlms + \tiltU \Ulmsig \right) \Plmnorm \left( \cos \col \right). 
\end{equation}

By expanding $\Ftidems$ and $\xilayerms$ in series of Hough functions, we transform \eq{Laplace_eq} into 

\begin{equation}
- \Lambdan \Ftiden = \frac{\Rpla^2 \ftide}{\Hoc} \xilayern,
\end{equation}
 
\noindent the $\Ftiden$ being given by

\begin{equation}
\Ftiden = \sum_{\llat \geq \mm} \Bln \left(  - \ggravi \tiltxi \sum_{\kk} \Akl \xilayerk + \tiltU \Ulmsig  \right). 
\end{equation}

In practice, the series of Hough functions and associated Legendre functions given by \eqs{legendre_series}{Hough_var} are truncated in numerical calculations. We denote by $\Nhough$ the number of functions used to approximate series, which is arbitrarily chosen large enough to be associated with negligible overlap coefficients. It follows that the $\xilayern$ of \eq{Hough_var} are the solutions of an algebraic system of the form

\begin{equation}
\begin{bmatrix}
\ftide \ftidedrag - \sig{1}{1}^2 & - \sig{1}{\nn}^2 & - \sig{1}{\Nhough}^2 \\
- \sig{\nn}{1}^2 & \ftide \ftidedrag - \sig{\nn}{\nn}^2 & - \sig{\nn}{\Nhough}^2 \\
- \sig{\Nhough}{1}^2 & - \sig{\Nhough}{\nn}^2 & \ftide \ftidedrag - \sig{\Nhough}{\Nhough}^2
\end{bmatrix}
\begin{bmatrix}
\xilayerind{1} \\
\xilayern \\
\xilayerind{\Nhough}
\end{bmatrix}
=
\begin{bmatrix}
\Fforceind{1} \\
\Fforcen \\
\Fforceind{\Nhough}
\end{bmatrix}.
\label{algebraic_system}
\end{equation}

In \eq{algebraic_system}, the $\sig{\nn}{\kk}$ are the complex characteristic frequencies defined by 

\begin{equation}
\sig{\nn}{\kk} \define \sqrt{\ggravi \Hoc \khori^2 \sum_{\llat \geq \mm} \tiltxi \Akl \Bln} ,
\label{fnk}
\end{equation}
 
\noindent where \smc{$\khori \define \sqrt{\Lambdan} / \Rpla$} designates the horizontal wavenumber of the degree-$\nn$ mode. The components $\Fforcen$ of the force vector in \eq{algebraic_system} are defined by

\begin{equation}
\Fforcen \define - \frac{\Hoc \Lambdan}{\Rpla^2} \sum_{\llat \geq \mm} \Bln \tiltU \Ulmsig. 
\label{Fforcen}
\end{equation}

\noindent We note that we have kept components of degrees greater than 2 in the preceding expression for the sake of generality. In reality, these components can be neglected since we assumed that $\Rpla \ll \smaxis$.

Finally, the components of the tidal displacement of the ocean surface in the basis of the normalised associated Legendre functions are simply deduced from the $\xilayern$ by using the overlap coefficients introduced in \eq{Hough_Plm},

\begin{equation}
\xilayerlms = \sum_{\nn} \Anl \xilayern. 
\label{xil}
\end{equation}

\section{Tidal torque created by eccentricity tides}
\label{sec:tidal_torque}

The modelling of the solid and oceanic tides detailed in the preceding section allows us to determine the tidal torque exerted on the planet. This is the object of this section. The obtained analytic formulae are used further to compute the evolution of non-synchronised rotation states of equilibrium with the planet eccentricity and ocean depth. 

\subsection{General case}

In the general case, the solid and oceanic tidal responses are coupled to each other by the gravitational force and ocean loading. In the quadrupolar approximation ($\Rpla \ll \smaxis$), the tidal torque is thus given by

\begin{equation}
\torquepla = \frac{3}{2} \Ggrav \Mstar^2 \frac{\Rpla^5}{\smaxis^6} \sum_{\ss = - \infty}^{+ \infty} \left[ \Hansen{\ss}{-3}{2} \left( \ecc \right)   \right]^2 \imag{ \kpquad } ,
\label{torque_pla}
\end{equation}

\noindent where $\kpquad$ is the quadrupolar component of the effective gravitational Love number of the planet, $\kp$, defined \smc{in the thin shell approximation} by 

\begin{equation}
\kp \define \kl + \left( 1 + \kloadl \right) \frac{3 \ggravi}{2 \llat + 1} \frac{\rhooc}{\rhocore} \frac{\xilayerlms}{\Ulmsig}.
\label{kpla}
\end{equation}

\noindent In this expression, the degree-$\llat$ component of the tidal gravitational potential $\Ulmsig$ and variation of ocean thickness $\xilayerlms$ are given by \eqs{Utidelms}{xil}, respectively. 

\subsection{Pure solid tidal response}

In the absence of oceanic layer, $\Rcore = \Rpla$ and the tidal torque exerted on the planet with respect to the spin-axis is simply expressed as \citep[][]{EW2009,Makarov2012,ME2013,Correia2014}

\begin{equation}
\torquesol = \frac{3}{2} \Ggrav \Mstar^2 \frac{\Rpla^5}{\smaxis^6} \sum_{\ss = - \infty}^{+ \infty} \left[ \Hansen{\ss}{-3}{2} \left( \ecc \right)   \right]^2 \imag{ \kquad } . 
\label{torque_sol}
\end{equation}

\noindent In this case, the tidal response of the planet is characterised by the solid gravitational Love number given by \eq{love_solid}. The expression of this parameter in the framework of the Andrade model may be found in -- for instance -- \cite{Makarov2012} (Eq.~6). As discussed in \sect{ssec:solid_part}, the Andrade and Maxwell times generally far exceed typical tidal periods. As a consequence the imaginary part of $\kquad$ may be approximated by 

\begin{equation}
\imag{\kquad} \sim - \sign \left( \ftide \right) \frac{3}{2} \frac{\Aquad}{\left(1+\Aquad \right)^2} \GammaF \left( 1 + \alphaA \right) \sin \left( \frac{\alphaA \pi}{2} \right) \left( \abs{\ftide} \tauA \right)^{-\alphaA}, 
\label{k2sol_asymp}
\end{equation}

\comments{à verifier, mais il manque sans doute un carré au dénominateur et un facteur devant le k2 pour être raccord avec Efroimsky 2012, Eq 70.}

\noindent with 

\begin{equation}
\Aquad \define \frac{ 38 \pi \Rcore^4 \mupla}{ 3 \Ggrav \Mcore^2},
\label{Aquad}
\end{equation}

\noindent except in the zero-frequency limit, where tidal periods become comparable with the Andrade and Maxwell times in order of magnitude. 



\subsection{Pure oceanic tidal response}
\label{ssec:pure_oceanic_response}

The tidal response of the oceanic layer is more complex than that of the solid part. To examine it, we have to consider the particular case where the coupling with the solid part vanishes. In this simplified case, the rigidity of the solid body is supposed to be infinite ($\mupla \rightarrow + \infty$), so that the Love numbers defined by \eq{love_solid} annihilate. Besides, the Cowling approximation \citep[e.g.][]{Cowling1941,Unno1989} is assumed, which means that the term in $\rhooc/\rhocore$ resulting from the self-attraction of the ocean in $\Ftide$ is ignored. In this case, the oceanic tidal torque may be written similarly as that of the solid part (\eq{torque_sol}),

\begin{equation}
\torqueoc = \frac{3}{2} \Ggrav \Mstar^2 \frac{\Rpla^5}{\smaxis^6} \sum_{\ss = - \infty}^{+ \infty} \left[ \Hansen{\ss}{-3}{2} \left( \ecc \right)   \right]^2 \imag{ \kocquad},
\label{torque_ocean}
\end{equation}

\noindent where $\kocquad$ designates the degree-2 tidal gravitational Love number of the ocean.

Owing to the Cowling approximation, $\sig{\nn}{\kk} = 0$ for $ \nn \neq \kk$. The matrix of the algebraic system given by \eq{algebraic_system} is thus diagonal, which leads to the solutions obtained in early studies \citep[e.g.][]{Webb1980,ADLML2018}, 

\begin{equation}
\xilayern = \frac{\Fforcen}{\left( \ftide - \focnmoins \right) \left( \ftide - \focnplus \right) },
\label{xilayern}
\end{equation}

\noindent the notation $\focnmoins$ and $\focnplus$ designating the complex eigenfrequencies of the degree-$\nn$ Hough mode, defined by \citep[see e.g.][Eq.~(2.12)]{Webb1980}

\begin{equation}
\focnpm \define \inumber \frac{\fdrag}{2} \pm \sqrt{\ggravi \Hoc \khori^2 - \left( \frac{\fdrag}{2} \right)^2}. 
\end{equation}
 
In this case, the quadrupolar Love number introduced in \eq{torque_ocean} is simply expressed as 

\begin{equation}
\kocquad = - \frac{4 \pi}{5} \frac{\Ggrav \Hoc \rhooc}{ \Rpla} \sum_{\nn} \Cnquad \frac{\Lambdan}{\left( \ftide - \focnmoins \right) \left( \ftide - \focnplus \right)}.
\end{equation}


\noindent Using the mean density of the planet ($\rhopla = 3 \Mpla / \left( 4 \pi \Rpla^3 \right) \approx \rhoc $), it may also be put into the form

\begin{equation}
\kocquad = \frac{3}{5} \frac{\rhooc}{\rhopla} \sum_{\nn} \frac{\Cnquad}{1 - \frac{\heqn}{\Hoc}},
\end{equation}


\noindent which highlights the fact that a resonance occurs when the equivalent depth of a given mode $\heqn = \Rpla^2 \ftide^2 / \left( \Lambdan \ggravi \right)$ equalises the ocean depth in the absence of friction.

The friction of tidal flows with the oceanic floor affects both the overlap coefficients $\Cnquad$ and the equivalent depths of Hough modes. Since these parameters are complex in the general case, characterising the dependence of the tidal torque on $\fdrag$ is not straightforward. However, this may be done in the quasi-adiabatic asymptotic regime ($\fdrag \ll \abs{\ftide}$), where the imaginary part of $\Cnquad$ is negligible, and for g-modes, as these laters are associated with positive $\Lambdan$. 

By introducing the characteristic frequency of the degree-$\nn$ surface gravity wave,

\begin{equation}
\focn \define \frac{\sqrt{\Lambdan \ggravi \Hoc}}{\Rpla},
\label{focn}
\end{equation}

\noindent we express the imaginary part of the oceanic Love number as 

\begin{equation}
\imag{\kocquad} = - \frac{3}{5} \frac{\rhooc}{\rhopla} \sum_{\nn} \Cnquad \frac{\frac{\fdrag \ftide}{\focn^2}}{\left( 1 - \frac{\ftide^2}{\focn^2} \right)^2 + \frac{\fdrag^2 \ftide^2}{\focn^4}  }. 
\label{koc_quadiadiab}
\end{equation}

 
The preceding expression characterises the shape of a resonant peak in the frequency-spectra of the tidal torque. The frequency at which the resonant peak of a mode reaches a maximum is expressed as

\begin{equation}
\fpeakn = \pm \frac{\focn}{\sqrt{6}} \left( 2 - \left( \frac{\fdrag}{\focn} \right)^2 + \sqrt{\left[ 2 - \left( \frac{\fdrag}{\focn} \right)^2 \right]^2 + 12 } \right)^{\frac{1}{2}}.
\label{fpeakn1}
\end{equation}

\noindent In the quasi-adiabatic regime ($\fdrag \ll \abs{\ftide} $), it reduces to

\begin{equation}
\fpeakn =  \focn \left[ 1 - \frac{1}{8} \left( \frac{\fdrag}{\focn}  \right)^2 \right].
\label{fpeakn}
\end{equation}

\noindent The corresponding maximum of $\imag{\kocquad} $ for the resonance associated with the degree-$\nn$ Hough mode is denoted by $\left. \imag{\kocquad} \right|_{\ipeak ; \nn}$ and obtained by substituting $\ftide$ by \eq{fpeakn} in \eq{koc_quadiadiab}. If the contributions of non-resonant components are neglected, we thus end up with 

\begin{equation}
\left. \imag{\kocquad} \right|_{\ipeak ; \nn} \approx - \frac{3}{5} \frac{\rhooc}{\rhopla} \Cnquad \left( \frac{\focn}{\fdrag} \right) ,
\label{maxpeak}
\end{equation}


\noindent which shows that, as a first approximation, the maximum of the peak associated with the degree-$\nn$ Hough mode varies with the ocean depth as $\left. \imag{\kocquad} \right|_{\ipeak ; \nn}  \propto \focn \propto \sqrt{\Hoc}$ and is inversely proportional to the typical frequency characterising the Rayleigh drag, \smc{in agreement with scaling laws derived using simplified Cartesian models \citep[e.g.][]{ADMLP2015}}. 

When the frequency associated with an eccentricity term becomes equal to the eigenfrequency of the resonant mode, this eccentricity term is enhanced by a factor corresponding to the ratio of the peak maximal value over the level of the non-resonant background. It can thus generate a torque strong enough to compensate the predominating contribution of the semidiurnal component in the low-eccentricity regime. 

\begin{figure}[htb]
   \centering
   \includegraphics[width=0.48\textwidth,trim = 1.5cm 0.5cm 0.5cm 0cm,clip]{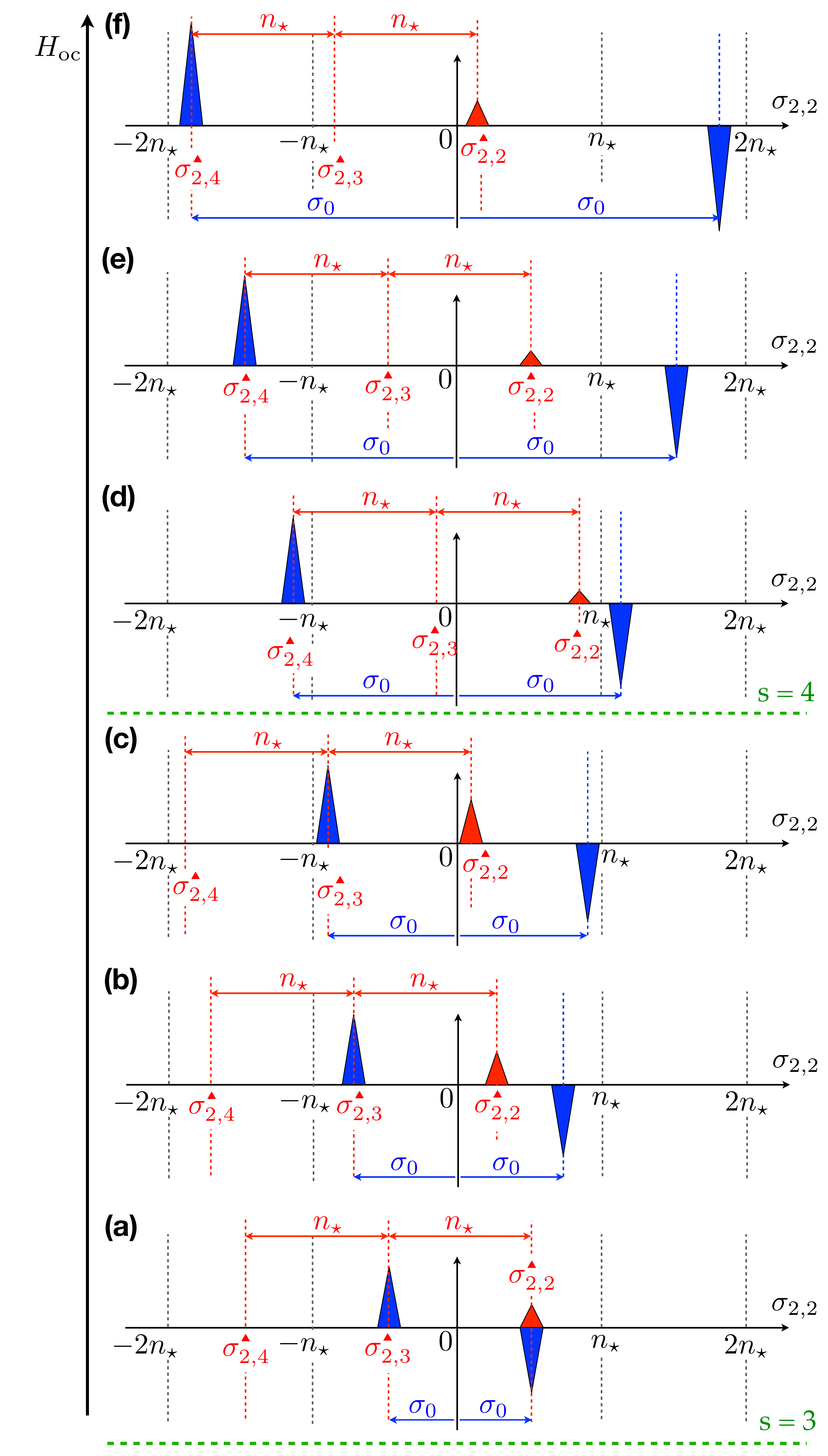}
   \caption{Positive tidal torque generated by oceanic eccentricity tides in the super-synchronous frequency range ($\spinrate > \norb$) for increasing oceanic depths (a-f). An oceanic mode is characterised by a resonance of eigenfrequency $\focn$ (blue peaks). We focus here on the degree-0 mode, of frequency $\foco$, \jlc{in the non-rotating approximation (Coriolis effects are ignored)}. An eccentricity term of frequency $\ftidequads = 2 \spinrate - \ss \norb$ generates a positive torque, which is enhanced by the resonance when $ \left| \ftidequads \right| = \foco $ (red peak). The semidiurnal ($\ftide_{2,2}$) and first eccentricity frequencies ($\ftide_{2,3}$ and $\ftide_{2,4}$) \jlc{associated with} the peak thus created are superscripted \jlc{by the symbol $\blacktriangleup$. The peak raised by the eccentricity tide occurs when $\ftide_{2,2} = \ftide_{2,2}^\blacktriangleup$, which also corresponds to $\ftide_{2,3} = \ftide_{2,3}^\blacktriangleup $ and $\ftide_{2,4} = \ftide_{2,4}^\blacktriangleup $. The direction of peaks indicates the sign of the tidal torque. In the super-synchronous regime ($\ftide_{2,2}>0$), downward blue peaks tend to drive the planet towards tidal locking in spin-orbit synchronous rotation ($\ftide_{2,2} = 0$), while upward red peaks tend to drive it away from this state of equilibrium. } }
       \label{fig:pics_ecc}%
\end{figure}

This mechanism is illustrated by \fig{fig:pics_ecc}, which represents the main features of the tidal torque as functions of the semidiurnal frequency, $\ftidequadsemid = 2 \left( \spinrate - \norb \right)$, in the super-synchronous regime ($\spinrate > \norb$). In this figure, the tidal response of the solid part is ignored to focus on the oceanic tidal torque, which is reduced to the contribution of the degree-0 Hough mode for simplification. For a pedagogical purpose, we ignore the effect of rotation, which induces Rossby modes through the Coriolis terms in the momentum equation. \ebc{This allows us to assume that the overlap coefficients and eigenvalues associated with the considered Hough mode do not vary with the tidal frequency, for a fixed ocean depth.}

\ebc{The simplified oceanic tidal response is thus solely due to the \smc{prograde and retrograde} degree-0 Hough modes. The corresponding frequencies of these resonant modes are $\pm \foco $ and the resulting tidal torque contributions, \smc{which both drive the system towards synchronism, are} represented schematically by blue cones in \fig{fig:pics_ecc}. As \eq{focn} shows, the \padc{frequencies of these intrinsic modes of the ocean} increase with the ocean depth $\Hoc$. \smc{Indeed, we can see} in \fig{fig:pics_ecc} that the blue cones' positions shift towards the higher frequencies when $\Hoc$ increases from panels a to f.}

\ebc{One of these peaks is located in the super-synchronous frequency range ($\ftidequadsemid > 0$) and the other one in the sub-synchronous frequency range ($\ftidequadsemid < 0$), symmetrically (this symmetry with respect to the synchronisation is due to the non-rotating approximation assumed in \fig{fig:pics_ecc}). We emphasise the fact that the peak of positive torque appearing in the super-synchronous regime results from the forcing of the resonance located in the sub-synchronous frequency range by a prograde eccentricity tidal potential ($\ftidequads = - \foco$). This configuration corresponds for instance to case~1 of \fig{fig:eccentricity_tides}, \padc{where the degree-3 eccentricity potential is propagating eastward.}}

\padc{\smc{In this framework,} \fig{fig:pics_ecc} \smc{simultaneously} shows the forcing frequencies $\ftidequads = 2 \spinrate - \ss \norb$ of the semidiurnal mode ($\ss = 2$) and of two eccentricity modes ($\ss = 3$ and $\ss = 4$) in the cases where a peak of positive torque is generated by one of these eccentricity modes in the interval $0 < \ftidequadsemid  < \norb $. These particular frequencies are superscripted by the symbol $\blacktriangleup$, and the created peak is designated by a red cone. Hence, the contribution of one of the eccentricity components is potentially leading to asynchronous equilibrium states located at $\ftidequadsemid \approx \ftidequadsemid^{\blacktriangleup}$. This configuration occurs when the peak is sufficiently important to counterbalance the semidiurnal tidal torque, whose resonances are represented by blue cones as we plot the schematic torques as a function of $\ftidequadsemid$. }

\smc{For the cases a to c of \fig{fig:pics_ecc}, the assumption made about $\ftidequadsemid^{\blacktriangleup} $ \padc{(i.e. $0< \ftidequadsemid^{\blacktriangleup} < \norb $)} leads to the fact that only the \padc{degree-3} eccentric term can coincide with the prograde resonant frequency of the ocean (sub-synchronous blue cone), \padc{that is $\ftide_{2,3} = - \foco = \ftide_{2,3}^\blacktriangleup$}. When the depth of the ocean increases (from d to f), keeping $\ftidequadsemid^{\blacktriangleup} $ in the range 0  - $\norb$ leads to an excitation of the ocean resonant mode only with the \padc{degree-4 eccentricity mode (for $\ftide_{2,4} = - \foco= \ftide_{2,4}^\blacktriangleup$).} This illustrates how increasing the depth of the ocean leads to the excitation of higher eccentricity modes in the tidal potential. }

Configuration (a) represents the case where $\foco = \norb / 2$. In this case, the peaks created by the resonance of the degree-3 and degree-2 (semidiurnal) component are exactly superposed. The semidiurnal one is the strongest in the low eccentricity regime since $\Utidelms{2}{2}{3} \ll \Utidelms{2}{2}{2}$. As a consequence, the oceanic tide leads the planet toward synchronisation ($\spinrate  = \norb$) in this configuration, and cannot generate an asynchronous rotation state of equilibrium in the range $ - \norb < \ftidequadsemid < \norb$.

While the ocean depth increases (configurations (b) and (c)), the resonance moves away from the synchronous rotation and becomes stronger, its eigenfrequency and maximum value scaling as $\propto \sqrt{\Hoc}$ (see \eqs{focn}{maxpeak}). The peak generated by the degree-3 eccentricity component increases as well in intensity and gets closer to synchronisation in the meantime. This means that the resulting asynchronous state of equilibrium -- if it exists -- gets also closer to synchronisation until being annihilated by the solid tidal torque, which outweighs the oceanic torque when $\tautide \sim \tauM$. 

Configuration (d) shows the switch from the degree-3 term to the degree-4 term that occurs while the ocean depth keeps increasing.  The peak of positive tidal torque now results from the amplification of the degree-4 eccentricity term by the resonance associated with the tidal oceanic mode. It is smaller than that generated by the degree-3 term since is scales quadratically with the forcing tidal potential (see \eqs{U22s}{torque_ocean}), which decays while the degree-$\ss$ component increases in the low-eccentricity regime \citep[for $\ecc \gtrsim \ecctrans$, non-linearities lead eccentricity components of higher degrees to predominate; see e.g.][Fig.~3]{Ogilvie2014}.

Similarly as the peak created by the degree-3 component, the peak associated with the degree-4 eccentricity term moves towards the synchronisation while the ocean depth increases, as shown by configurations (e) and (f). When $\foco$ becomes greater than $2 \norb$, the term generating the peak of positive tidal torque switches from the degree-4 to the degree-5 eccentricity term, and so on, as discussed at the end of \sect{ssec:perturbing_potential} (see \eq{condition_desynch}).

The mechanism highlighted by \fig{fig:pics_ecc} for the degree-0 Hough mode can be generalised to other modes, each of them being able to amplify an eccentricity term provided that the associated resonance dominates the non-resonant background level. This picture is also completed by the action of solid tides, which enable the existence of tidally-locked asynchronous rotation states of equilibrium for the rotation rates $\spinrateeqs \approx \left( \ss / 2 \right) \norb$, as discussed in \sect{sec:critical_eccentricity}. 

\smc{In reality, the simplified oceanic tidal response is not solely due to the degree-0 Hough mode of frequency $\foco$ since $\abs{\spinrate} \sim \norb  \sim \abs{\ftidequads} $ in the frequency range of interest. The real tidal response is composed of g- and r-modes (i.e. the gravity modes modified by rotation and the Rossby modes, respectively), these laters being restored by the spin rotation of the planet. Therefore, the resulting behaviour is much more complex than that described in \fig{fig:pics_ecc} although it exhibits the main highlighted features.}

In the quasi-adiabatic regime, the maximum of the tidal torque given by \eq{maxpeak} can be arbitrarily high depending on the value of the Rayleigh drag frequency. When $\fdrag \rightarrow 0$, this maximum tends to infinity. This is an artefact of the linear theory. In reality, the large amplitudes of tidal fields associated with a resonance in the quasi-adiabatic regime violate the small perturbation approximation. As a consequence, the tidal response becomes non-linear and the maximum is attenuated with respect to that predicted by the linear theory. 

In the case of the Earth, the Rayleigh drag frequency used to model the friction with the oceanic floor was estimated to $\fdrag \approx 10^{-5}$\units{s^{-1}} \citep[e.g.][]{Webb1980}, which is comparable with typical tidal frequencies. In this case, the enhanced oceanic tidal response resulting from a resonance remains in the framework of the linear theory, and the associated amplification of the tidal torque generally does not exceed a decade in logarithmic scale \citep[see][Fig.~5]{ADLML2018}.

As the effective Rayleigh drag frequency accounts here for the amount of energy dissipated through the interaction of tidal waves with the Earth complex topography, we assume that $ 10^{-5}$\units{s^{-1}} is an upper estimation for $\fdrag$. It seems likely that $\fdrag$ takes smaller values in the case of a planet hosting a uniform ocean. To take these discrepancies into account, we  consider the values $10^{-7}$ and $10^{-6}$\units{s^{-1}} in addition to $10^{-5}$\units{s^{-1}} in numerical calculations. 

In spite of the limitations mentioned above, studying the quasi-adiabatic regime is a useful step to understand how the resonances characterising the oceanic tidal response can enhance the desynchronising effect of eccentricity tides. This is thus the framework that we adopt to derive a theoretical estimation of the critical eccentricity in \sect{sec:critical_eccentricity}. 

\subsection{Frequency behaviour of the tidal torque}
\label{ssec:frequency_behaviour}

The tidal torque resulting from the planet tidal response is characterised by a complex frequency behaviour, which slightly diverges from the idealised picture showed by \fig{fig:pics_ecc}. This behaviour results both from the gravitational and surface coupling between the solid and oceanic layers, and from the dependence of the oceanic tidal response on Coriolis effects in the frequency range of interest.

It seems helpful to focus on a given case to visualise how the tidal torque exerted on the planet varies with the tidal frequency before going further in the analysis. As shown previously, the amplification of eccentricity terms by resonances associated with the oceanic tide most likely takes place at short orbital periods ($\Porb \sim 1 - 10$~days) since this regime corresponds to $\Hoc \sim 1-10$~km. This regime is typically encountered in tightly-packed systems such as that hosted by the TRAPPIST-1 ultra-cool dwarf star \citep[][]{Gillon2017,Grimm2018}, where planet e exhibits a 6.10~days orbital period \citep[][]{Gillon2017}.

We thus consider along the whole study the case of an Earth-sized planet orbiting TRAPPIST-1 \citep[$\Mstar = 0.09 \, \Msun $, see][]{VanGrootel2018} with a 6-days orbital period. \rec{We note that the mass of the host star is the only parameter of the system here, since the internal structure and surface conditions of the planet are fixed.} For illustrative purpose, the planet is assumed to host a 0.2~km-deep uniform ocean, which corresponds to the resonance of one of the dominating oceanic modes for the degree-3 eccentricity term in the range $0 < \ftidequadsemid < \norb$. We set the typical frequency of the Rayleigh friction to $\fdrag = 1.0\expdec{-6}$\units{s^{-1}}. Finally, the response of the solid part is described by the values of Andrade parameters given by \tab{tab:andrade_para}, as mentioned in \sect{ssec:solid_part}. 

\begin{figure}[htb]
   \centering
   \includegraphics[width=0.46\textwidth,trim = 0.1cm 0.8cm 1.2cm 1.1cm,clip]{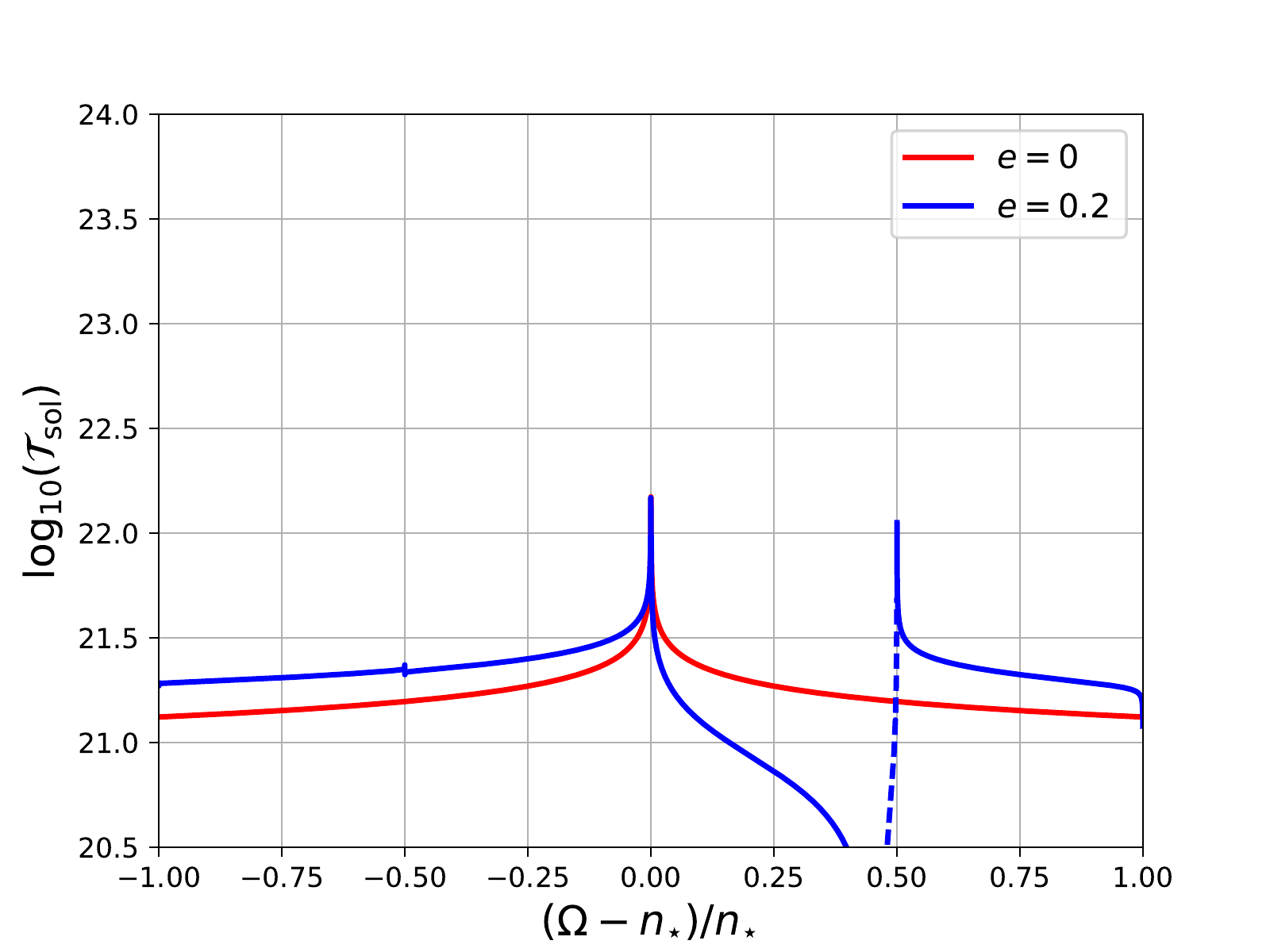} \\
   \includegraphics[width=0.46\textwidth,trim = 0.1cm 0.8cm 1.2cm 1.1cm,clip]{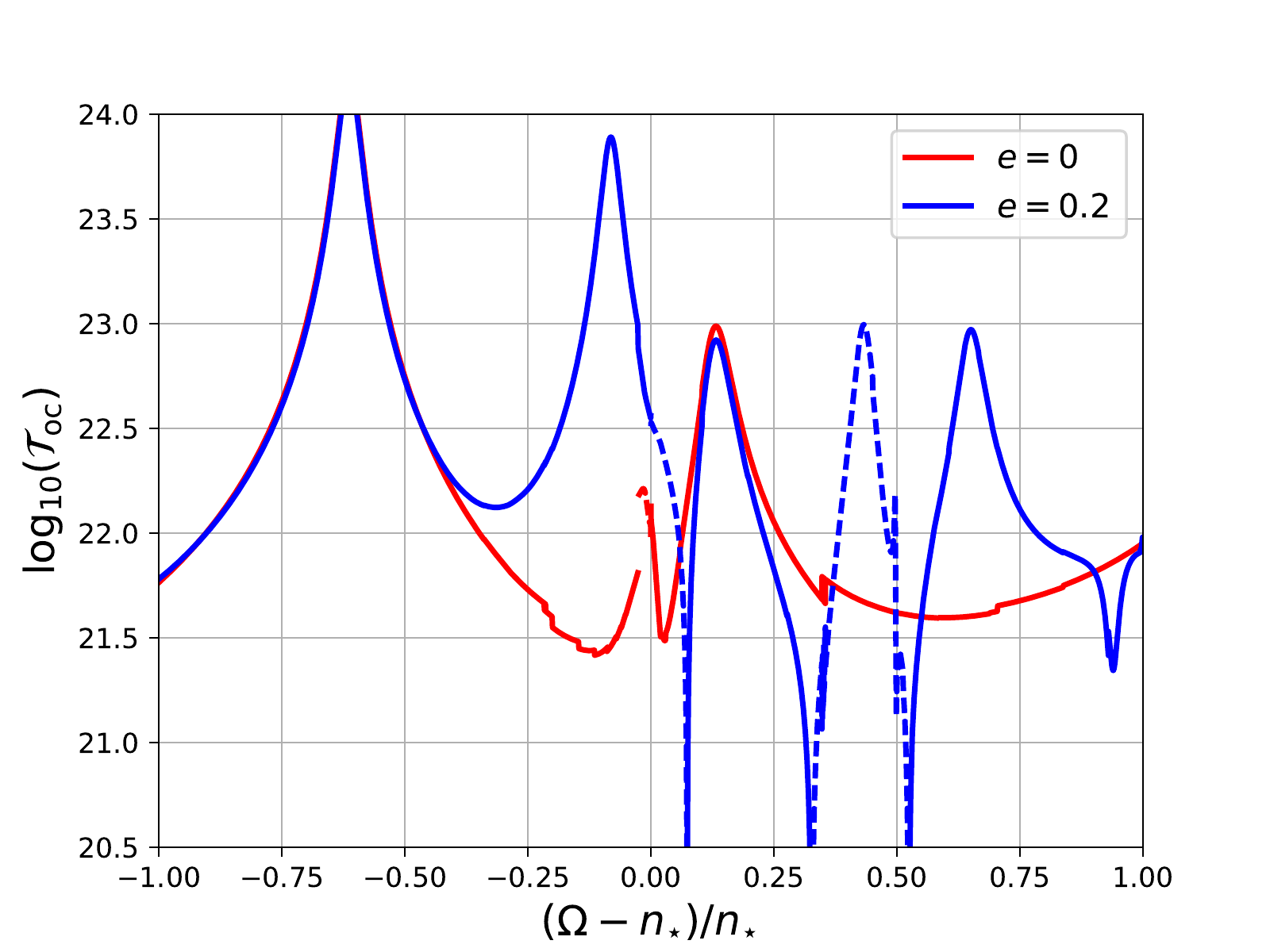} \\
   \includegraphics[width=0.46\textwidth,trim = 0.1cm 0.0cm 1.2cm 1.1cm,clip]{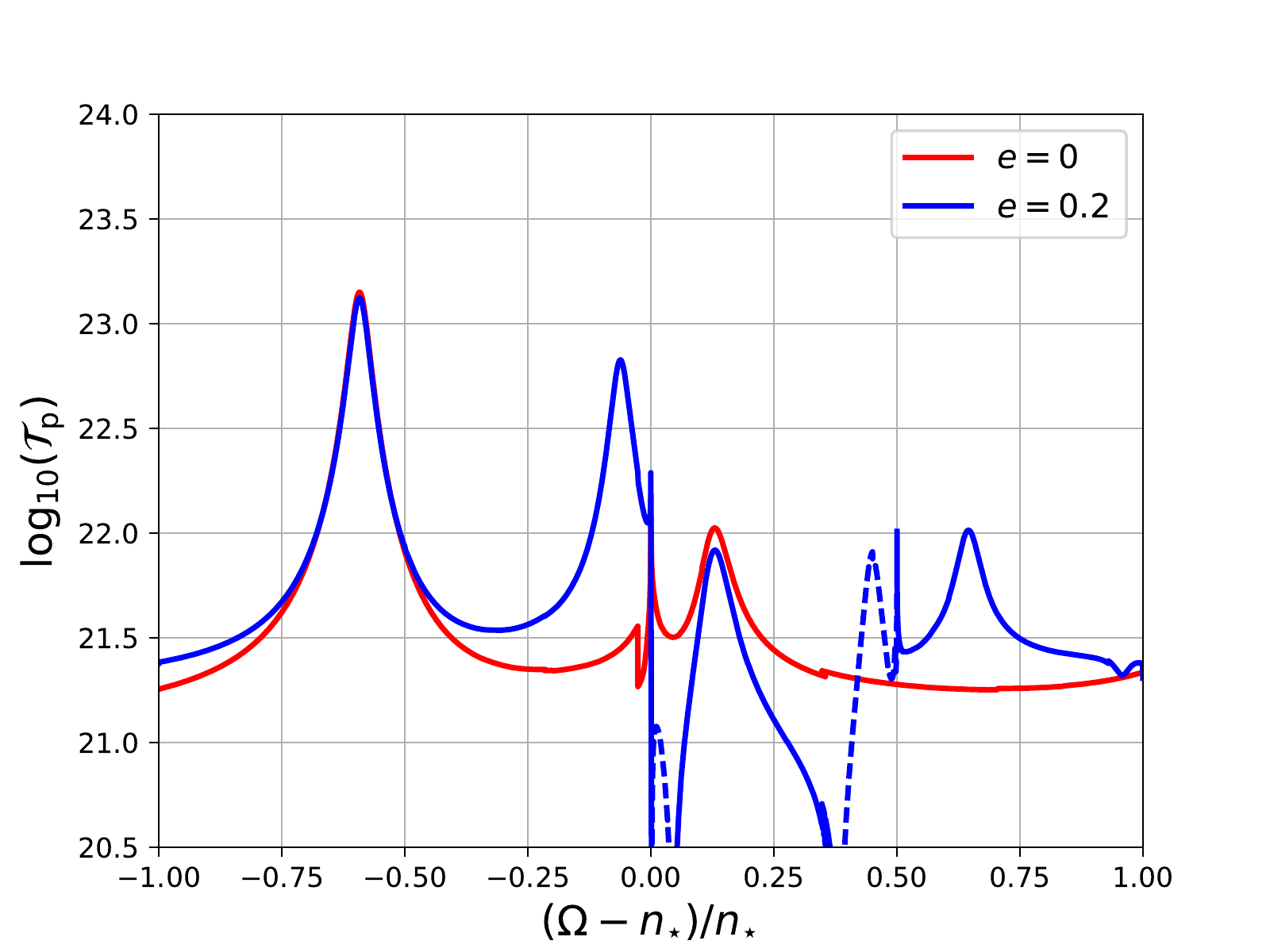}
   \caption{Logarithm of the tidal torque exerted on the planet \jlc{in the cases of the pure solid (top panel) and oceanic (middle panel) tidal responses, and in the general case (bottom panel),} as functions of the normalised semidiurnal frequency $\ftidequadsemid / \left( 2 \norb \right) = \left( \spinrate - \norb \right) / \norb$ for a circular orbit (red line) and an orbit of eccentricity $\ecc = 0.2$. The torques \jlc{$\torquesol$, $\torqueoc$, and $\torquepla$} are computed using the expression given by \eq{torque_pla}, \ebc{and the values given by \tab{tab:andrade_para} for an Earth-sized planet orbiting TRAPPIST-1 ($\Mstar = 0.09 \, \Msun $) with a 6-days orbital period. The Rayleigh drag frequency is set to $\fdrag = 1.0\expdec{-6}$\units{s^{-1}}.} Solid (dashed) lines designate the regimes where the tidal torque drives the planet towards (away from) spin-orbit synchronous rotation.}
       \label{fig:spectra}%
\end{figure}

The tidal torque exerted on the planet is plotted in \fig{fig:spectra} as a function of the normalised semidiurnal tidal frequency $\ftidequadsemid / \left( 2 \norb \right)  = \left( \spinrate - \norb \right) / \norb$, for circular ($\ecc = 0$) and eccentric ($\ecc = 0.2$) orbits. Panels from top to bottom correspond to \jlc{the case of a pure solid tide ($\Hoc = 0$), the case of a pure oceanic tide ($\mupla = + \infty$), and the general case, respectively}. In the three cases, the tidal torque is calculated using the expression given by \eq{torque_pla}, with $\Hoc = 0$ or $\mupla = + \infty$ to obtain the pure solid and oceanic response, respectively. To highlight the effect of eccentricity components, regimes where the planet is driven towards (away from) the spin-orbit synchronous rotation state of equilibrium are designated by solid (dashed) lines. 

First, we consider the torque generated by a pure solid tide \jlc{(\fig{fig:spectra}, top panel)}. We retrieve here the behaviour investigated by \cite{ME2013} and \cite{Correia2014}. In the absence of eccentricity, the semidiurnal component solely generates a torque, which varies with the frequency following the Andrade frequency-dependence illustrated by \fig{fig:andrade_maxwell}. For tidal periods smaller than the Maxwell and Andrade times, we recover the scaling law $\torquesol \propto \left| \ftidequadsemid \right|^{-\alphaA}$ given by \eq{k2sol_asymp}. This torque drives the planet towards the spin-orbit synchronisation. A sufficiently large eccentricity induces additional asynchronous rotation states of equilibrium associated with the rotation rates $\spinrateeqs \approx \left( \ss / 2 \right) \norb$ \citep[][]{ME2013,Correia2014}, which are discussed in \sect{sec:critical_eccentricity}.

We then move to the pure oceanic tidal response \jlc{(\fig{fig:spectra}, middle panel)}, where the rigidity of the solid part is supposed to be infinite. In this case, the tidal torque is composed of a batch of resonances associated with the main Hough modes coupled with the tidal gravitational potential. In the circular configuration, two dominating peaks may be observed. For $\ecc = 0.2$, additional peaks appear. Each of these peaks is generated by the enhanced contribution of an eccentricity term. For instance, the peak noticed for $\left( \spinrate - \norb \right) / \norb \approx 0.4$ clearly corresponds to the configuration illustrated by \fig{fig:pics_ecc} (panel (b)), where the frequency associated with the degree-3 eccentricity term meets the eigenfrequency of one of the dominating modes. 

\padc{When the typical tidal periods of predominating tidal components exceed the characteristic timescale of inertial waves and are less than or comparable with the energy decay \ebc{timescale} of the ocean in the meantime ($ \fdrag \lesssim \abs{\ftide} < \abs{2 \spinrate}$), the tidal forcing tends to couple with an infinite number of Hough modes as $\abs{\spinpar}$ increases \citep[this corresponds to the regime of sub-inertial waves, defined by $\abs{\spinpar} > 1$, and discussed by, e.g.][]{LS1997}. The method adopted to solve the Laplace's tidal equation (see \eq{Laplace}) in this study cannot treat correctly this phenomenon because the number of computed modes is fixed by the dimension of the truncated matrix used in calculations \citep[][]{Wang2016}. The effect of truncation becomes more important as $\abs{\spinpar} $ increases.} 

\padc{As a result, the oceanic tidal response is poorly described by the theory for $\abs{\spinpar} \gg 1$, which degrades the estimation of the induced tidal torque and leads to the unrealistic discontinuous change of sign observed for the circular case in the vicinity of the synchronisation (\fig{fig:spectra}, \padc{middle panel}).} This feature of the model prevents us to explore the region of the parameter space characterised by very small drag frequencies. Fortunately, the action of the drag annihilates the distortion caused by the spin rotation by making Hough functions converge towards the associated Legendre functions, as discussed in \sect{sec:ocean_response}. The change of sign thus does not occur from the moment that $\abs{2 \spinrate} \lesssim \fdrag$, which is the case in this study if $\fdrag \gtrsim 10^{-6}$\units{s^{-1}}. 

We finally consider the tidal torque in the general case, where the tidal responses of the solid and oceanic layers are coupled together by gravitational and pressure forces \jlc{(\fig{fig:spectra}, bottom panel)}. We mainly retrieve here the frequency-resonant behaviour of the ocean, but the tidally dissipated energy is attenuated by one order of magnitude. This is due to the \padc{elastic} adjustment of the \padc{deformable} solid part, which tends to compensate the horizontal gradient of mass distribution associated with the elevation of the ocean surface level. 

As we now better visualise the dependence of the tidal torque on the forcing frequency, we can proceed to an analytical estimation of the critical eccentricity beyond which asynchronous rotation states of equilibrium may exist. 

\section{Analytical estimation of the critical eccentricity for an ocean planet}
\label{sec:critical_eccentricity}

In the previous sections, we detailed the formalism that describes the tidal response of an ocean planet by including the interactions between the solid part and the oceanic layer, and derived the expressions of the planetary tidal torque and second order Love number. First, we use here these solutions to establish the conditions that have to be satisfied to end up with tidally-locked asynchronous rotation states of equilibrium. Second, we derive an analytical estimation of the critical eccentricity beyond which such states can exist in the low eccentricity regime. The obtained results are used in \sects{sec:final_rotation}{sec:critical_ecc}.

 In the non-resonant case, eccentricity tides may drive the planet away from synchronisation only if the associated components of the tidal gravitational potential are comparable with or greater than the component associated with the semidiurnal tide, that is for \padc{$\ecc \gtrsim \ecctrans \sim 0.25$} typically, as showed in \sect{sec:tidal_dynamics}. The critical eccentricity $\eccasync$ may be strongly lowered when a resonance occurs, since this later generates an important increase of the tidally dissipated energy. This mechanism enables the existence of asynchronous rotation states of equilibrium at smaller eccentricities. \padc{In this section, the planet is considered as an idealised \rec{rotationally symmetric} body, and the triaxial torque due to triaxiality is thus ignored. The action of triaxiality regarding the trapping of the planet in spin-orbit resonances is investigated in \sect{ssec:capture}}.

If we assume that $\ecc \ll 1$, three conditions must be satisfied at the same time to lock the planet into a non-synchronised rotation rate: (i) the forcing frequency associated with one of the eccentricity components of the response (see \eq{torque_ocean}) must be equal to one of the eigenfrequencies of the oceanic surface gravity modes, given by \eq{focn}, (ii) the resonant eccentricity component must drive the planet away from synchronisation as discussed in \sect{sec:tidal_dynamics}, and (iii) the resonance must dominate the non-resonant background, which is the sum of non-resonant terms. We examine these conditions for eccentricity components of degrees $\ss \geq 3$, as they predominate.

Condition (ii) is expressed by \eq{condition_desynch}. In this configuration, the degree-$\ss$ forcing frequencies are negative while the semidiurnal forcing frequency is positive. Assuming condition (i), we suppose that the degree-$\nn$ mode is resonant. This leads to 

\begin{equation}
\ftidequads  = - \focn ,
\label{egalite_reso}
\end{equation}

\noindent where $\ftidequads = 2 \spinrate - \ss \norb$ is the degree-$\ss$ eccentricity frequency defined by \eq{ftide2s} and $\focn$ the eigenfrequency of the degree-$\nn$ Hough mode given by \eq{focn}. By combining \eqs{condition_desynch}{egalite_reso}, and introducing the notation $\spinparquads \define \spinpar \left( \ftidequads \right)$, we hence derive from (i) and (ii) a first condition on the ocean depth \rec{($\Hoc$)}, which must satisfy the inequality

\begin{equation}
\Hoc \leq \left( \ss - 2 \right)^2 \left( \frac{\Rpla^2 \norb^2}{\Lambdanquads \ggravi} \right) .
\label{Hoc_cond1}
\end{equation}

This inequality determines the range of depths for which a resonance may occur. We observe that the range of $\Hoc$ widens with $\ss$, meaning that the smaller the ocean depth, the larger the number of eccentricity components that may be excited by the resonance. Similarly, the upper limit of $\Hoc$ increases with $\norb$, which favours resonances in the case of close-in terrestrial planets. Conversely, this limit is inversely proportional to the eigenvalue $\Lambdan$ of the mode. In the quasi-adiabatic regime, the $\Lambdan$ are sorted in ascending order \citep[e.g.][]{LS1997}. Thus, the upper limit of $\Hoc$ is lower for high degrees $\nn$ than for low degrees. \padc{Basically}, this means that the resonance of the degree-$0$ mode will express preferentially with respect to those associated with other modes. A high-degree mode cannot enter into resonance except in the case of a very thin ocean. 

Establishing an analytical formulation of condition (iii) is challenging -- if not impossible -- in the general case because of the coupling between the solid and fluid layers. We thus choose to ignore this coupling as well as the effect of the variation of the planet self-attraction (this is the Cowling approximation mentioned above). Moreover, as we focus on the case of small eccentricities ($\ecc \ll 1$), the major part of the non-resonant background level results from the semidiurnal component (see \eq{U222}), which tends to drive the planet towards synchronisation. We thus neglect other terms and assume that the non-resonant part of the response reduces to the semidiurnal tide. 

The case of a purely solid planet has been studied by \cite{ME2013} and \cite{Correia2014}, who used Maxwell \citep[e.g.][]{Greenberg2009} and Andrade \citep[e.g.][]{Efroimsky2012} rheologies to model the frequency behaviour of the body, respectively. \ebc{These early works show} that eccentricity tides drive the planet towards final rotation rates that are multiples of $\norb / 2$ in the regime where $\tautide \ll \tauM$ and for sufficiently high eccentricities. In this case the spin equilibria occur when the torque associated with one of the eccentricity components \rec{becomes resonant} (see \fig{fig:andrade_maxwell}). These positions approximately correspond to the rotation rates for which the forcing frequencies annihilate, that is $\spinrateeqs \approx \left( \ss / 2 \right) \norb$. For small parameters $\alphaA$ of the Andrade rheology (typically $\alphaA = 0.25$), the order of magnitude of the tidal torque does not vary much with the tidal frequency. As a consequence, asynchronous spin equilibria may not exist at small eccentricities, if the departure between the frequencies of the semidiurnal and degree-3 components (that is $\norb$) is not large enough. 

The sharp variations of the tidal torque associated with the resonances characterising the ocean tidal response enable asynchronous rotation state of equilibrium to exist at \padc{small} eccentricities. If we neglect the core-ocean coupling resulting from gravitational and surface forces, condition (iii) may be simply written as

\begin{equation}
  \left[ \Hansen{\ss}{-3}{2} \left( \ecc \right)   \right]^2 \abs{\imag{\kocquads}} > \abs{\imag{\kpquadsemid}},
\label{condition3}
\end{equation}

\noindent where $\ftidequads = 2 \spinrate - \ss \norb = - \focn$ (see \eq{egalite_reso}) designates the resonant forcing frequency of the degree-$\ss$ quadrupole component (the resonance being associated with the degree-$\nn$ Hough mode), and $\ftidequadsemid = 2 \left( \spinrate - \norb \right)$ the semidiurnal frequency.

First, we consider the case where the oceanic tide always predominates over the solid tide, which amounts to assuming that the solid part of the planet is of infinite rigidity ($\mupla \rightarrow + \infty$). In this case, $\kpquad \approx \kocquad$. The associated Legendre function $\Plmnormquad$ is generally well coupled with the degree-0 Hough mode, which means that $ \Cnquad \ll 1 $ except for $\nn = 0$. As a consequence, the total tidal torque is dominated by the contribution of the degree-0 Hough mode outside of the resonances associated with other modes \citep[][]{ADLML2018}, and may be reduced to this component as a first approximation. 

It follows that the condition given by \eq{condition3} is simply expressed as 

\begin{equation}
 \Coquadsemid \frac{\fdrag \ftidequadsemid \foco^2}{ \left( \foco^2 - \ftidequadsemid^2 \right)^2 + \fdrag^2 \ftidequadsemid^2} < \left[ \Hansen{\ss}{-3}{2} \left( \ecc \right)   \right]^2 \Coquads \left( \frac{\foco}{\fdrag} \right). 
\label{Coquadsemid}
\end{equation}

\noindent To obtain this expression, we have assumed that the degree-$\ss$ eccentricity component is resonant, that is $\ftidequads = - \foco$ and $\ftidequadsemid = \left( \ss - 2 \right) \norb - \foco$ (\eq{egalite_reso}), made use of \eqs{koc_quadiadiab}{maxpeak} for the degree-0 Hough mode, and introduced the notation $\spinparquadsemid \define \spinpar \left( \ftidequadsemid \right)$.

The condition on the eigenfrequency $\foco$ resulting from the preceding expression is implicit in the general case since the overlap coefficients ($ \Coquadsemid$ and $\Coquads$) and eigenvalue ($\Lambdaoquads$) associated with the degree-0 Hough mode both depend on the forcing frequencies $\ftidequads$ and $\ftidequadsemid$. \rec{This dependence can be neglected in the regime of rapid rotation ($\abs{\spinrate} \gg \norb$) since spin parameters determining the solutions of Laplace's tidal equation \rec{(\eq{Laplace})} hardly vary in this case. In the general case ($\abs{\spinrate} \sim \norb$), the dependence of the spin parameters on the forcing frequencies is both stronger and much more complex, as shown for example by \cite{LS1997} and \cite{Townsend2003}. It is fully taken into account in the numerical calculations performed in \sectsto{sec:final_rotation}{sec:evolution_timescale}.}

\rec{ Nevertheless, the fact that analytic solutions of the Laplace's tidal equation do not exist in the general case leads us to ignore this dependence here, and we} obtain thereby that condition~(iii) is satisfied for 

\begin{equation}
\Hoc < \frac{1}{4} \left( \ss - 2 \right)^2 \left( \frac{ \Rpla^2 \norb^2 }{ \Lambdaoquads \ggravi} \right) \left( 1 - \sqrt{\frac{\gamCn}{1 + \gamCn}} \right)^2,
\label{Hoc_cond2}
\end{equation}

\noindent or

\begin{equation}
\Hoc > \frac{1}{4} \left( \ss - 2 \right)^2 \left( \frac{ \Rpla^2 \norb^2 }{ \Lambdaoquads \ggravi} \right) \left( 1 + \sqrt{\frac{\gamCn}{1 + \gamCn}} \right)^2,
\label{Hoc_cond3}
\end{equation}

\noindent where we have introduced the supposed constant parameter 

\begin{equation}
\gamCn \define \left[ \frac{\fdrag}{ 2 \left( \ss - 2 \right) \norb \Hansen{\ss}{-3}{2} \left( \ecc \right)}  \right]^2 \frac{\Coquadsemid}{\Coquads}.
\label{gamCn}
\end{equation}

\eqs{Hoc_cond2}{Hoc_cond3} determine the interval of $\Hoc$ in which the semidiurnal torque dominates the torque induced by the degree-$\ss$ eccentricity component, and thus leads to synchronisation. The width of this interval depends on the parameter $\gamma$, which compares the Rayleigh friction frequency to a characteristic frequency proportional to the degree-$\ss$ eccentricity component of the tidal gravitational potential. By reminding us \eq{Hoc_cond1}, we hence observe that the degree-$\ss$ eccentricity term cannot drive the planet away from spin-orbit synchronous rotation if $\gamCn \gg 1$ since the torque induced by the semidiurnal component is systematically stronger. 

Asynchronous rotation rates may exist only if $\gamCn \lesssim 1$. In the asymptotic limit ($\gamCn \rightarrow 0$), the eccentricity component always predominates when its forcing frequency is equal to the eigenfrequency of the degree-$0$ Hough mode ($\foco$). In this case, the only condition that has to be satisfied to make possible the existence of asynchronous final rotation rates is the condition given by \eq{Hoc_cond1}. This highlights the criterion for which resonances may enable eccentricity terms to predominate at small eccentricities, 

\begin{equation}
\fdrag \lesssim 2 \left( \ss - 2 \right) \norb \Hansen{\ss}{-3}{2} \left( \ecc \right).
\label{fdrags}
\end{equation} 

\noindent When this criterion is satisfied, the amplifying effect of the resonance compensates the smallness of the eccentricity forcing component with respect to the semidiurnal one. 

By focusing on the degree-3 component, which is the stronger in the low frequency limit \citep[e.g.][]{Ogilvie2014}, and using \eq{U223}, we immediately derive from \eq{fdrags} an analytical estimation of the critical frequency beyond which tidally-locked asynchronous states of equilibrium can exist, 

\begin{equation}
\ecc \gtrsim \eccasync = \frac{\fdrag}{7 \norb}. 
\label{eccasync_ocean}
\end{equation}

\noindent This expression shows that the critical eccentricity is actually determined by only two parameters when the tidal response of the planet is reduced to the oceanic tide: the orbital and Rayleigh drag frequencies. For a typical orbital period $\Porb = 6$\days, $\norb = 1.21 \times 10^{-5}$\units{s^{-1}}. This leads to $\eccasync \approx 0.1$ for $\fdrag = 10^{-5}$\units{s^{-1}} and $\eccasync \approx 0.01$ for $\fdrag = 10^{-6}$\units{s^{-1}}. We retrieve these orders of magnitude in the numerical results detailed in \sect{sec:final_rotation}.

We investigated above the case where the oceanic tidal torque always predominates over that exerted on the solid part. We now consider the case where the solid tidal torque predominates outside of the resonances of the oceanic tidal response, the non-resonant component of the oceanic tidal torque being assumed negligible compared to the solid tidal torque. Condition (iii) is thus expressed as 

\begin{equation}
\left[ \Hansen{\ss}{-3}{2} \left( \ecc \right)   \right]^2 \abs{\imag{\kocquads}} > \abs{\imag{\kquadsemid}}.
\end{equation}

By assuming that $ \abs{\ftidequadsemid} \gg \max \left\{ \tauM^{-1}, \tauA^{-1} \right\}$, considering that the degree-$\ss$ is amplified by the resonance associated with the degree-0 Hough mode, and combining together \eqs{k2sol_asymp}{maxpeak}, we put the preceding equation into the form

\begin{equation}
\left( \ss - 2 - \Xfreq \right) \Xfreq^{1/\alphaA} > \Aconst,
\label{poly_sol}
\end{equation}

\noindent where we have introduced the normalised frequency $\Xfreq = \foco / \norb$ and the dimensionless constant

\begin{equation}
\Aconst \define  \left\{ \frac{5}{2} \frac{\rhopla}{\rhooc} \frac{\Aquad}{\left(1+\Aquad \right)^2}  \frac{\GammaF \left( 1 + \alphaA \right) \sin \left( \frac{\alphaA \pi}{2} \right)}{ \Coquads  \left[ \Hansen{\ss}{-3}{2} \left( \ecc \right)   \right]^{2}} \frac{\fdrag}{\norb} \left( \tauA \norb \right)^{-\alphaA}  \right\}^{\frac{1}{\alphaA}}.
\label{Aconst_eccasync}
\end{equation}

\comments{Attention, \eq{k2sol_asymp} a un impact sur ce résultat.}


Similarly as what we did to establish \eq{eccasync_ocean}, we suppose that the oceanic tidal response behaves as in the \padc{non-rotating} case, and ignore the implicit dependence of the overlap coefficient $\Coquads$ on the spin rotation ($\spinrate$) and forcing ($\ftidequads$) frequencies through the spin parameter ($\spinparquads$). We thus assume that $\Coquads \approx 1$. Moreover, as we are interested in the low-eccentricity regime, we focus on the effect of the degree-3 eccentricity term. 

In the general case, the two roots defined by the inequality given by \eq{poly_sol} cannot be derived analytically since $\alphaA$ is real. However, the calculation of the maximum of the polynomial function corresponding to the left-hand member of the equation is straightforward. Denoting by $\ualpha$ this maximum, we obtain for $\ss = 3$,

\begin{equation}
\ualpha  = \alphaA \left( 1 + \alphaA \right)^{-\left( 1 + \alphaA \right) / \alphaA}.
\label{ualpha}
\end{equation}

\noindent Since the inequality given by \eq{poly_sol} can be satisfied only if $\ualpha > \Aconst $, it follows that an estimation of the critical eccentricity can be derived by substituting $\Hansen{3}{-3}{2} = \left( 7 / 2 \right) \ecc $ in this later inequality. We end up with 

\begin{equation}
\eccasync = \frac{2}{7} \sqrt{\frac{5}{2} \frac{\rhopla}{\rhooc}  \frac{\Aquad}{\left( 1+\Aquad \right)^2} \GammaF \left( 1 + \alphaA \right) \sin \left( \frac{\alphaA \pi}{2} \right) \frac{\fdrag}{\norb} \left( \norb \tauA \ualpha \right)^{-\alphaA} }.
\label{eccasync_sol}
\end{equation}

\comments{Attention, \eq{k2sol_asymp} a un impact sur ce résultat.}


The comparison between the two obtained expressions of $\eccasync$, given by \eqs{eccasync_ocean}{eccasync_sol}, highlight the role played by the tidal response of the solid part, which significantly attenuates the sensitivity of the studied amplification mechanism to the strength of the drag. This sensitivity is important in the regime where the oceanic tide is the predominating source of the tidally dissipated energy. In this regime, the critical eccentricity for asynchronous states of equilibrium scales as $\eccasync \scale \fdrag$. When the solid tide determines the non-resonant background level, this scaling law switches to $\eccasync \scale \fdrag^{1/2}$. As the non-resonant background level of the oceanic component of the tidal torque scales as $\scale \fdrag$ \citep[see e.g.][Fig.~5]{ADLML2018}, this means that the critical eccentricity ceases to decay proportionally with $\fdrag$ from the moment that $\fdrag $ becomes smaller than a critical value.  

The above analysis provides a reference basis to interpret numerical results. It enables us to go deeper into the exploration of the parameter space with the calculation of the planet final rotation state as a function of its eccentricity and ocean depth.


\begin{figure*}[t]
   \centering
 \begin{flushleft}
  \includegraphics[height=0.015\textheight]{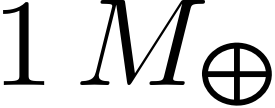} 
   \hspace{0.11\textwidth} 
  \includegraphics[height=0.018\textheight]{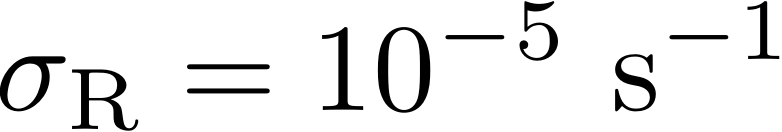} \hspace{0.13\textwidth} 
  \includegraphics[height=0.018\textheight]{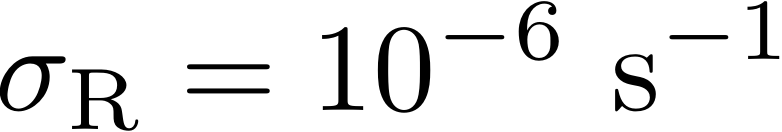}  \hspace{0.13\textwidth} 
  \includegraphics[height=0.018\textheight]{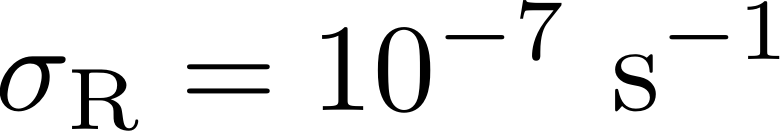} \\
 \end{flushleft}
 \vspace{-0.2cm}
 \raisebox{1.2cm}{\includegraphics[width=0.025\textwidth]{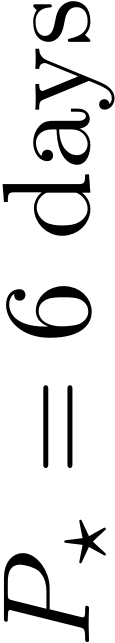}}
 \hspace{0.2cm}
  \raisebox{1.5cm}{\includegraphics[width=0.018\textwidth]{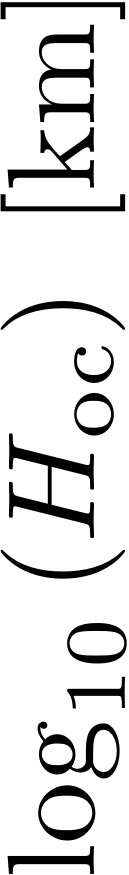}}
   \includegraphics[width=0.28\textwidth,trim = 2.8cm 6.5cm 6.4cm 3.5cm,clip]{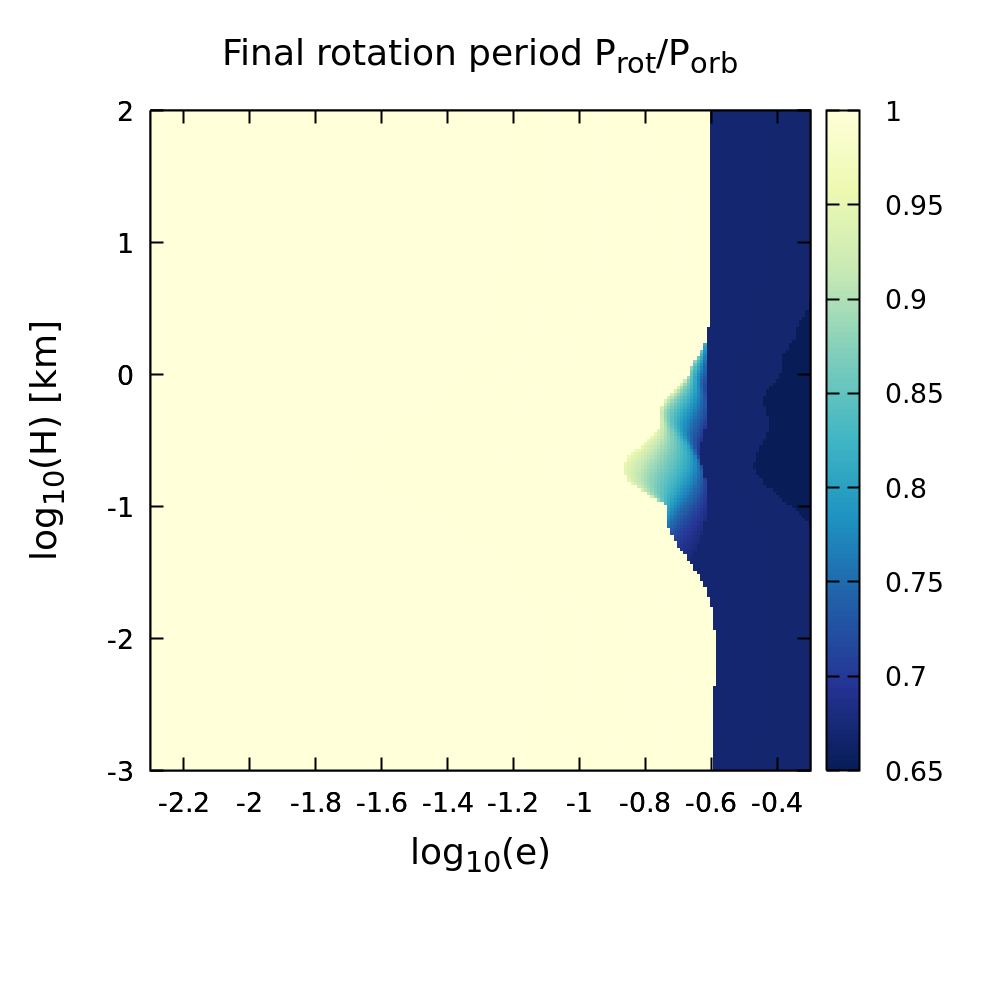} 
    \includegraphics[width=0.28\textwidth,trim = 2.8cm 6.5cm 6.4cm 3.5cm,clip]{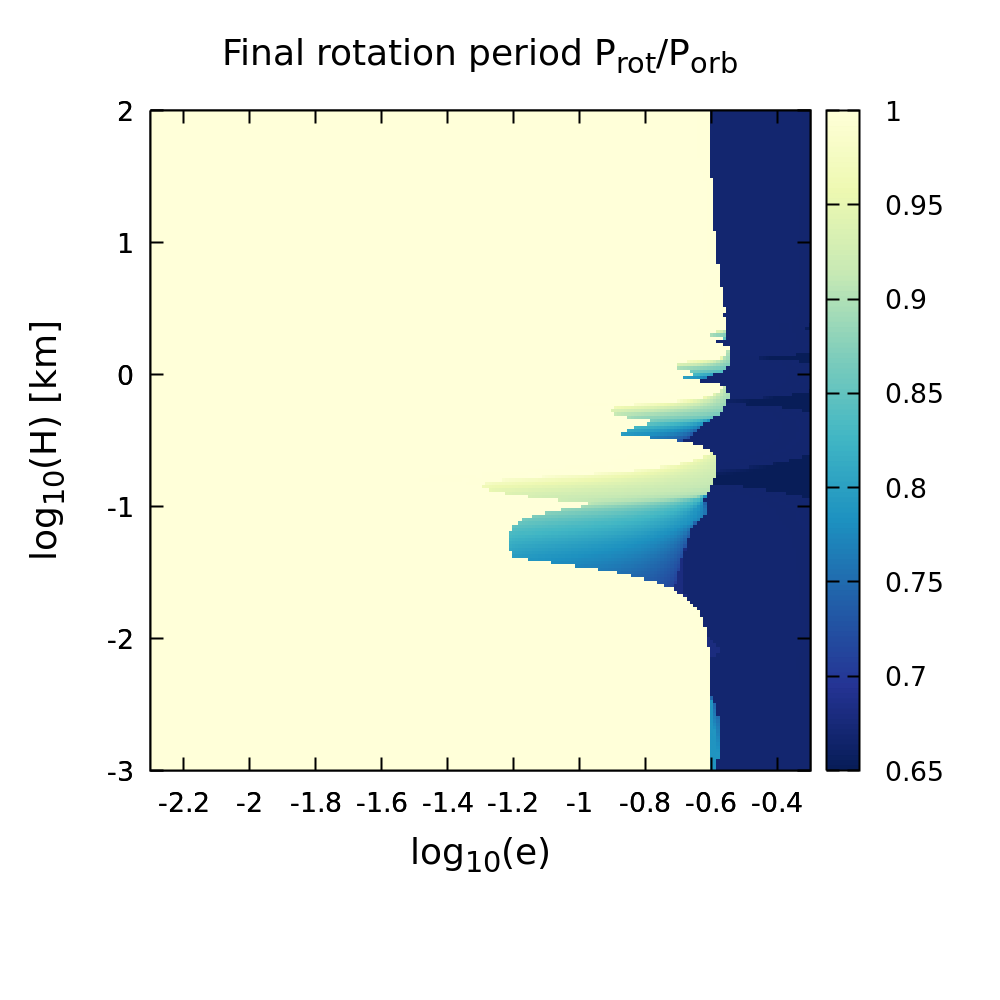}
     \includegraphics[width=0.28\textwidth,trim = 2.8cm 6.5cm 6.4cm 3.5cm,clip]{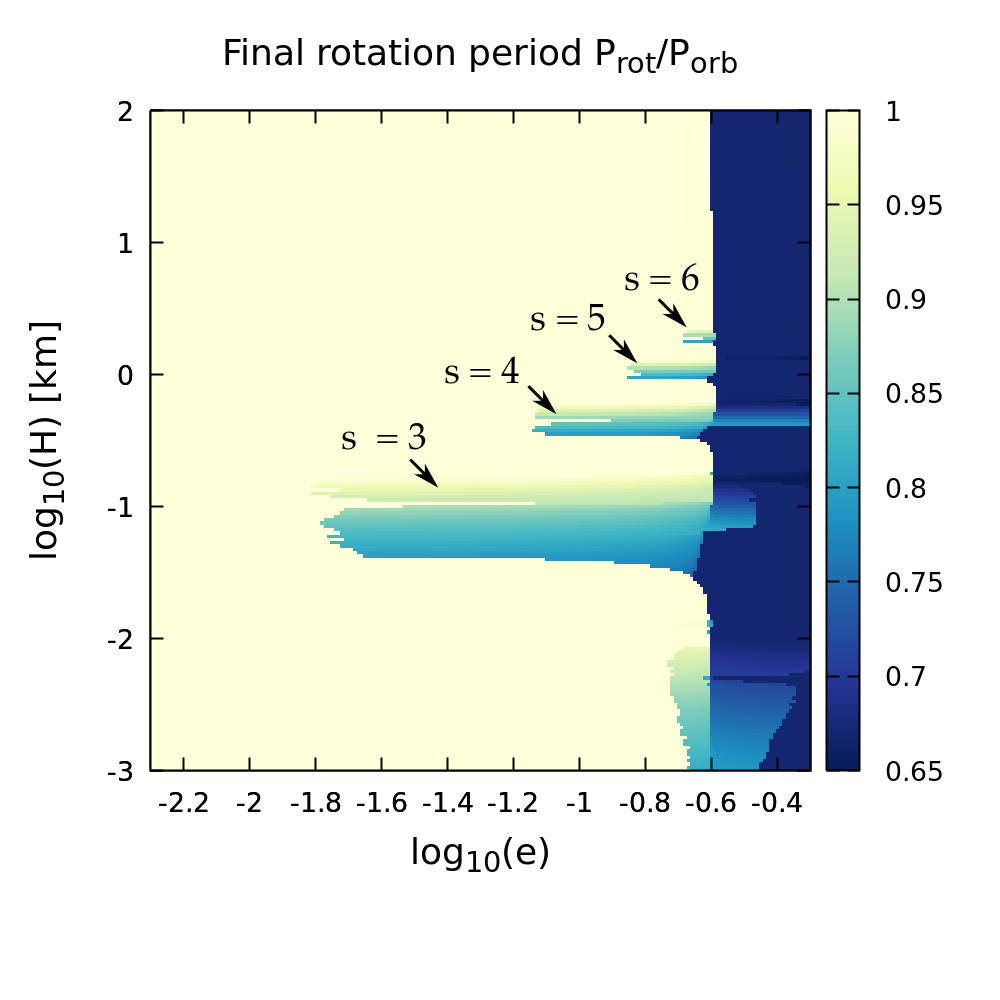}
    \includegraphics[width=0.046\textwidth,trim = 29cm 6.5cm 2.0cm 3.5cm,clip]{auclair-desrotour_fig6h.png} \\
   \raisebox{1.2cm}{\includegraphics[width=0.025\textwidth]{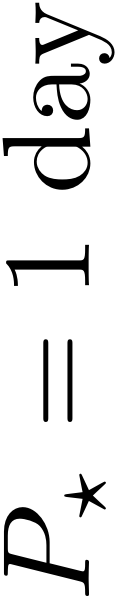}}
    \hspace{0.2cm}
   \raisebox{1.5cm}{\includegraphics[width=0.018\textwidth]{auclair-desrotour_fig6f}}
   \includegraphics[width=0.28\textwidth,trim = 2.8cm 6.5cm 6.4cm 3.5cm,clip]{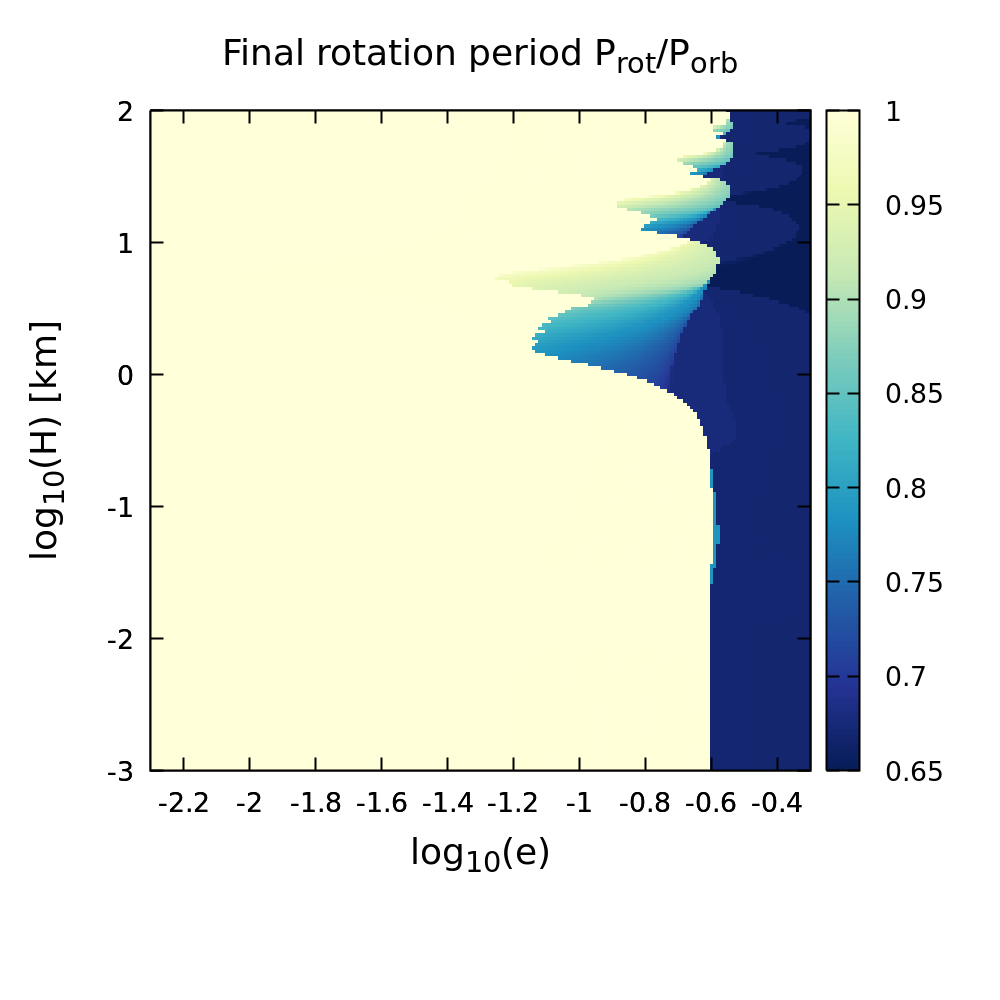} 
    \includegraphics[width=0.28\textwidth,trim = 2.8cm 6.5cm 6.4cm 3.5cm,clip]{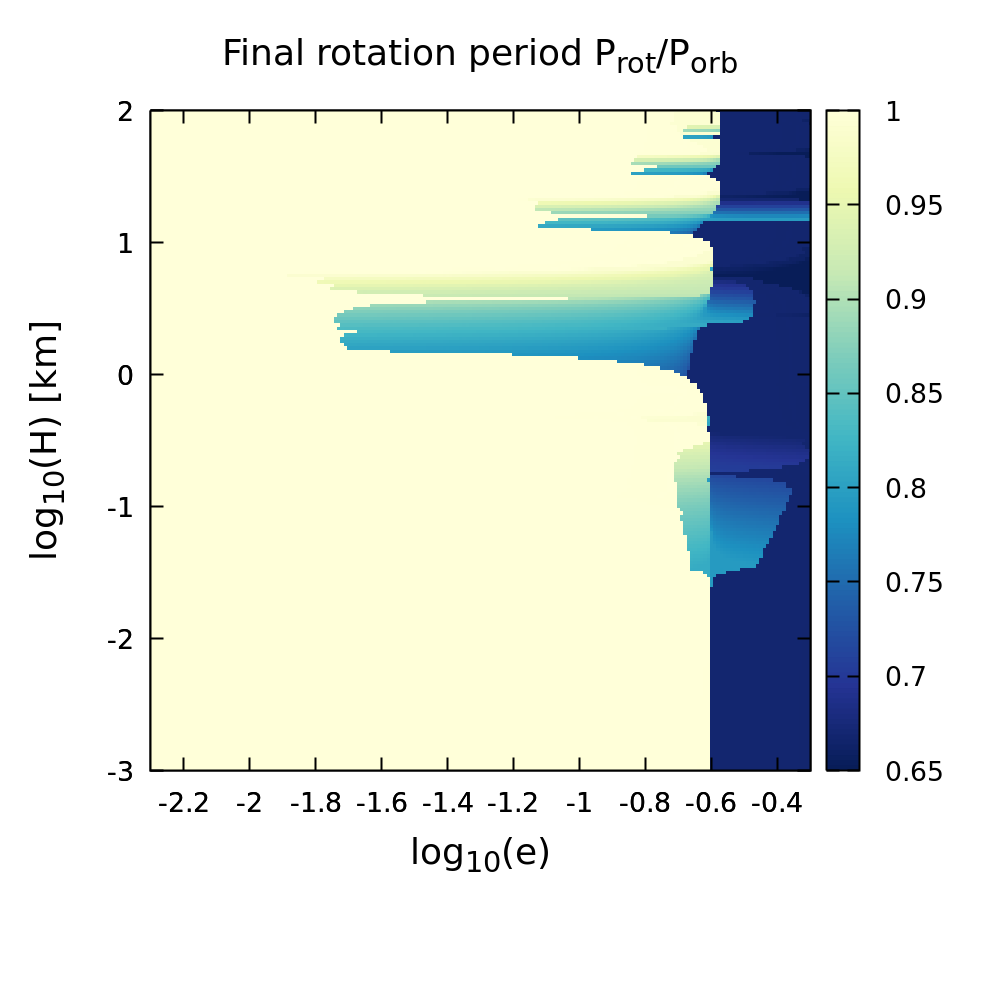}
    \includegraphics[width=0.28\textwidth,trim = 2.8cm 6.5cm 6.4cm 3.5cm,clip]{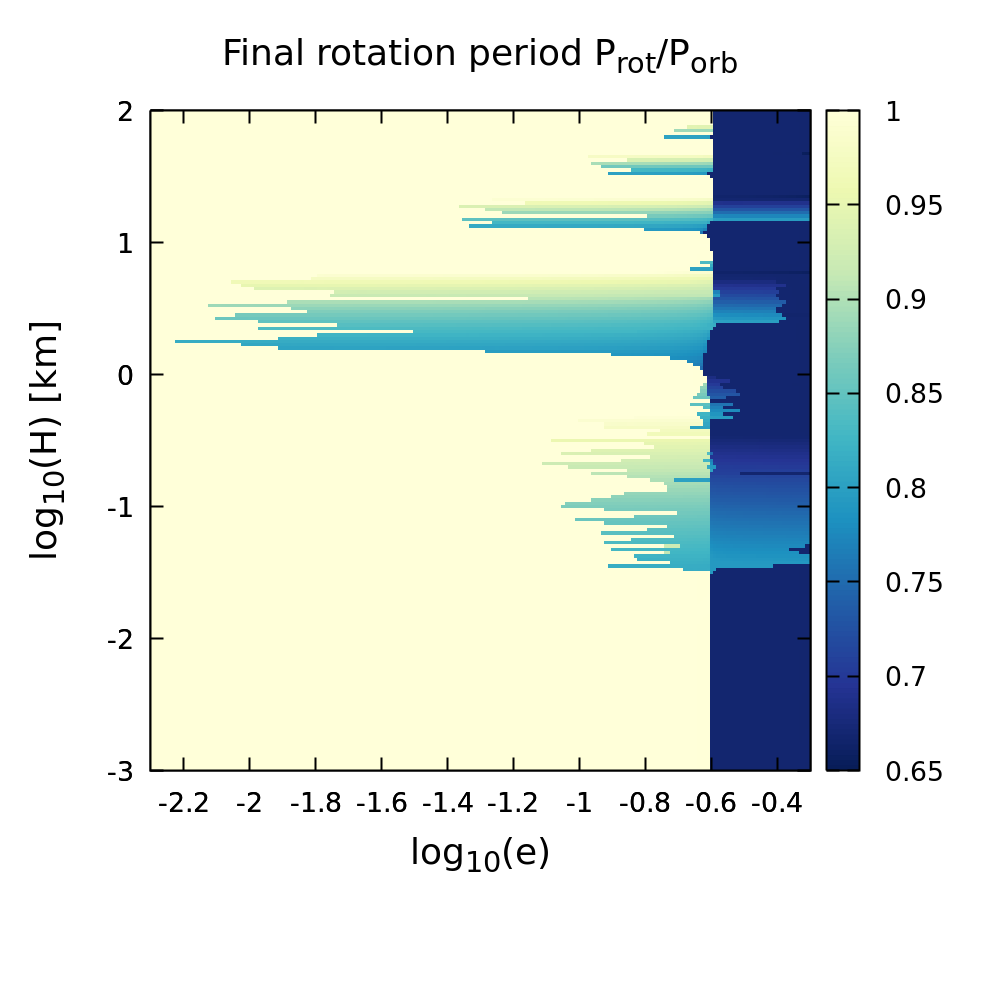}
    \hspace{0.046\textwidth} 
  
    \begin{flushleft}
\includegraphics[height=0.015\textheight]{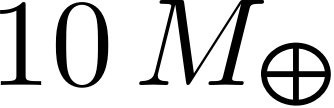}  \hspace{0.1\textwidth}
 \end{flushleft} 
  \vspace{-0.8cm}
   \raisebox{1.7cm}{\includegraphics[width=0.025\textwidth]{auclair-desrotour_fig6e}}
    \hspace{0.2cm}
   \raisebox{1.9cm}{\includegraphics[width=0.018\textwidth]{auclair-desrotour_fig6f}}
   \includegraphics[width=0.28\textwidth,trim = 2.8cm 4.5cm 6.4cm 3.5cm,clip]{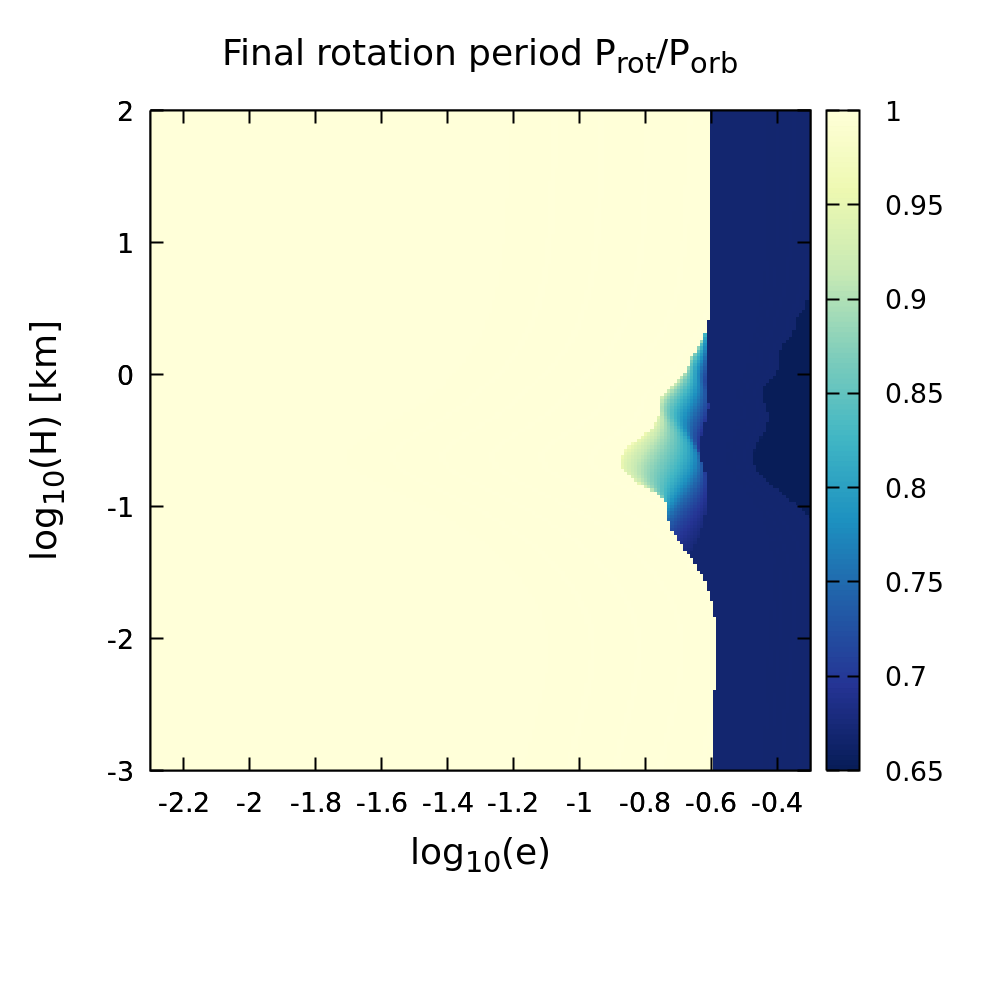} 
    \includegraphics[width=0.28\textwidth,trim = 2.8cm 4.5cm 6.4cm 3.5cm,clip]{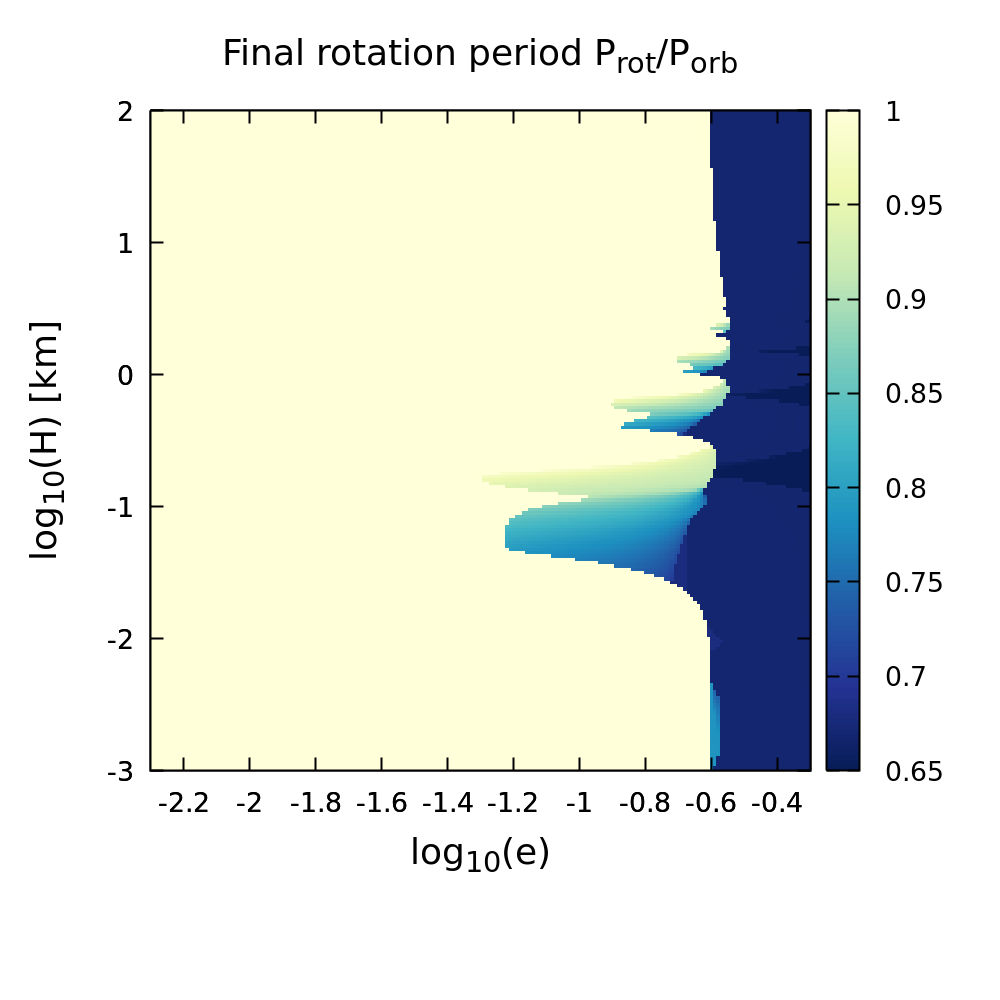}
    \includegraphics[width=0.28\textwidth,trim = 2.8cm 4.5cm 6.4cm 3.5cm,clip]{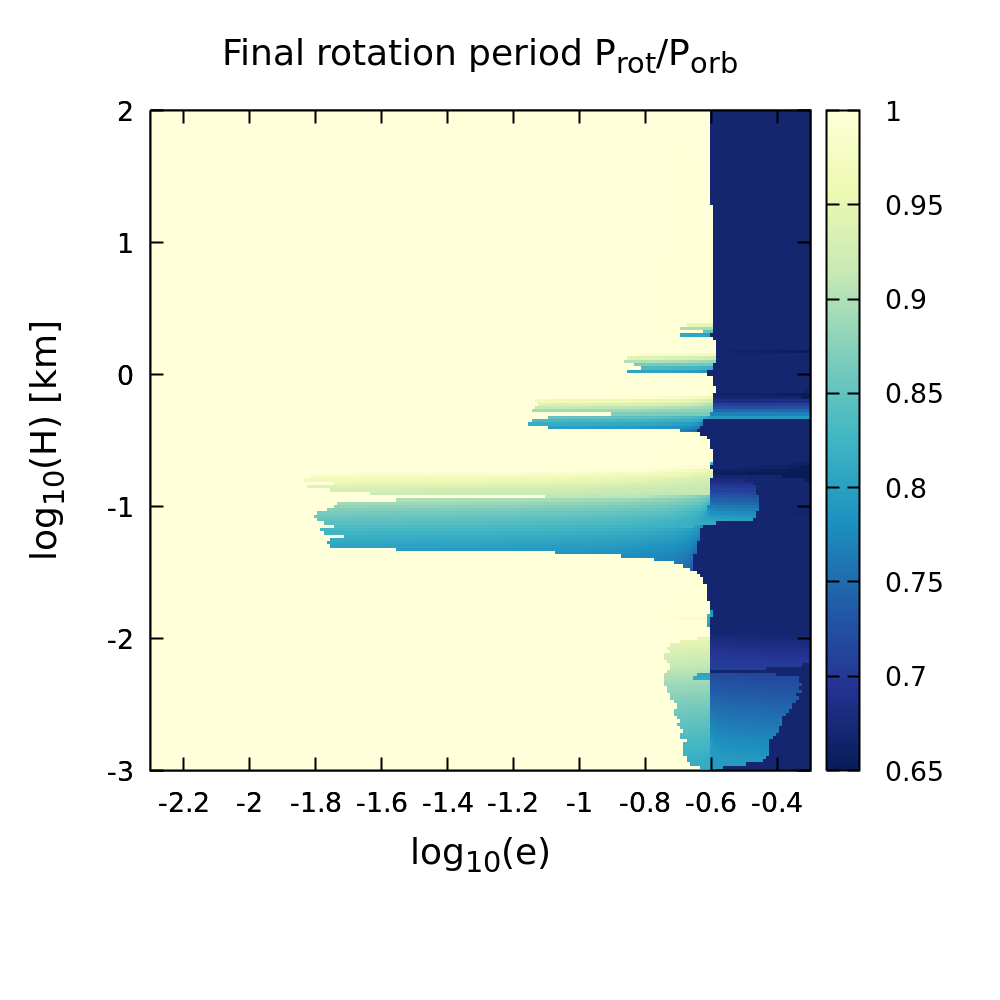}
    \hspace{0.046\textwidth}
   \caption{Normalized final rotation period $\Prot / \Porb$ as a function of the eccentricity (horizontal axis) and ocean depth (vertical axis) in logarithmic scales. Two planets are considered: an Earth-sized planet with 6~days (top panels) and 1~day (middle panels) orbital periods, and a $10 \, \Mearth$-super-Earth with a 6~days orbital period (bottom panels). From left to right, the Rayleigh drag frequency decays, taking the values $\fdrag = 10^{-5}, 10^{-6}, 10^{-7}$\units{s^{-1}}. Values of parameters used for the solid part are given by \tab{tab:andrade_para}. In all cases, the stellar mass is $\Mstar = 0.09 \ \Msun$, the initial rotation rate is $\spinrate = \left( 5/4 \right) \norb$ (middle of the interval $ 0 < \ftidequadsemid < \norb$, see \fig{fig:pics_ecc}), and 10~Hough modes are taken into account in the calculation of the tidal response. \ebc{The degrees of eccentricity modes causing resonances are indicated in the case of the 6 days-period Earth-sized planet with a weak drag (top right panel), consistently with the discussion on \fig{fig:pics_ecc}.}  }
       \label{fig:final_rotation}%
\end{figure*}

\section{Evolution of the planet final rotation with the eccentricity and ocean depth}
\label{sec:final_rotation}

In this section, we investigate how the final rotation rate of a planet evolves with its eccentricity and ocean depth. \jlc{First, we focus on the equilibria determined by the tidal torque. Second, we discuss the existence of pseudo-synchronous equilibria close to the synchronisation by considering the role played by the triaxial torque -- that is the torque due to the planet inherent \smc{and permanent} triaxiality -- and using the theory of capture in spin-orbit resonances \citep[][]{Goldreich1966,GP1966}.  }

\subsection{\jlc{Tidal locking in asynchronous states of equilibrium}}
\label{ssec:tidal_locking}

We treat the cases of an Earth-sized planet and of a 10-$\Mearth$ super-Earth orbiting the TRAPPIST-1 dwarf star ($\Mstar = 0.09 \ \Msun$). In the first case, we consider two orbital periods: $\Porb = 1$~day and $\Porb = 6$~days. In the second case, the \padc{super-Earth} orbits the star with a 6~days orbital period. As the Rayleigh drag frequency cannot be accurately specified for an exoplanet of unknown topography, we perform the calculations for $\fdrag = 10^{-7}, 10^{-6}, 10^{-5}$\units{s^{-1}}, the lower and upper values corresponding to weak and strong drags, respectively. The values of Andrade parameters used for the two planets are those specified in \tab{tab:andrade_para}.

For given eccentricity and ocean depth, the final rotation of a planet is determined by successive iterations from an initial value, specified thereafter, with a frequency step scaling as $\scale \fdrag$ in order to take the dependence of the width of resonance peaks on $\fdrag$ into account. At each step of the research, the sign of the tidal torque exerted on the planet is computed with TRIP \citep[][]{GL2011} using the expression derived in the general case, and given by \eq{torque_pla}. \rec{As the research evolves following the direction defined by the sign of the tidal torque, the} final rotation state of equilibrium corresponds to a change of sign of $\torquepla$. When this change of sign is encountered, the final rotation state is determined with an arbitrary precision using a combination of the dichotomy and secant methods. \jlc{As mentioned in the introduction, we ignore for the moment the torque due to the planet triaxiality and consider that the final state of equilibrium is solely determined by the annihilation of the tidal torque exerted on the planet.}

We focus on the existence of asynchronous rotation states in the interval $0 < \ftidequadsemid < \norb$ (see \fig{fig:pics_ecc}), which corresponds to the super-synchronous frequency range ($\spinrate > \norb$). In this interval, the tidal torque generated by the solid part in the absence of rotation reaches its maximal values for $\abs{\ftidequadsemid} \sim \tauM^{-1}$ and $\abs{\ftidequad{3}} \sim \tauM^{-1} $, that is in the vicinity of the interval bounds. As a consequence, if the initial value of the semidiurnal tidal frequency is too close from 0 or $\norb$, the solid tide predominates and the final state of equilibrium is either the spin-orbit synchronous rotation or the asynchronous rotation state induced by the degree-3 eccentricity term, $\spinrateeqsnum{3} \approx \left( 3 / 2 \right) \norb $ (see discussion in \sect{sec:critical_eccentricity}). 

As a consequence, the initial semidiurnal frequency has to be set sufficiently far from the bounds of the studied frequency interval \smc{to observe the effect of oceanic tides on the final rotation state of equilibrium}. In this work, we arbitrarily start the research at the middle of the interval, that is with the initial semidiurnal frequency $\ftidequadsemid = \norb /2$. This corresponds to the initial rotation rate $\spinrate = \left( 5/4 \right) \norb$.

Results are plotted in \fig{fig:final_rotation}. Each panel of the figure represents the normalised final rotation period $\Prot / \Porb$ as a function of the eccentricity and ocean depth in logarithmic scales for a given case defined by the type of planet (Earth or super-Earth), the orbital period ($\Porb$), and the Rayleigh drag frequency ($\fdrag$). \jlc{We note that, although global oceans of depths $\lesssim 0.1$~km are very unlikely \padc{owing to the extremely flat topography it would suppose}, the lower bound of $\logdix \left( \Hoc \right) $ is set to $\logdix \left( \Hoc \right) = - 3$ in order to emphasise the asymptotic regime of dry planets}. The yellow colour designates the region of the parameter space where the final state is the spin-orbit synchronous rotation, while shades of blues indicate asynchronous rotation states of equilibrium such that $\norb < \spinrateeq < \left( 3/2 \right) \norb$. The six top and middle panel correspond to the Earth, the three bottom ones to the 10-$\Mearth$ super-Earth. 

Numerical calculations highlight the mechanism discussed in the previous sections: a resonance associated with an oceanic mode can decrease the lower eccentricity of the region where asynchronous final states exist. This region approximately corresponds to $\logdix \ecc \gtrsim -0.6 $ \padc{(i.e. $\ecc \approx 0.25$)} in absence of ocean, that is in the asymptotic limit $\Hoc \rightarrow 0$. \padc{We note that this value is approximately the transition eccentricity $\ecctrans$ at which the degree-3 eccentricity tidal potential becomes equal to the semidiurnal tidal potential (see \sect{ssec:perturbing_potential}).} 

As the depth of the ocean increases, the eigenfrequency of the predominant oceanic mode first equalises the frequency associated with the degree-3 eccentricity term, thus leading to the most important decay of the critical eccentricity. It then meets the frequency of the degree-4 eccentricity term, and so on, as illustrated by \fig{fig:pics_ecc}. This generates a series of peaks located at the corresponding resonant depths, denoted by $\Hocs$. Moreover, the asynchronous rotation state of equilibrium resulting from a resonance tends to get closer to the synchronisation while $\Hoc$ increases, since it follows the decay of the resonant eigenfrequency (see \fig{fig:pics_ecc}).

Owing to the scaling law $ \Hocs \scale \left (\ss - 2 \right)^2 \norb^2 $ given by \eq{Hoc_cond1}, the resonant ocean depths increase when the orbital period switches from 6 to 1~day. \smc{The smaller the orbital period, the deeper the ocean likely to be subject to resonances}. A decay of the Rayleigh drag frequency (from left to right) accentuates the observed features, and particularly the decay of the critical eccentricity in resonant configurations as discussed in \sect{sec:critical_eccentricity} (see \eqs{eccasync_ocean}{eccasync_sol}). In these cases, the value of the critical eccentricity drops from \padc{$\eccasync \approx 0.25$} to $\eccasync \approx 0.015$ (see e.g. the top right and middle panels of \fig{fig:final_rotation}). In the most extremal treated case ($\Porb = 1$\units{day} and $\fdrag = 10^{-7}$\units{s^{-1}}, middle right panel), the lowest critical eccentricity reached is $\eccasync \approx 0.006$. However, this later case is unlikely since it clearly goes beyond the scope of the linear theory. 

By considering the cases defined by a 6-days orbital period and the Rayleigh frequencies $10^{-5}$ and $10^{-6}$\units{s^{-1}} (top left and middle panels), we observe the transition from the frictional regime ($\fdrag \sim \abs{\ftide}$) to the quasi-adiabatic regime ($\fdrag \ll \abs{\ftide}$). In the first configuration, the ocean depth minimising the critical eccentricity is affected by the drag, while it is not the case for $\fdrag \lesssim 10^{-6}$\units{s^{-1}}. 

Finally, we note that the results obtained for the super-Earth are practically identical to those obtained for the Earth. This is mainly due to the fact that the oceanic tidal response generally predominates over the tidal response of the solid part, except in a very small vicinity of the lower and upper bounds of the studied frequency interval. \ebc{Thus, the final rotation of the planet is not very sensitive to the properties of the solid part, although it is strongly affected by its tidal response (see \fig{fig:spectra})}. Besides, the scale factors $\sqrt{\ggravi / \Rpla^2 }$ and $\Rpla^3 / \Mpla$ associated respectively with the eigenfrequency of a mode -- given by \eq{focn} -- and the corresponding maximum of the tidal torque -- given by \eq{maxpeak} -- are approximately the same for the two planets. As a consequence, variations related to the change of mass and radius from a case to another are small. 

In the dry asymptotic limit ($\Hoc \rightarrow 0$), one might expect that the critical eccentricity depends on the Andrade parameters characterising the solid part of the planet, which would lead to two different values for the Earth and the super-Earth. This is not the case however, given that $\eccasync \approx 0.25$ for both planets. To explain this feature, we come back to the values given by \tab{tab:andrade_para}. The Maxwell and Andrade timescales of both planets far exceed typical tidal periods. As a consequence, the imaginary part of the degree-2 Love number is well approximated by its high-frequency asymptotic functional form, given by \eq{k2sol_asymp}. This allows us to factorise all eccentricity terms (\eq{torque_sol}) by

\begin{equation}
\Csol =  \frac{\Aquad}{\left( 1+\Aquad \right)^2} \GammaF \left( 1 + \alphaA \right) \sin \left( \frac{\alphaA \pi}{2} \right) \left( \tauA \right)^{-\alphaA},
\label{Csol}
\end{equation}

\comments{Attention, \eq{k2sol_asymp} a un impact sur ce résultat.}

\noindent and write the tidal torque exerted on the solid body as 

\begin{equation}
\torquesol = - \frac{9}{4} \Ggrav \Mstar^2 \frac{\Rpla^5}{\smaxis^6} \Csol \sum_{\ss = - \infty}^{+ \infty} \left[ \Hansen{\ss}{-3}{2} \left( \ecc \right)   \right]^2  \sign \left( \ftidequads\right) \abs{\ftidequads}^{-\alphaA}. 
\end{equation}

As may be noticed, the sign of the torque in this equation only depends on the rheological exponent $\alphaA$, set to $0.25$ for both planets, and the eccentricity tidal frequencies and Hansen coefficients. Thus, the critical eccentricity is not affected at all by the other parameters of the solid body, except in the close vicinity of the resonances associated with eccentricity terms. This explains why no difference can be detected between the Earth and the super-Earth in the dry asymptotic regime ($\Hoc \rightarrow 0$) in \fig{fig:pics_ecc} (top and bottom panels).

The above observations suggest that final rotation states of equilibrium distinct from spin-orbit resonances ($\spinrateeqs = \left( \ss / 2 \right) \norb$) are not very sensitive to most of the parameters of the solid part (size, mass, bulk rigidity, Maxwell and Andrade times), although they are affected by the rheological exponent ($\alphaA$) in the high-eccentricity regime. The important parameters for these states are the ocean parameters (depth, density, Rayleigh drag frequency) and the orbital period.


\begin{figure*}[htb]
   \centering
   \includegraphics[width=0.325\textwidth,trim = 3.6cm 2.5cm 4.8cm 2.0cm,clip]{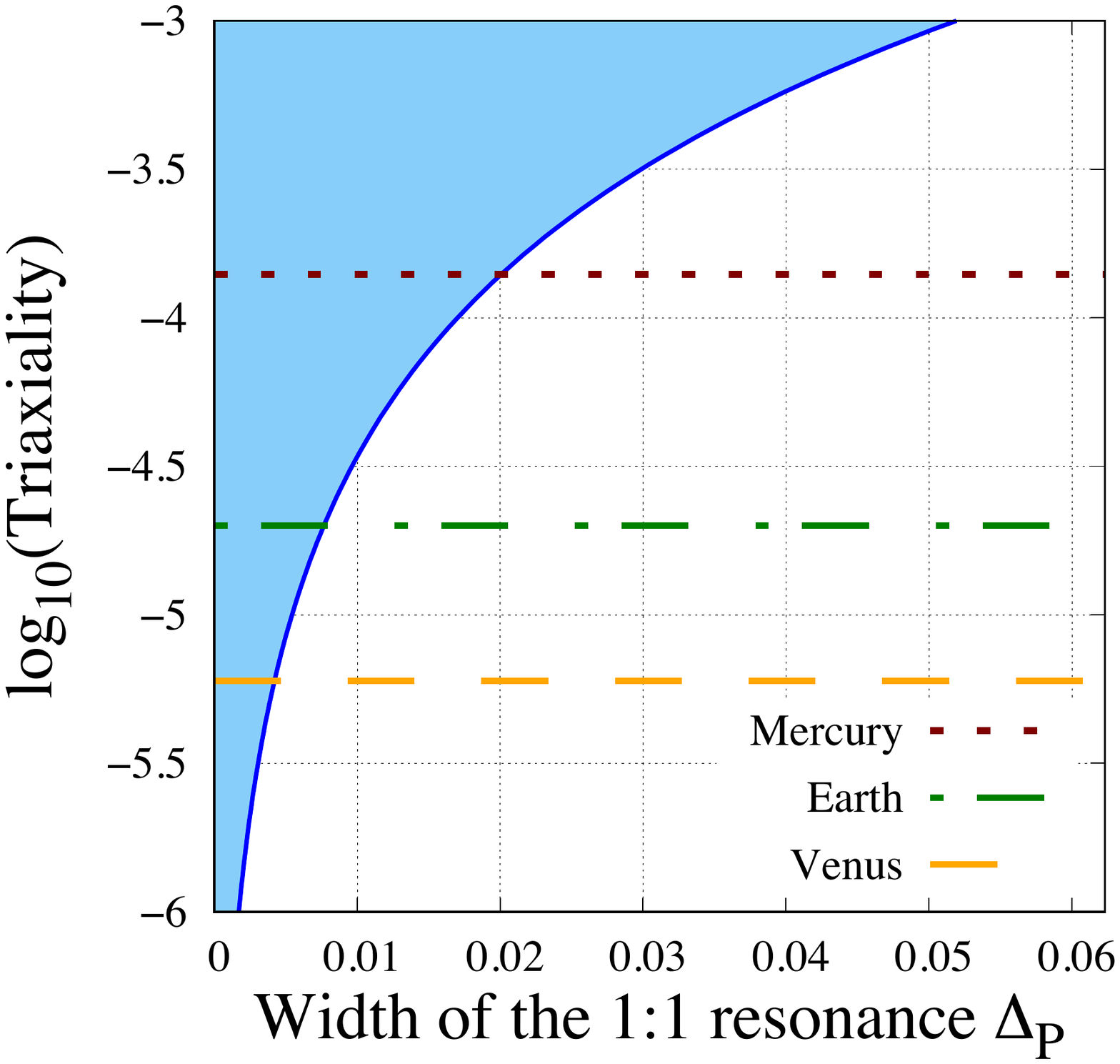} 
    \includegraphics[width=0.32\textwidth,trim = 1.0cm 4.5cm 3.0cm 2.0cm,clip]{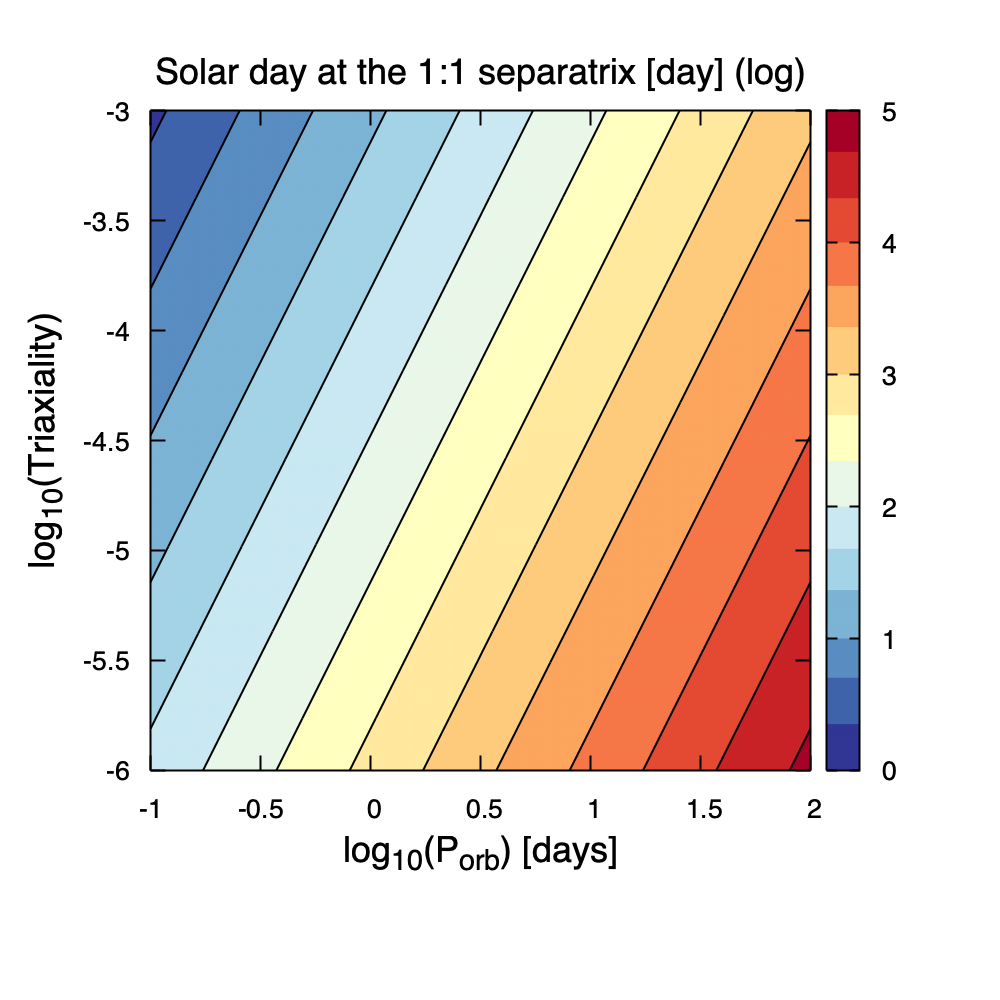}
    \includegraphics[width=0.32\textwidth,trim = 1.0cm 4.5cm 3.0cm 2.0cm,clip]{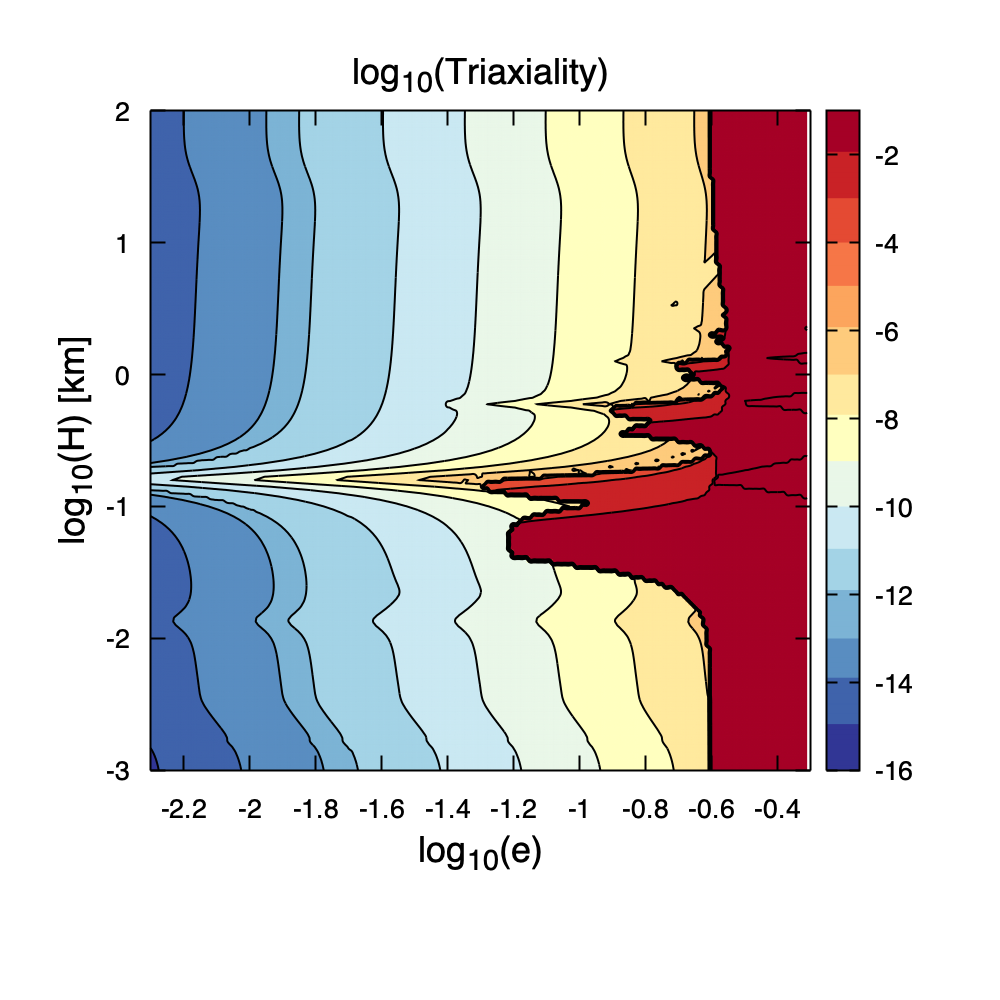} 
   \caption{\jlc{{\it Left panel}: normalized width of the 1:1 spin-orbit resonance $\DeltaProt = 1 - \Protres / \Porb$ as a function of the triaxiality of the planet. The blue zone show the area  covered by librations inside the resonance, which is the resonance width. Dotted and dashed brown, green and orange lines correspond to the triaxialities of Mercury \citep[${\sim} 1.4 \times 10^{-4}$ ;][]{Smith2012,Ribas2016}, Earth (${\sim} 2 \times 10^{-5}$), and Venus \citep[${\sim} 6 \times 10^{-6}$;][]{Yoder1995}, respectively. {\it Middle panel}: logarithm of the Solar day of the planet at the bound of the 1:1 resonance as a function of its orbital period in Earth days (horizontal axis) and triaxiality (vertical axis) in logarithmic scales. {\it Right panel}: logarithm of the triaxiality such that the final asynchronous state of equilibrium of the planet corresponds to the width of the 1:1 spin-orbit resonance as a function of the eccentricity (horizontal axis) and ocean depth (vertical axis, km) of the planet in logarithmic scales. The treated case corresponds to the top middle panel of \fig{fig:final_rotation} ($\Mpla = \Mearth$, $\Porb = 6$~days, and $\fdrag = 10^{-6}$\units{s^{-1}}).}}
       \label{fig:resonance_width}%
\end{figure*}

\subsection{\jlc{Capture in the 1:1 spin-orbit resonance}}
\label{ssec:capture}
In reality, the final rotation state of equilibrium of the planet is not determined by the tidal torque solely, but also depends on its triaxiality, that is its permanent, non-axisymmetric deformation. As highlighted by \cite{Goldreich1966} and \cite{GP1966}, the triaxiality of the planet creates an additional gravitational torque, which tends to trap it in spin-orbit resonances. 

In the framework of the coplanar case, the torque due to triaxiality -- denoted $\torquetri$ -- adds to the tidal torque -- superscripted '$\itide$' here to avoid confusion with the total torque -- in the right-hand member of the equation of the spin evolution,

\begin{equation}
\Cinertpla \Dt{\spinrate} = \torquetri + \torquepla^{\rm tid}.
\label{spin_evolution_eq}
\end{equation}

\noindent  \smc{This torque} is expressed \rec{in the general case} as \rec{(see \append{app:triaxial_torque})}

\begin{equation}
\torquetri \define -\frac{3}{2} \left( \Binertpla - \Ainertpla \right) \norb^2 \sum_{\ss = - \infty}^{+ \infty}   \Hansen{\ss}{-3}{2} \left( \ecc \right)  \sin \left( 2 \anglequads \right).
\label{torquetri}
\end{equation}

 \rec{In the preceding equations,} the symbol $~\Dt{}~$ is the time derivative, $\Ainertpla$, $\Binertpla$, and $\Cinertpla$ the principal moments of inertia of the planet (in increasing magnitude), and $\anglequads \define  \spinangle - \left( \ss / 2 \right) \meana$ the angles associated with the eccentricity frequencies $\ftidequads $, the notation $\spinangle$ and $\meana$ designating the planet rotation angle and mean anomaly, respectively. The triaxiality, $\left( \Binertpla  - \Ainertpla \right) / \Cinertpla$, is the dimensionless parameter that quantifies the degree of permanent non-\rec{rotationally symmetric} deformation of the planet. It annihilates for a spherically symmetric body, where $\Ainertpla = \Binertpla = \Cinertpla$. 

In the vicinity of the spin-orbit resonance $\ss / 2$, the torque due to triaxiality is dominated by the degree-$\ss$ term in \eq{torquetri} \rec{when averaged over an orbital period \citep[][]{Goldreich1966,GP1966,Ribas2016}} and can thus be approximated by\footnote{\rec{In the vicinity of the spin-orbit resonance $\ss/2$, the angle $\anglequads$ is small. Therefore, integrating \eq{torquetri} over an orbital period makes that all eccentricity terms vanish except the degree-$\ss$ term, which reduces \eq{spin_evolution_eq} to the equation of motion of a simple pendulum in the absence of tides \citep[see e.g., Eq.~(5) in][]{Goldreich1966}. However, we note that $\sin \left( \anglequads \right) $ vanishes at the resonance ($\spinangle / \meana = \ss / 2 $).   }}

\begin{equation}
\torquetri \approx - \frac{3}{2} \left( \Binertpla - \Ainertpla \right) \norb^2  \Hansen{\ss}{-3}{2} \left( \ecc \right)  \sin \left( 2 \anglequads \right),
\end{equation}

\noindent \rec{which is oscillatory and  leads to libration when the planet is in spin-orbit resonance. This approximation allows us to} rewrite the equation of the spin evolution as

\begin{equation}
\Cinertpla \Dtt{\anglequads}  = \torquetri \left( \anglequads \right) + \torqueplatid \left( \Dt{\anglequads} \right) ,
\label{spin_evolution_angle}
\end{equation}

\noindent where we have made use of the relationship $\ftidequads = 2 \Dt{\anglequads} $ and denoted the second order time derivative by $~\Dtt{} ~$. Assuming that the triaxial torque predominates in the vicinity of the resonance, we neglect the tidal torque ($\left| \torquepla^{\rm tid} \right| \ll \left|  \torquetri  \right|$). It follows that the maximum absolute value that $\Dt{\anglequads}$ can reach inside the resonance is given by \citep[][Eq.~(17)]{GP1966}

\begin{equation}
\Deltares \define \norb \sqrt{3 \frac{\Binertpla - \Ainertpla}{\Cinertpla} \Hansen{\ss}{-3}{2} \left( \ecc \right)}.
\label{Deltares}
\end{equation}

Following \cite{Ribas2016}, $\Deltares$ is called the width of the resonance in the following, since the separatrix between the librating (trapped) and circulating states is defined as 

\begin{equation}
\Dt{\anglequads} \define \Deltares \cos \anglequads. 
\label{separatrix}
\end{equation}

\noindent The width of the resonance defines the frequency range where a capture may occur, that is the range where the planet may be driven towards the exact spin-orbit ratio of the resonance by the triaxial torque. In this range, the existence of rotation equilibria distinct from the spin-orbit resonance depends on the combination of the triaxial and tidal torques. As a consequence, asynchronous equilibria defined by the annihilation of the tidal torque and plotted in \fig{fig:final_rotation} may not exist if $ \left| \ftidequads \right| < 2 \Deltares $. 

In addition to this criterion, \cite{GP1966} derived a probability of capture in the case of weak tidal torques. Their theory, generalised by \cite{Makarov2012}, established that this probability only depends on the ratio between the even and odd components of the tidal torque with respect to the $\ss/2$ resonance, defined as

\begin{align}
\label{torqueeven}
\torqueeven \left( \Dt{\anglequads}  \right) \define & \dfrac{1}{2} \left[ \torqueplatid \left( \Dt{\anglequads} \right) + \torqueplatid \left( - \Dt{\anglequads} \right) \right] , \\
\label{torqueodd}
\torqueodd \left( \Dt{\anglequads}\right) \define & \dfrac{1}{2} \left[ \torqueplatid \left( \Dt{\anglequads} \right) - \torqueplatid \left( - \Dt{\anglequads} \right) \right]. 
\end{align}

\noindent The even part of the torque tends to make the planet traverse the resonance, while the odd part drives it back towards this configuration. Thus the capture probability of the $ \ss/2$ spin-orbit resonance is defined for any tidal torque as \citep[][]{GP1966,Makarov2012,Ribas2016}

\begin{equation}
\captureprob \define 2 \left[ 1 + \frac{\integ{\torqueeven \left( \Dt{\anglequads} \right) }{\anglequads}{-\pi/2}{\pi/2}}{\integ{\torqueodd \left( \Dt{\anglequads} \right) }{\anglequads}{-\pi/2}{\pi/2}} \right]^{-1},
\label{captureprob}
\end{equation}  

\noindent where the integrals should be performed over the separatrix between the librating and circulating states, given by \eq{separatrix}. We note that $\captureprob >1$ when the integral of the odd component is greater than that of the even component, meaning that the capture always occurs in this case.

We focus on the $1{:}1$ spin-orbit resonance. In this case, the mechanism described in this section has strong repercussions on the climate and surface conditions of the planet since it determines whether the body is tidally locked in synchronous rotation or in a non-synchronous state. For comparisons with results  obtained in previous sections, we introduce the normalised width of the $1{:}1$ resonance, 

\begin{equation}
\DeltaProt \define 1 - \frac{\Protres}{\Porb} \define \frac{\Deltares}{\Deltares + \norb},
\label{DeltaProt}
\end{equation}

\noindent where $\Protres \define 2 \pi / \left( \Deltares + \norb \right)$ designates the rotation period at the bound of the resonance, that is such that $\spinrate - \norb = \Deltares$. 

This normalised width is plotted as a function of the logarithm of the triaxiality of the planet (\fig{fig:resonance_width}, left panel), with indicative levels corresponding to the triaxialities of several rocky planets of the Solar system (in decreasing orders of magnitude): ${\sim} 1.4 \times 10^{-4}$ for Mercury, as derived by \cite{Ribas2016} from the gravity moments measured by \cite{Smith2012}, ${\sim}  2 \times 10^{-5} $ for the Earth, and ${\sim} 6 \times 10^{-6}$ for Venus \citep[][]{Yoder1995}. Similarly, the length of the Solar day at the bound of the resonance ($\Dt{\anglequads} = \Deltares$) is plotted as a function of the orbital period and triaxiality in logarithmic scales (\fig{fig:resonance_width}, middle panel). In this two plots, we assume $\Hansen{2}{-3}{2} \left( \ecc \right) =  1$, since this Hansen coefficient hardly varies for $\ecc \lesssim 0.3$.

The first plot shows that typical values of the normalised width of the $1{:}1$ spin-orbit resonance fall \padc{within the interval $0.5 - 3 \%$} of the orbital period, which corresponds to the yellow areas of the parameter space in \fig{fig:final_rotation}. The Solar day associated with these values, $\Psolres \define \Protres \Porb / \left( \Porb - \Protres \right) = 2 \pi / \Deltares $ varies from ${\sim} 50$ to ${\sim} 3000$ Earth days in the range $\Porb \sim 1-10 $ Earth days. As a first approximation, such day lengths may be considered as upper estimations of those that can be reached by asynchronous rotation equilibria.

Using this criterion, we calculate the triaxiality necessary to make the width of the synchronisation coincide with the final state determined by the tidal torque,

\begin{equation}
\frac{\Binertpla - \Ainertpla}{\Cinertpla} = \frac{1}{3 \Hansen{\ss}{-3}{2} \left( \ecc \right)} \left( \frac{\spinrateeq}{\norb} - 1 \right)^2,
\end{equation}

\noindent where $\spinrateeq$ designate the rotation rate of the final state derived from the tidal torque. This triaxiality is plotted in logarithmic scale in the case of an Earth-sized planet of 6~day orbital period with $\fdrag = 10^{-6}$\units{s^{-1}} (\fig{fig:final_rotation}, top middle panel) as a function of the logarithms of the eccentricity and ocean depth (\fig{fig:resonance_width}, right panel). 

The very small triaxialities obtained outside of oceanic tidal resonances for $\ecc \lesssim 0.25$ (blue to orange areas) reveal that the planet is likely to end in tidally locked spin-orbit synchronous rotation in this region of the parameter space, instead of being driven towards slightly asynchronous states of equilibrium. Conversely, the high triaxialities observed for asynchronous states induced by the oceanic tide (red areas) suggest that the corresponding final rotation rates are not affected by the triaxial torque \ebc{unless triaxiality is very high}. 

 We characterised in this section the evolution of final rotations with the eccentricity and the ocean depth \jlc{in light of the theory of capture in spin-orbit resonance}, and recovered the features described by the analytical theory detailed in \sect{sec:critical_eccentricity}. We now focus on the critical eccentricity below which the planet is driven towards the spin-orbit synchronous rotation. We aim to identify the regions of the parameter space where this eccentricity is the smallest.

\begin{figure*}[t]
   \centering
 \begin{flushleft}
    \hspace{0.13\textwidth} 
  \includegraphics[height=0.015\textheight]{auclair-desrotour_fig6a} 
   \hspace{0.07\textwidth} 
    \includegraphics[height=0.018\textheight]{auclair-desrotour_fig6b} \hspace{0.13\textwidth} 
  \includegraphics[height=0.018\textheight]{auclair-desrotour_fig6c}  \hspace{0.13\textwidth} 
 \end{flushleft}
 \vspace{-0.2cm}
  \raisebox{1.5cm}{\includegraphics[width=0.018\textwidth]{auclair-desrotour_fig6f}}
   \includegraphics[width=0.28\textwidth,trim = 2.8cm 6.5cm 6.4cm 3.5cm,clip]{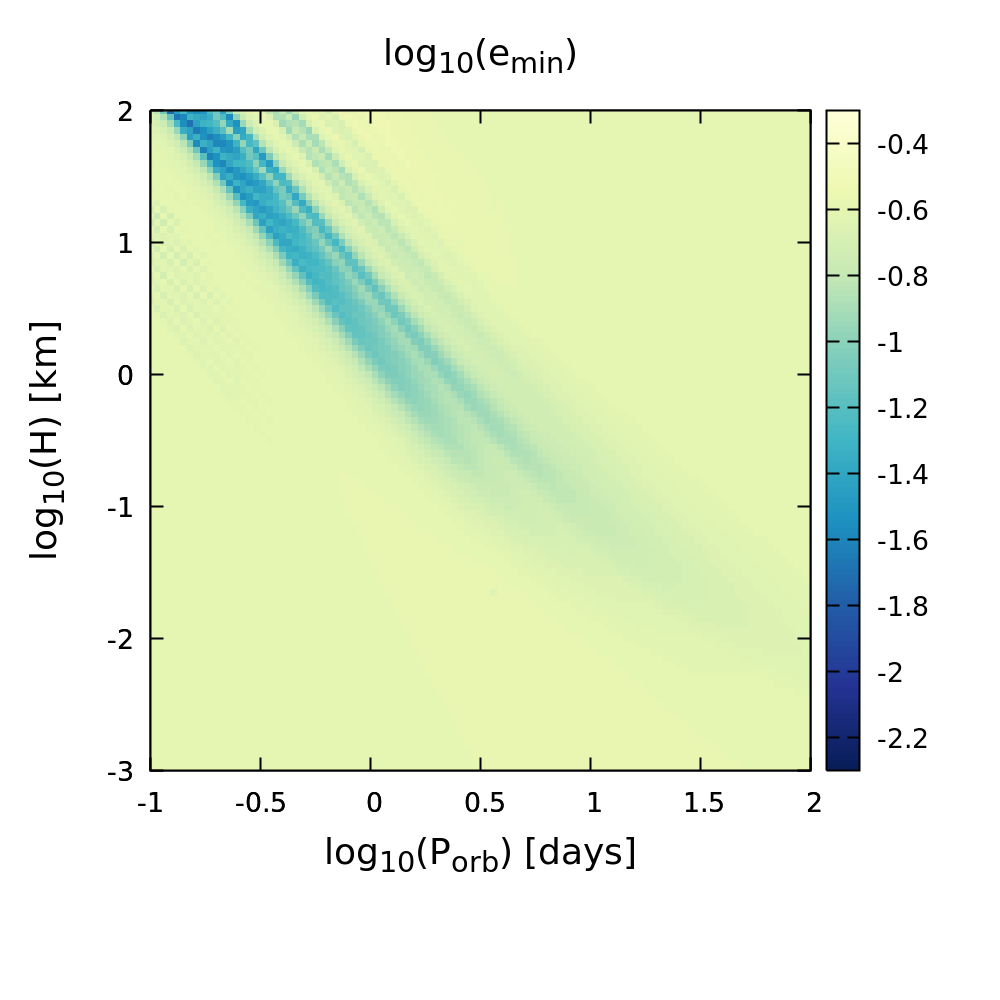} 
    \includegraphics[width=0.28\textwidth,trim = 2.8cm 6.5cm 6.4cm 3.5cm,clip]{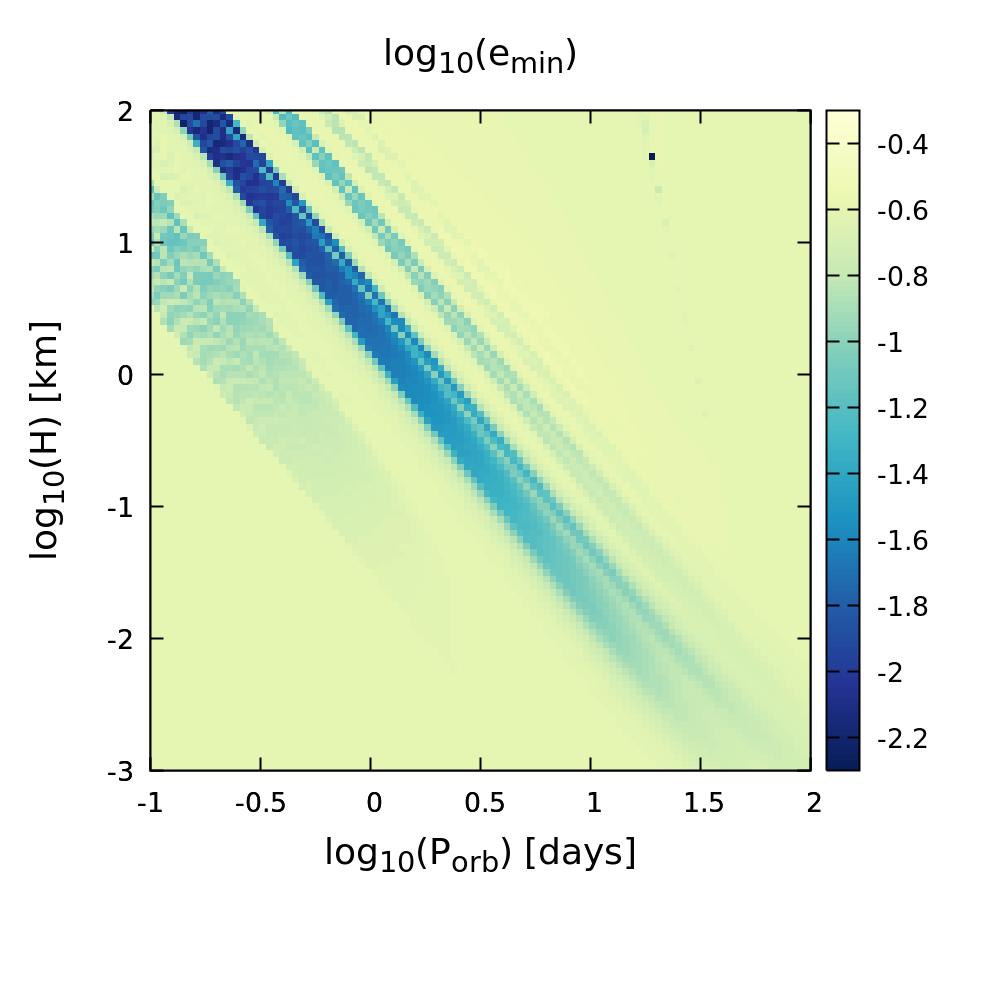}
    \includegraphics[width=0.046\textwidth,trim = 29cm 6.5cm 2.0cm 3.5cm,clip]{auclair-desrotour_fig8a.png} 
    \hspace{0.2cm}
    \raisebox{1.7cm}{\includegraphics[width=0.025\textwidth]{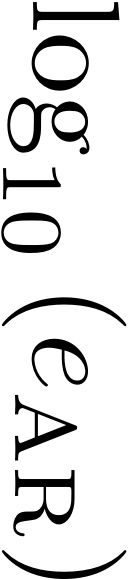}} \\[-0.4cm]
   \raisebox{1.5cm}{\includegraphics[width=0.018\textwidth]{auclair-desrotour_fig6f}}
   \includegraphics[width=0.28\textwidth,trim = 2.8cm 4.5cm 6.4cm 3.5cm,clip]{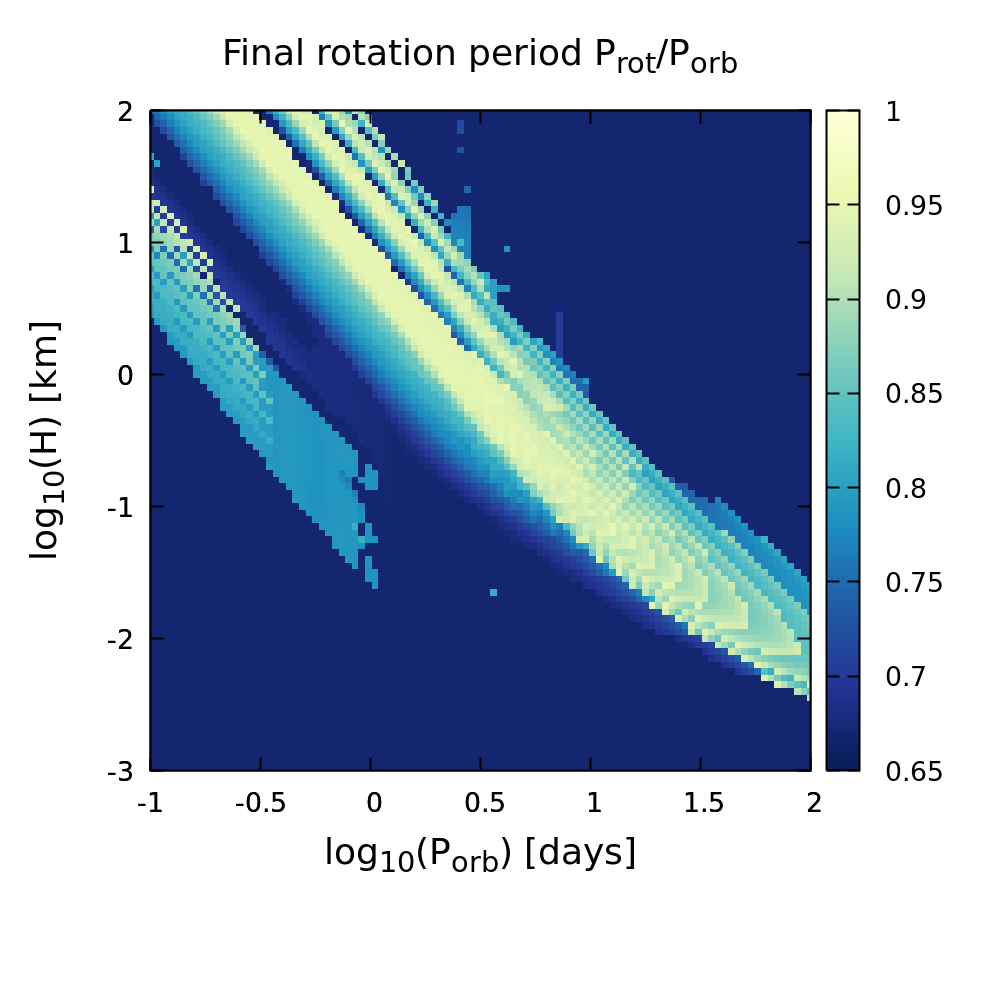} 
    \includegraphics[width=0.28\textwidth,trim = 2.8cm 4.5cm 6.4cm 3.5cm,clip]{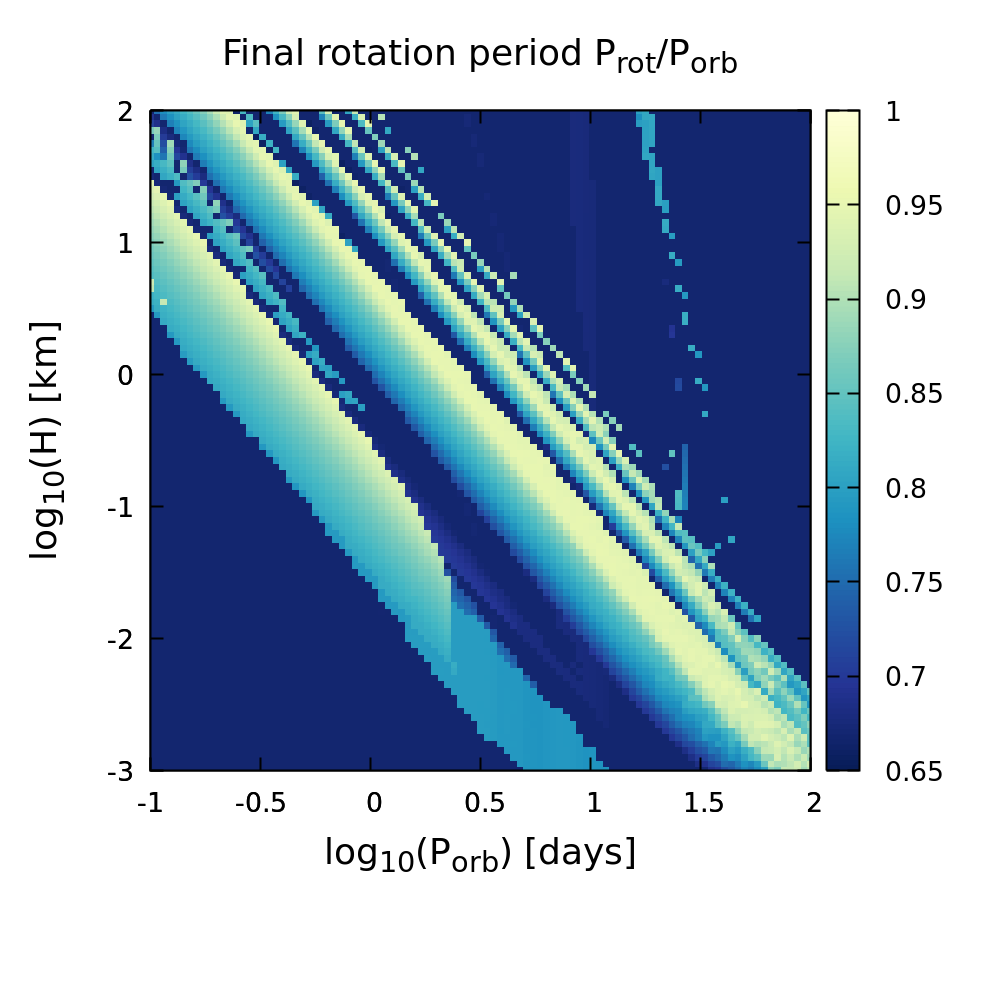}
\includegraphics[width=0.046\textwidth,trim = 29cm 4.5cm 2.0cm 1.2cm,clip]{auclair-desrotour_fig8d.png} 
\hspace{0.2cm}
\raisebox{0.5cm}{\includegraphics[width=0.025\textwidth]{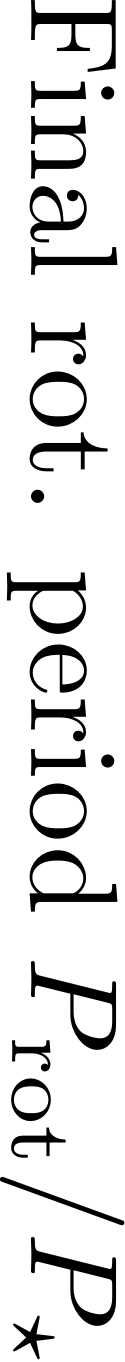}}

   \caption{Logarithm of the critical eccentricity $\eccasync$ (top panels) and associated normalised final rotation period $\Prot/\Porb$ (bottom panels) as functions of the logarithms of the orbital period (horizontal axis) and ocean depth (vertical axis), for $\fdrag = 10^{-5}$\units{s^{-1}} (left) and $\fdrag = 10^{-6}$\units{s^{-1}} (right). The calculated critical eccentricity is such that the final rotation period satisfies the condition $\Prot/\Porb \leq 0.95$ to retain solely asynchronous states clearly separated from the synchronisation. In the eccentricity maps (top panels), the yellow colour designates the region of the non-resonant regime, where the critical eccentricity is high ($\logdix \eccasync \approx -0.6$). Conversely, the blue shades designate the regions where $\eccasync$ is low owing the resonances of the oceanic tidal response. In the final rotation maps, the yellow colour designates states that are close to synchronisation, while the dark blue colour corresponds to the 3/2 spin-orbit resonance, induced by the degree-3 eccentricity term. Values of parameters used for the solid part are given by \tab{tab:andrade_para}. In both cases, the stellar mass is $\Mstar = 0.09 \ \Msun$, the initial rotation rate is $\spinrate = \left( 5/4 \right) \norb$, and 10~Hough modes are taken into account in the calculation of the tidal response.}
       \label{fig:critical_ecc}%
\end{figure*}

\section{Evolution of the critical eccentricity with the orbital period and ocean depth}
\label{sec:critical_ecc}

The critical eccentricity is the parameter that determines whether asynchronous rotation states of equilibrium may exist outside of spin-orbit resonances \citep[$\spinrateeq {:} \norb =  3{:}2,2 {:} 1,5 {:} 2,\ldots $; e.g.][]{ME2013,Correia2014} or not. It provides a qualitative information on the rotation state at which a planet can be found. Thus, we examine in this section how it depends on the orbital period of the planet and its ocean depth. Since we obtained similar results for the Earth and super-Earth studied in the preceding section, we consider here the Earth case solely. 

As highlighted by \fig{fig:final_rotation}, oceanic tides can generate asynchronous states in the vicinity of the synchronisation ($\spinrateeq \approx \norb$). \jlc{In these states, the planet may be tidally locked in the 1:1 spin-orbit resonance by the torque due to triaxiality. Hence, we are interested in the asynchronous states that are outside of the resonance.} For this reason, we define in this section the critical eccentricity as the minimal eccentricity such that the final rotation period satisfies the condition $\Prot / \Porb \leq 0.95$. This means that the departure between the final rotation period and the orbital period of the planet has to be greater than $5 \%$ of the orbital period, \jlc{which is the upper estimation of the width of the $1{:}1$ resonance obtained for a triaxiality ${\sim} 10^{-3}$ (see \fig{fig:resonance_width}, left panel)}.

In \fig{fig:critical_ecc}, we plot the logarithm of the critical eccentricity calculated using the above definition (top panels) and the associated normalised final rotation period $\Prot/\Porb$ (bottom panels) as functions of the logarithms of the orbital period (horizontal axis) and ocean depth (vertical axis), for a strong drag ($\fdrag = 10^{-5}$\units{s^{-1}}, left) and a weak drag ($\fdrag = 10^{-6}$\units{s^{-1}}, right). To obtain these maps, we performed the calculations detailed in \sect{sec:final_rotation} for each of the sampled orbital periods with the same parametrisation. 

We first consider the maps of the critical eccentricity (top panels). In these maps, yellow-green areas designate the non-resonant regime, where the critical eccentricity is high ($\logdix \eccasync \approx -0.6$). This corresponds to the asymptotic configuration of a dry rocky planet. Resonances of the oceanic tidal response induce blue diagonal bands, where the critical eccentricity is decreased with respect to the reference value. Each of these bands, from left to right, is associated with an eccentricity degree in ascending order, the main one being due to the degree-3 eccentricity term. The diagonal pattern follows the scaling law derived analytically for the resonant ocean depths (see \eq{Hoc_cond1}), that is

\begin{equation}
\Hocs \scale \left( \ss - 2 \right)^2 \Porb^{-2}.
\end{equation}

Moreover, the contrast between the non-resonant and resonant regimes becomes stronger as the Rayleigh drag frequency and orbital period decay, in agreement with the scaling laws given by \eqs{eccasync_ocean}{eccasync_sol}, that is 

\begin{equation}
\begin{array}{ll}
\eccasync \scale \fdrag \Porb & \mbox{(ocean-dominated equilibrium tide),} \\[0.2cm]
\eccasync \scale \fdrag^{1/2} \Porb^{\left(1 + \alphaA\right)/2} & \mbox{(solid-dominated equilibrium tide),}
\end{array}
\end{equation}

\noindent respectively. We remind here that these scaling laws where derived in the framework of approximations such as the quasi-adiabatic ($\fdrag \ll \abs{\ftide}$) and \padc{non-rotating} (no Coriolis effects) approximations, which leads to differences with the results of numerical calculations, and particularly in the strong drag configuration ($\fdrag = 10^{-5}$\units{s^{-1}}).

We now move to the maps representing the final rotation rates associated with the computed critical eccentricity (bottom panels). Yellow areas designate sates of equilibrium close to synchronisation, and blue areas the $3{:}2$ spin-orbit resonance. In order to avoid confusion when looking at these maps, one should bear in mind that they show both the low-eccentricity regime, which is the subject of the analytical theory detailed in \sects{sec:tidal_torque}{sec:critical_eccentricity}, and the high-eccentricity regime, where the derived results do not apply since it is beyond the scope of this work. 

We shall also stress here a somehow counter-intuitive feature of the behaviour of the final spin rate that we already emphasized in \fig{fig:pics_ecc} and discussed in \sect{sec:critical_eccentricity}: as the ocean depth increases, the peak generated by a resonant eccentricity term moves towards the spin-orbit synchronisation, and so does the associated asynchronous state of equilibrium. Thus, the asynchronous state generated by a strong eccentricity term is not necessarily far from the synchronisation. We retrieve this feature in \fig{sec:critical_ecc} (bottom panels). While no correlation can be observed between the final rotation and the critical eccentricity in the low-eccentricity regime, the displacement of the state of equilibrium induces a colour gradation from blue to yellow along the direction of ascending ocean depths. 

Up to now, we examined the final rotation state of equilibrium of the planet without considering how much time is necessary to reach this state. The evolution timescale of the planet rotation rate is related to the tidally dissipated energy, which strongly depends on the forcing frequencies associated with tidal components \citep[e.g.][]{ADLPM2014}. It is thus important to complete the present work with a study of the evolution of the planet spin. This is the object of the next section.

\begin{figure*}[t]
   \centering
 \begin{flushleft}
  \includegraphics[height=0.015\textheight]{auclair-desrotour_fig6a} 
   \hspace{0.11\textwidth} 
  \includegraphics[height=0.012\textheight]{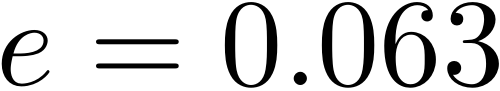} \hspace{0.2\textwidth} 
  \includegraphics[height=0.012\textheight]{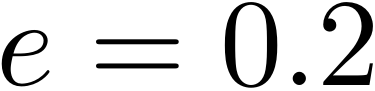}  \hspace{0.2\textwidth} 
  \includegraphics[height=0.012\textheight]{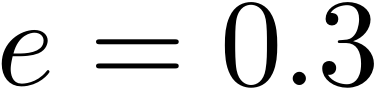} \\
 \end{flushleft}
 \vspace{-0.2cm}
 \raisebox{1.2cm}{\includegraphics[width=0.025\textwidth]{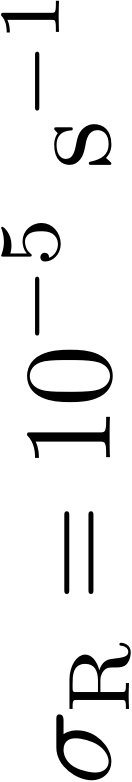}}
 \hspace{0.2cm}
  \raisebox{0.5cm}{\includegraphics[width=0.018\textwidth]{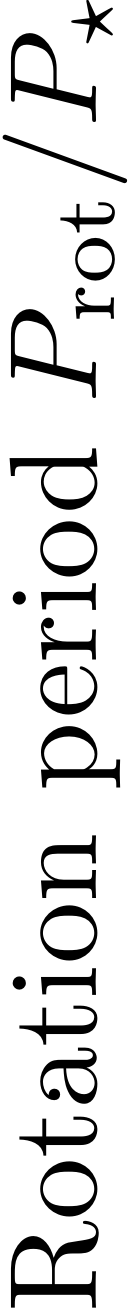}}
   \includegraphics[width=0.27\textwidth,trim = 3.5cm 4.0cm 8.0cm 2.1cm,clip]{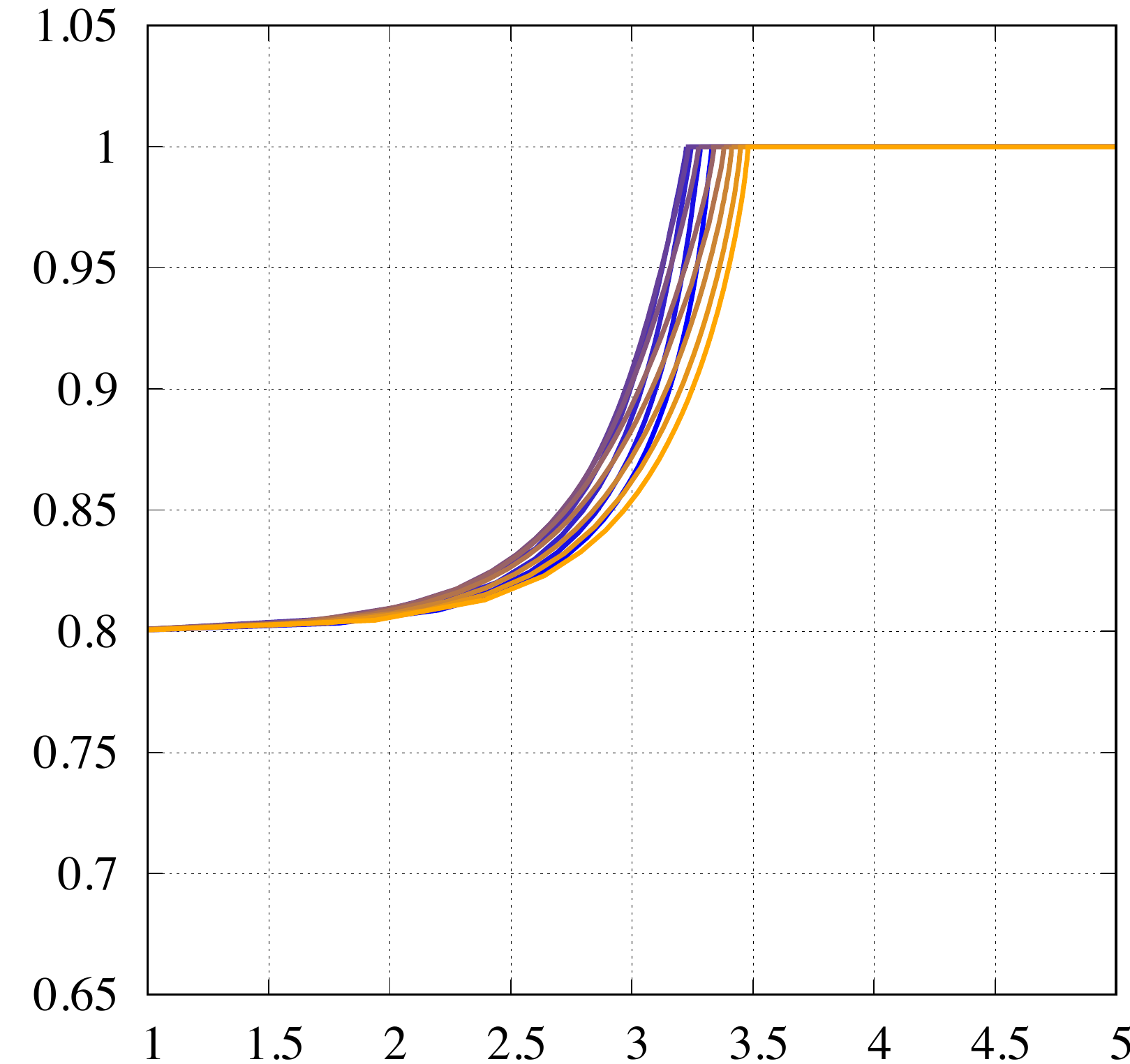} 
    \includegraphics[width=0.27\textwidth,trim = 3.5cm 4.0cm 8.0cm 2.1cm,clip]{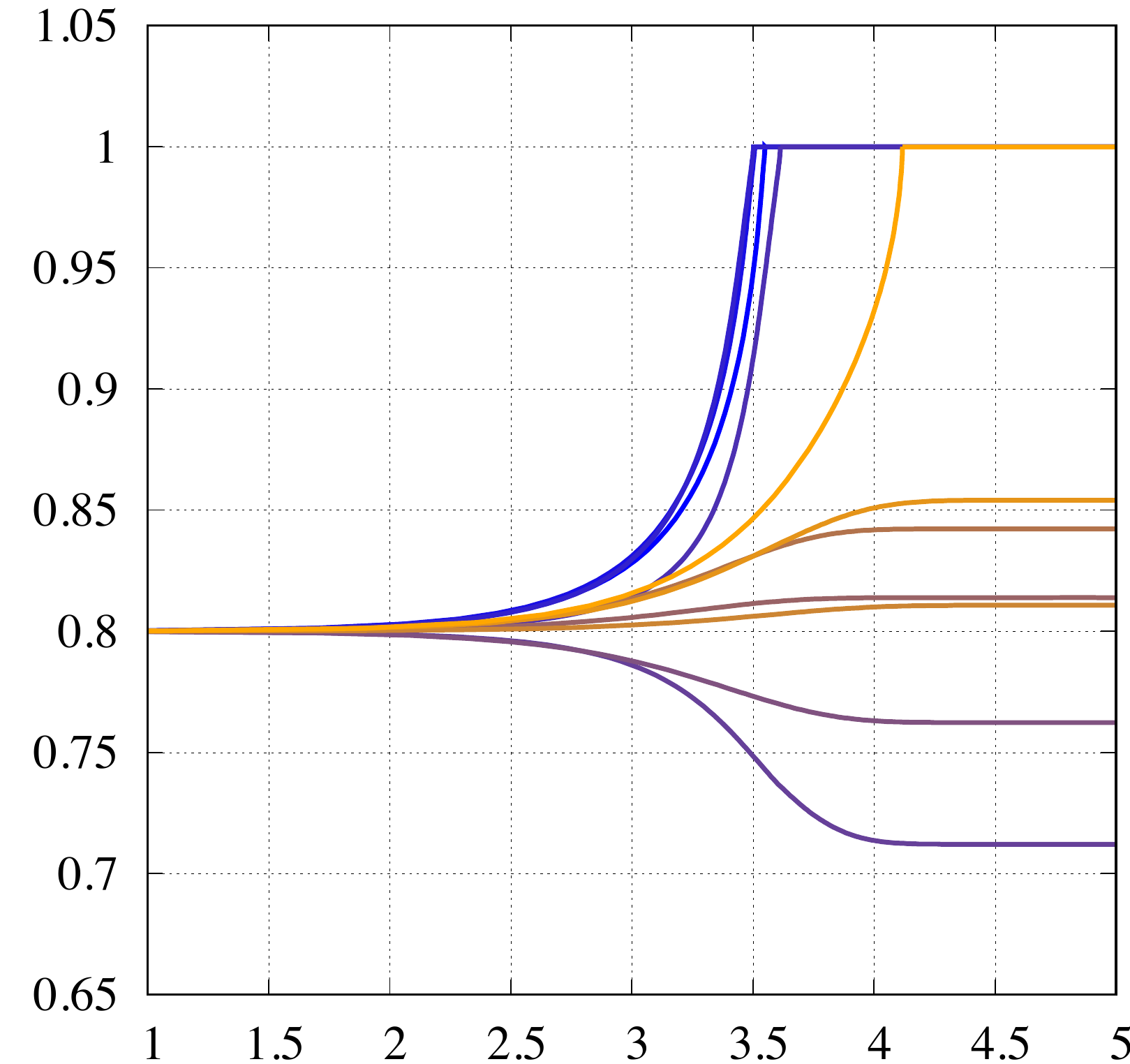}
     \includegraphics[width=0.27\textwidth,trim = 3.5cm 4.0cm 8.0cm 2.1cm,clip]{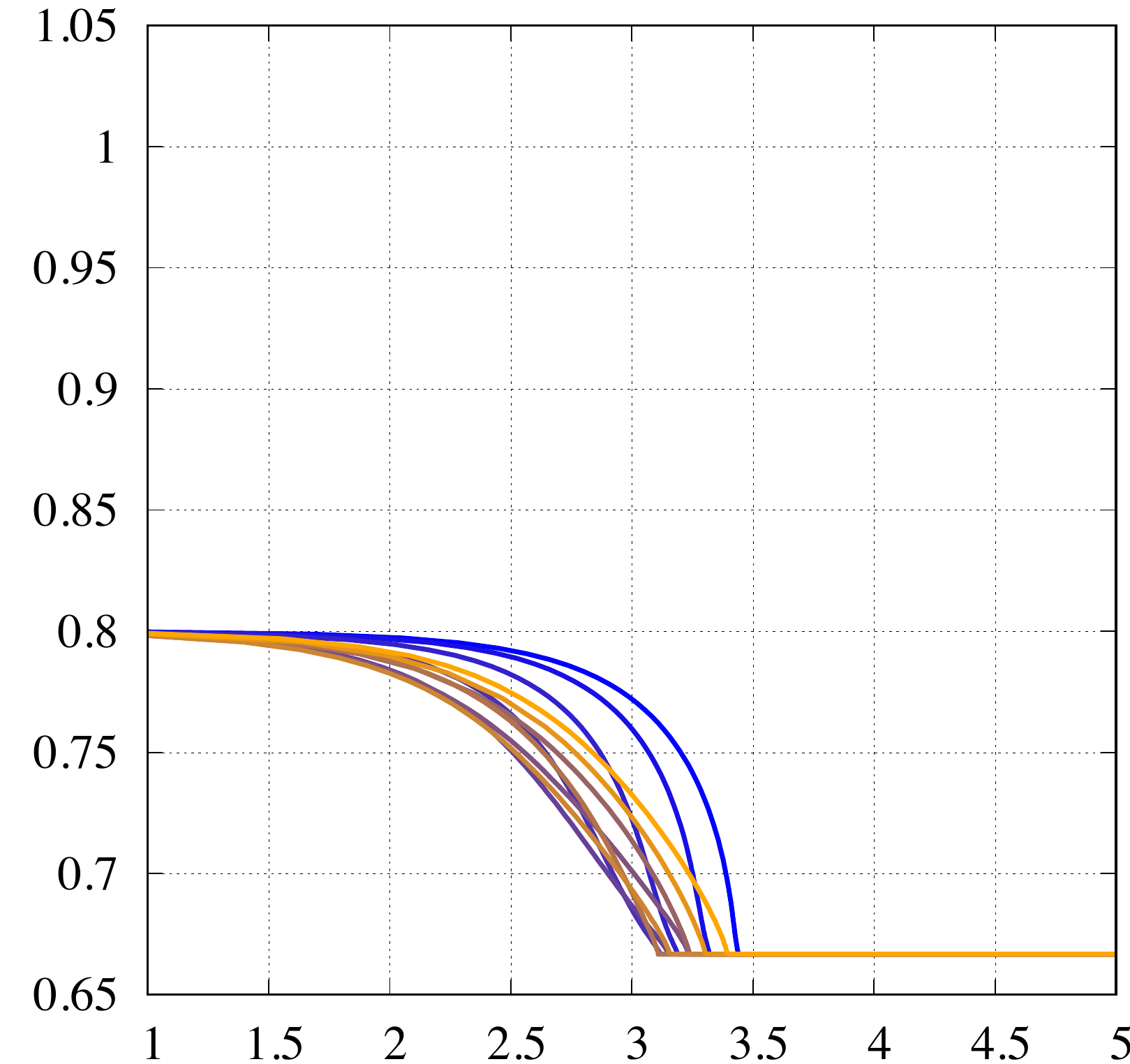}
    \includegraphics[width=0.1\textwidth,trim = 21cm 6.0cm 1.5cm 2.1cm,clip]{auclair-desrotour_fig9f} \\
   \raisebox{1.2cm}{\includegraphics[width=0.025\textwidth]{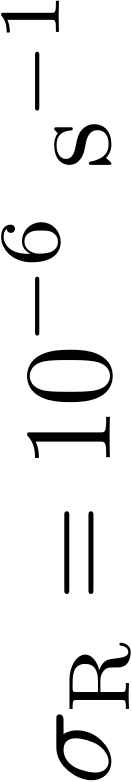}}
    \hspace{0.2cm}
   \raisebox{1.0cm}{\includegraphics[width=0.018\textwidth]{auclair-desrotour_fig9e}}
   \includegraphics[width=0.27\textwidth,trim = 3.5cm 2.5cm 8.0cm 2.1cm,clip]{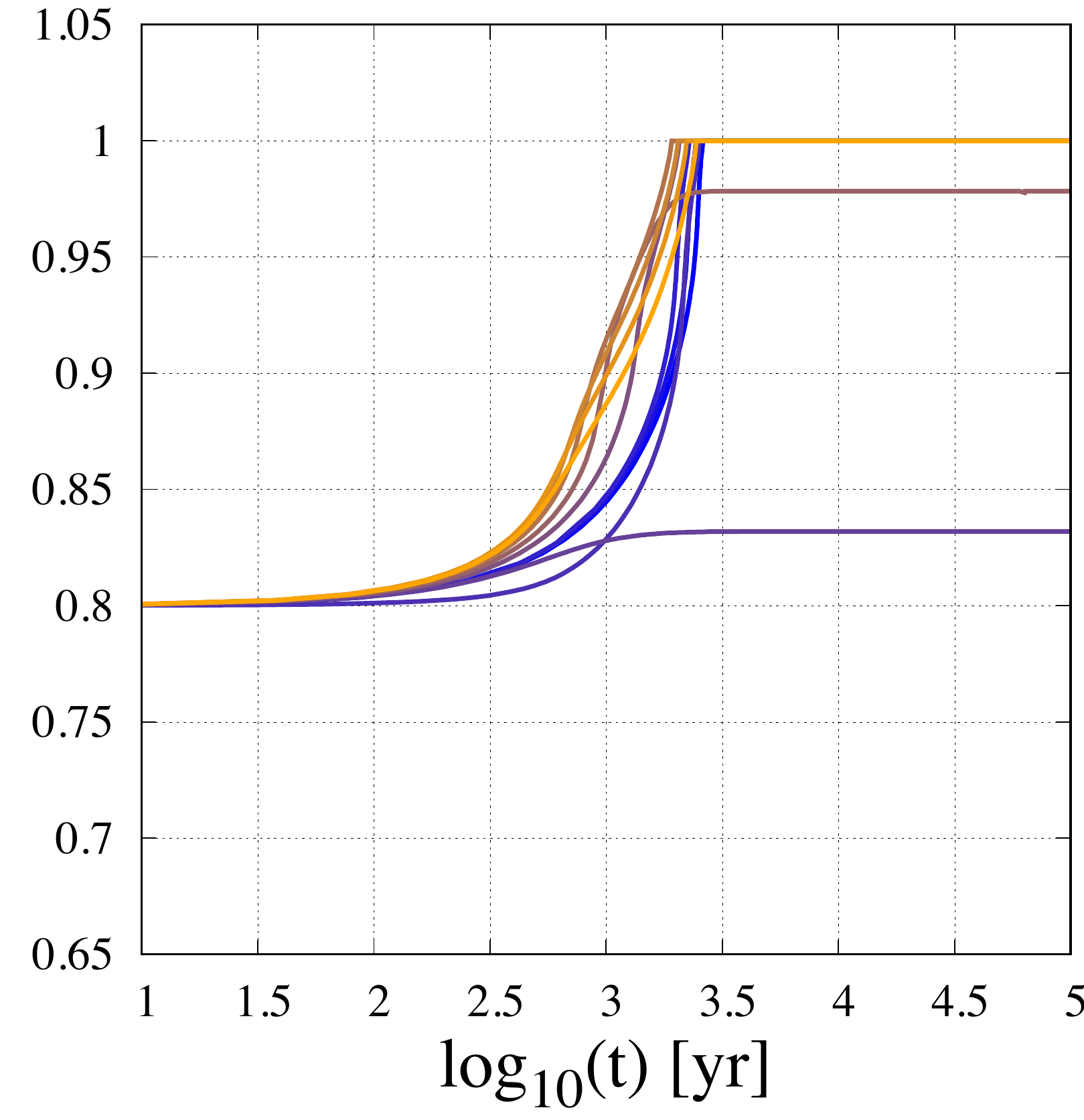} 
    \includegraphics[width=0.27\textwidth,trim = 3.5cm 2.5cm 8.0cm 2.1cm,clip]{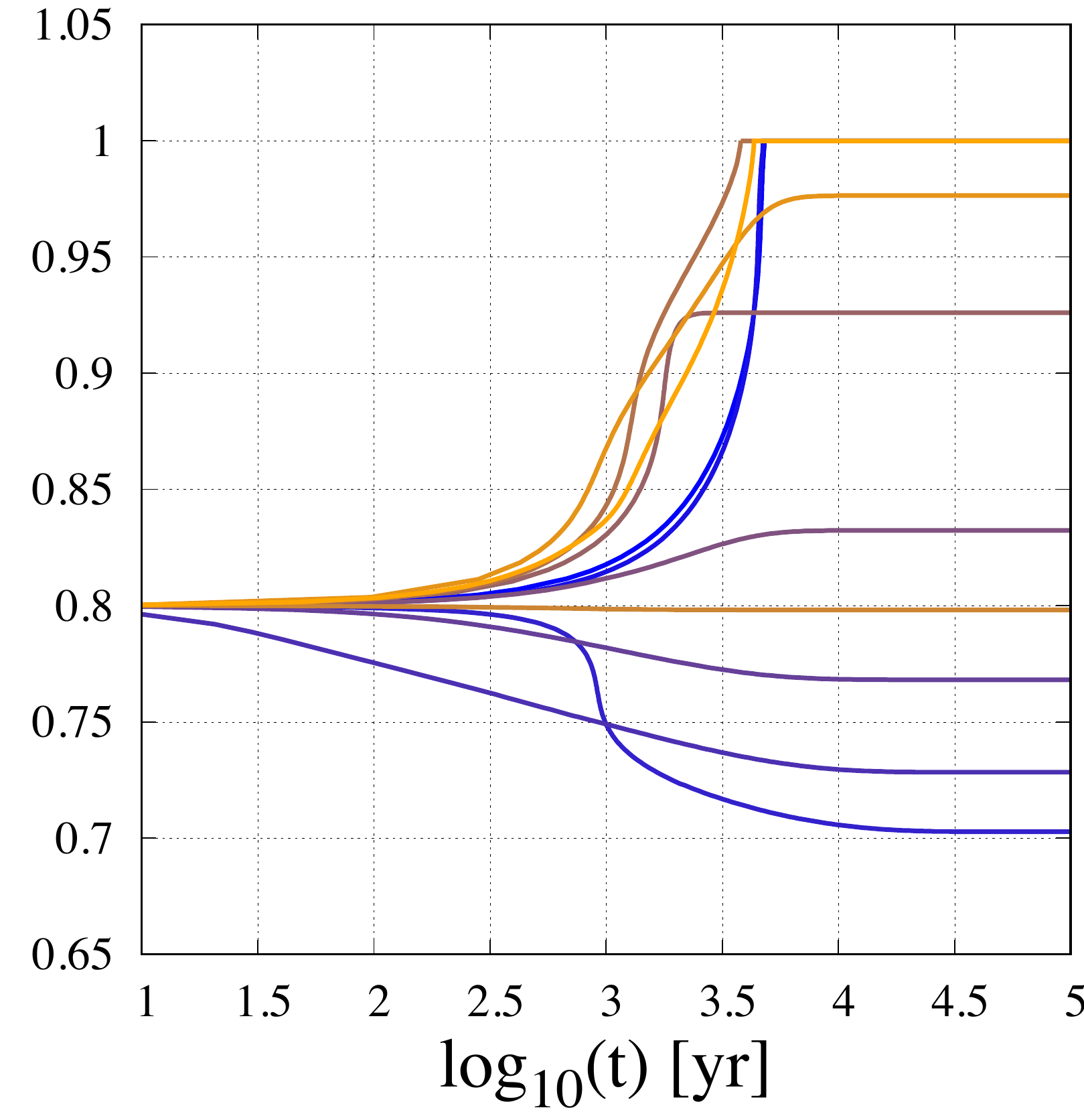}
    \includegraphics[width=0.27\textwidth,trim = 3.5cm 2.5cm 8.0cm 2.1cm,clip]{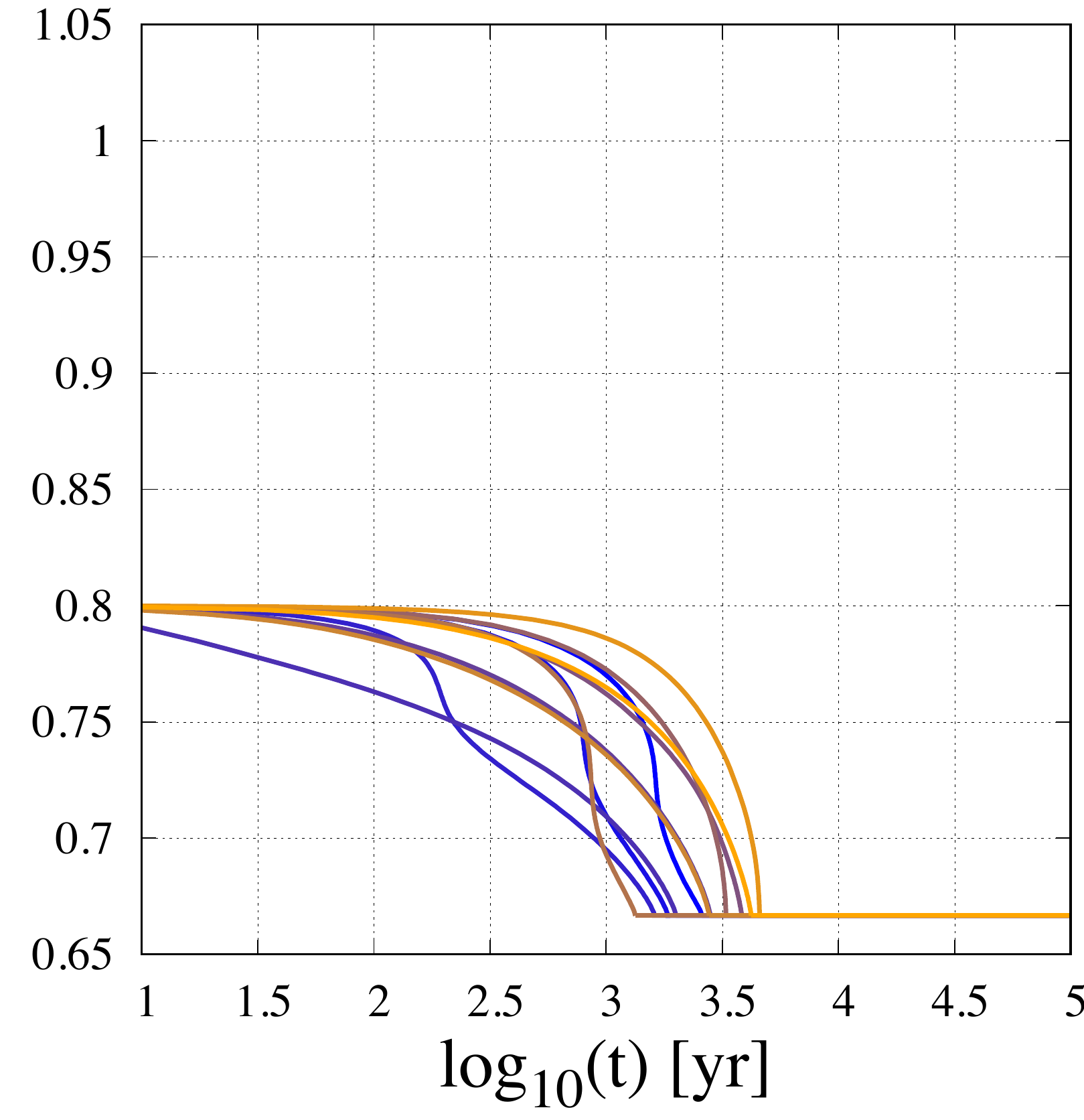}
    \hspace{0.1\textwidth} 
   \caption{Evolution of the normalised rotation period $\Prot / \Porb$ as a function of time (yr) in logarithmic scale in the case of an Earth-sized planet orbiting the TRAPPIST-1 star with a 6-days orbital period. The Rayleigh drag frequency of the ocean is set to $ \fdrag = 10^{-5} $\units{s^{-1}} for cases of strong drag (top panels), and to $\fdrag = 10^{-6}$\units{s^{-1}} for cases of weak drag (bottom panels). The planet eccentricity takes the values $\ecc = 0.063$, 0.2, and 0.3 from left to right. In each panels, a set of planets with various ocean depths is studied. These configurations are parametrised by $\gamma = \logdix \left( \Hoc \right)$, where the ocean depth is given in km. The parameter $\gamma$ is sampled from -2 (driest case of the sampling, orange solid line) to 0 (most humid case, blue solid line). Values used for the planet parameters are given by \tab{tab:andrade_para}.}
       \label{fig:evolution}%
\end{figure*}

\section{Evolution timescale}
\label{sec:evolution_timescale}

We return to the case of the Earth-sized planet orbiting the TRAPPIST-1 star with a 6-days orbital period. The features of the planet and of its solid part in particular are determined by the parameters given by \tab{tab:andrade_para}. As done in \sect{sec:final_rotation}, we consider the strong and weak drag configurations characterised by $\fdrag = 10^{-5}$\units{s^{-1}} and $\fdrag = 10^{-6}$\units{s^{-1}}, respectively. Three orbits of various eccentricity are studied. In the first one, $\ecc = 0.063$, which is the value of the critical eccentricity predicted by the theory in the weak drag configuration (see \fig{fig:final_rotation}, top middle panel). In the second orbit, $\ecc $ is set to 0.2, that is just below the value of the critical eccentricity obtained in the absence of ocean. In the third orbit, it is set to $0.3$, which is slightly greater than this critical eccentricity.

In each configuration we compute the evolution of the spin rotation rate of a set of planets characterised by ocean depths sampled in the interval $-2 \leq \gamma \leq 0$, where $\gamma \define \logdix \left( \Hoc \right) $ with $\Hoc$ given in kilometres. \jlc{As mentioned in \sect{sec:final_rotation}, the range of values $\Hoc \lesssim 0.1$~km does not correspond to realistic cases \ebc{since the global ocean approximation is no longer valid for such small depths. The} chosen lower bound $\gamma = -2$ \ebc{is} intended to emphasise the continuous transition between the asymptotic regime of dry planets ($\Hoc \rightarrow  0$) in our ocean planet tidal model and the case of a dry solid body described by the Andrade model.} 

The initial value of the spin rotation rate is set to $\spinrate = \left( 5/4 \right) \norb$, which corresponds to the middle of the interval of forcing frequencies defined by the $1{:}1$ and $3{:}2$ spin-orbit resonances. \jlc{If we ignore the librating motion of the planet and the associated triaxial torque,} the evolution of the spin angular velocity is determined by the equation

\begin{equation}
\Cinertpla \Dt{\spinrate}= \torquepla \left( \Omega \right),
\end{equation}

\noindent which is integrated over time using a variable step size Bulirsch-Stoer method \citep[][]{press2007numerical}. The normalised moment of inertia $\inertieplanorm = \inertiepla / \left( \Mpla \Rpla^2 \right)$ is set to $\inertieplanorm = 0.3308 $ \citep[][]{Bullen2012} and the tidal torque exerted on the planet is calculated using the expression given by \eq{torque_pla}.

We plot in \fig{fig:evolution} the normalised rotation period $\Prot/\Porb$ (vertical axis) as a function of time in logarithmic scale (horizontal axis) for each of the above defined configurations: the strong (top panels) and weak (bottom panels) drag regimes with eccentricities $\ecc = 0.063$, 0.2, and 0.3 (from left to right). Colours of solid lines varies from orange to blue as the ocean depth increases.

The order of magnitude of the evolution timescale $\tauevol$ can be estimated by considering the contribution of the solid part in the absence of ocean in the circular case. By assuming that $\ftidequadsemid \approx \norb/2 = \pi / \Porb $, we thus obtain 

\begin{equation}
\tauevol \sim \frac{\inertieplanorm}{48 \pi^3} \left( \frac{\Ggrav \Mpla \Porb^3}{\Rpla^3 \abs{\imag{\kquad \left( \pi / \Porb \right)}} } \right),
\label{tauevol}
\end{equation}

\noindent where the expression of $\imag{\kquad}$ as a function of the Andrade parameters is given by \eq{k2sol_asymp}. The substitution of parameters by their numerical values leads to $\tauevol \sim 10^3$\units{yr}, and we hence recover the order of magnitude given by numerical calculations.

The plots of \fig{fig:evolution} show the three possible regimes of the evolution of the planetary spin. In the first regime ($\ecc = 0.063$, left panels), the planet is driven towards the spin-orbit synchronous rotation in most of cases. Particularly, in the strong drag configuration, all of the planets are locked in this state of equilibrium after $\sim 5000$\units{yr} whatever the depth of their oceanic layer. For $\fdrag = 10^{-6}$\units{s^{-1}}, two of the studied cases avoid the synchronisation owing to the action of oceanic tides and evolve towards asynchronous rotation states located between the $1{:}1$ and $3{:}2$ spin-orbit resonances.

The second regime ($\ecc = 0.2$, middle panels) corresponds to the transition between low and high eccentricities. In this regime, a solid planet is driven towards the synchronisation while most of those hosting an ocean converge towards the non-synchronised rotation states of equilibrium identified in our exploration of the parameter space (see \fig{fig:final_rotation}, top left and middle panels). We note that trajectories of planets are affected by the variations of the tidal torque with the spin angular velocity. A resonant peak associated with an oceanic tidal mode generates a rapid evolution, which slows down when the tidal torques returns to its non-resonant level.  

In the third regime ($\ecc = 0.3$, right panels) the contribution of non-resonant eccentricity terms is so strong that it counterbalances the action of the semidiurnal component. As a consequence, intermediate rotation sates of equilibrium cannot exist and all of the planets converge towards the $3{:}2$ spin-orbit resonances where they are locked by solid tides. The ocean just slightly modifies the time necessary to reach the state of equilibrium in this case.


\section{Conclusions}
\label{sec:conclusions}

By combining the linear theory of oceanic tides with realistic models of the solid tide, we derived a self-consistent analytic model to characterise the response of an ocean planet undergoing eccentricity tides. Following early works \citep[e.g.][]{Tyler2011,Tyler2014,Chen2014,Matsuyama2014,Matsuyama2018}, this model assumes the shallow water approximation and includes the effects of ocean loading,  self-attraction, and deformation of the solid regions, which tightly couple the oceanic tidal response with that of the solid part. The energy dissipation induced by the interactions of tidal flows with the planet topography is taken into account with a Rayleigh drag, which enabled us to extend the usual theoretical treatments applied in adiabatic regimes to the dissipative case. 

The tidal response of the solid part is described in our approach by the Andrade model, with parameters computed from advanced models \padc{of the internal structure and tidal oscillations of solid bodies}. The Andrade model solves the inability of commonly used models -- such as the Maxwell model for instance -- to describe properly how the tidally dissipated energy scales with the tidal frequency, which usually leads to underestimate it by several orders of magnitude in the high-frequency regime. 

Firstly, we derived an analytic solution for the tidal torque exerted on the planet, which allowed us to characterise the main features of this torque in the quasi-adiabatic regime. Particularly, we identified the dependence of eigenfrequencies associated with oceanic modes on the planet parameters and quantified the maximum of resonant peaks. This led us to derive an analytic estimation of the critical eccentricity beyond which asynchronous rotation states of equilibrium can be induced by eccentricity tides. 

Secondly, we used the obtained expression of the tidal torque to explore numerically the parameter space in the case of an Earth and a super-Earth. We thus showed that resonances of oceanic modes are likely to decrease the critical eccentricity by one order of magnitude, enabling thereby the existence of asynchronous rotation states of equilibrium distinct from the spin-orbit resonances induced by tides in the case of solid bodies \citep[][]{ME2013,Correia2014}. Particularly, for typical configurations, the critical eccentricity can switch from $\sim \!  0.3$ to $\sim \! 0.06$ owing to the resonant amplification of the degree-3 (or eastward propagating) eccentricity term, which is the largest one in the low-eccentricity regime. Calculations allowed us to determine the region of the parameter space where the impact of resonances on the critical eccentricity is the most significant, and to establish scaling laws characterising its dependence on the system features (orbital period the planet, ocean depth and Rayleigh drag frequency).

Thirdly, we highlighted the action of resonances associated with oceanic modes on the history of the planetary rotation by coupling the time evolution equation of the spin with our tidal model. Results shows the three possible regimes of evolution: (i) the low-eccentricity limit ($\ecc \ll 0.2$), where planets are driven towards synchronisation except in case of strong resonance of the oceanic tidal response, (ii) the transition regime ($\ecc \approx 0.2$), where a large fraction of ocean planets converge towards asynchronous states while solid ones are driven towards the spin-orbit synchronous rotation, and (iii) the high-eccentricity regime ($\ecc \gtrsim 0.3$), where the eccentricity is strong enough to make planets converge towards the $3{:}2$ spin-orbit resonance induced by solid tides whatever the depth of their oceans. 

\padc{These results can be used to better constrain the rotation of discovered rocky planets. In the case of TRAPPIST-1 planets, eccentricities \ebc{should be limited to low values} \citep[][]{Luger2017,Grimm2018,Turbet2018}. \rec{Planets b and c are in a runaway greenhouse state so that they are too hot to sustain surface liquid water. Planet d, depending on the habitable zone model, could be able to sustain surface liquid water or it could be in a runaway greenhouse state. However, being relatively close-in, planet d could experience a strong tidal heating \citep[e.g.][]{Makarov2018}, which could trigger a tidal runaway greenhouse \citep[][]{Barnes2013} and prevent the presence of a global surface ocean. Planet f to h are too cold to sustain surface liquid water but could still host interior oceans below a layer of ice \citep[][]{Turbet2018}. Our formalism does not allow us to model such planets at this point.}

\rec{However, planet e has the highest probability of hosting a global ocean of liquid water. It could have retained its initial water content \citep[][]{Bourrier2017}, and if there is water, it could be liquid \citep[][]{Turbet2018}. And due to the small eccentricity of its orbit, it is likely to be tidally locked in synchronous rotation in our framework, unless the ocean has a very specific depth and tidal flows undergo a weak drag. Note that the strong tidal heating obtained by \cite{Makarov2018} could trigger a runaway greenhouse state for planet e as well. However they provide an upper estimate of the tidal heating and other studies have found much lower tidal heat fluxes for planet e \citep[][Bolmont et al., in prep.]{Barr2018}, which are compatible with surface liquid water.}

In the case of Proxima, the eccentricity could be as high as 0.1 \citep[][]{AE2016,Ribas2016}, meaning that the planet could be locked in a spin state different from the $1{:}1$ and $3{:}2$ spin-orbit resonances, depending on the depth of a supposed global ocean. Thus the synchronous rotation state is less probable in this case \rec{than in that of TRAPPIST-1 planets}.}

Our approach based on the linear theory of tides can help to better understand the action of eccentricity tides on the rotation of ocean planets, and provides solutions that can be implemented in evolutionary codes to compute the tidal Love numbers and torque in a realistic -- and nevertheless efficient -- way. Particularly, the linear analysis leads to a full characterisation of the frequency behaviour of these parameters, which is crucial in the case of oceanic tides. However, it presents limitations that make it complementary with \ebc{general circulation models}. 

The main limitation of our model is that it cannot be applied to large tidal distortions since these laters implies non-linear mechanisms. Thus, resonances violating the small perturbation approximation are beyond the scope of the linear theory and shall be studied \smc{by integrating the full  primitive equations of the ocean \citep[e.g.][]{Vallis2006}}. This limitation prevented us to quantify the critical eccentricity in the quasi-adiabatic regime, below $\fdrag \sim 10^{-6}$\units{s^{-1}}. A second limitation is related to the shallow water approximation, which eludes the role played by the ocean internal structure on the tidally dissipated energy. Although the formalism can easily be extended to stably stratified deep oceans if the effects of ocean loading, self-attraction, and deformation of the solid regions are ignored \citep[][]{ADLML2018}, taking these ingredients into account complicates the analytic treatment of the problem significantly. 

Finally, we would like to draw the attention of the reader on the importance of the Rayleigh drag frequency in this approach. The Rayleigh drag frequency accounts for the global effect of topography and ocean streams, which converts barotropic tidal flows into internal \padc{gravity} waves. For Earth, this parameter has been estimated from the energy dissipated by the semidiurnal Lunar tide \citep[e.g.][]{Webb1980} and satellite measurements of the ocean surface elevation \citep[e.g.][]{ER2001,ER2003}. Given the impact it has on the planet tidal response, it should be characterised as a function of the planet properties (spin rotation, topography, ocean depth) in order to improve the predictions of the linear theory from a quantitative point of view. Moreover, one should investigate in future studies how the presence of a partial or global ice cap may affect the results obtained here for a free-surface ocean planet, following the methodology adopted for icy satellites \citep[e.g.][]{Kamata2015,Matsuyama2018}.

\begin{acknowledgements}
\rec{The authors thank the anonymous referee for constructive comments that improved the manuscript.} \jlc{This project has received funding from the European Research Council (ERC) under the European Union’s Horizon 2020 research and innovation programme (grant agreements No. 679030/WHIPLASH and 771620/EXOKLEIN),} \ebc{and has been carried out within the framework of the NCCR PlanetS supported by the Swiss National Science Foundation.}  \smc{S.~Mathis acknowledges funding by the  European Research Council through the ERC grant No. 647383/SPIRE and by the PLATO CNES grant at CEA Saclay. E.~Bolmont and P.~Auclair-Desrotour are grateful to \ebc{O.~Grasset and G.~Tobie, whose models were used in this work to calculate the internal structure of the studied rocky planets and the tidal visco-elastic response of their solid part, respectively.}} \ebc{This research has made use of NASA's Astrophysics Data System.} 
\end{acknowledgements}

\bibliographystyle{aa}  
\bibliography{auclair-desrotour} 

\appendix

\section{Associated Legendre functions}
\label{app:legendre_functions}

Associated Legendre functions are defined for $x \in \left[ -1 , 1 \right]$ and $\llat \geq \abs{\mm}$ as \citep[][]{AS1972,Arfken2005} 

\begin{equation}
\Plm \left( x \right) \define \left(-1 \right)^\mm \left( 1 - x^2 \right)^{\mm/2}  \DDn{}{x}{\mm} \Pl \left( x \right),
\end{equation}

\noindent the $\Pl$ designating the Legendre polynomials, given by 

\begin{equation}
\Pl \left( x \right) \define \frac{1}{2^\llat \llat !} \DDn{}{x}{\llat} \left[ \left( x^2 - 1 \right)^\llat \right]. 
\end{equation}

The normalised associated Legendre functions $\Plmnorm$ are characterised by 

\begin{equation}
\integ{\Plmnorm \LegFnorm{\kk}{\mm}}{x}{-1}{1} = \kroind{\llat}{\kk}, 
\end{equation}

\noindent where the \padc{Kronecker symbol $\kroind{\llat}{\kk}$ is} such that $\kroind{\llat}{\kk} = 1$ if $\llat = \kk$ and $\kroind{\llat}{\kk} = 0$ otherwise. Hence they are expressed as 

\begin{equation}
\Plmnorm \left( x \right) \define \left[ \frac{\left( 2 \llat + 1 \right) \left( \llat - \mm \right) !}{2 \left( \llat + \mm \right) !} \right]^{\frac{1}{2}} \Plm \left( x \right).
\end{equation}

\section{Hansen coefficients}
\label{app:Hansen}

Hansen coefficients are \padc{derived from} the Fourier coefficients of the eccentricity ($\ecc$) and mean anomaly ($\meana$) function \citep[e.g.][]{Hughes1981,PS1990,Laskar2005}

\begin{equation}
\gHansen{\llat}{\mm} \left( \ecc , \meana \right) = \left[  \frac{\rstar \left( \ecc , \meana \right)}{a} \right]^\llat \expo{\inumber \mm \truea \left( \ecc , \meana \right)},
\label{gHansen}
\end{equation}

\noindent where $\truea$ designate the true anomaly and $\rstar$ the star planet distance, these two quantities being themselves expressed as 

\begin{align}
\rstar \left( \ecc , \meana \right) = & \ \smaxis \left( 1 - \ecc \cos \meana \right) , \\
 \truea \left( \ecc , \meana \right) = & \ \meana + 2 \ecc \sin \meana .
\end{align}

\noindent Hansen coefficients are thus defined by \citep[e.g.][]{Hughes1981}

\begin{equation}
\Hansen{\ss}{\llat}{\mm} \left( \ecc \right) \define \frac{1}{2 \pi} \integ{\gHansen{\llat}{\mm} \left( \ecc , \meana \right) \expo{- \inumber \ss \meana} }{\meana}{-\pi}{\pi}. 
\end{equation}

\noindent We note that the $\Xslm$ are real since $\gHansen{\llat}{\mm}$ is an even function of $\meana$. 

By changing the sign of $\ss$ and using the symmetry property $\Hansen{-\ss}{\llat}{-\mm} = \Hansen{\ss}{\llat}{\mm} $, we obtain

\begin{equation}
\gHansen{- \left( \llat + 1 \right) }{- \mm} \left( \ecc , \meana \right) = \sum_{s = - \infty}^{+ \infty} \Hansen{\ss}{- \left( \llat + 1 \right)}{\mm} \left( \ecc \right) \expo{- \inumber \ss \meana }, 
\label{ghansen1}
\end{equation}

\noindent which is convenient to put the perturbing tidal potential into the form given by \eq{Utide}. 

In practice, considering the fact that Hansen coefficients are the Fourier coefficients of the function $\gHansen{\llat}{\mm}$, one can efficiently get a numerical estimate of $2^N + 1$ coefficients for $- K \leq \ss \leq K $ with $K = 2^{N-1}$ by means of a fast Fourier transform (FFT) \citep[][]{ASH2010,Correia2014}. The integer $N$ shall be chosen appropriately so that $ \maxi{\abs{\Utidelms{\llat}{\mm}{\ss}}}{\ss = \pm K} \ll \maxi{\abs{\Utidelms{\llat}{\mm}{\ss}}}{-K \leq \ss \leq K} $, and increases with the eccentricity as the spectrum of forcing terms widens \citep[see Fig.~3 in][]{Ogilvie2014}.

\section{Torque due to triaxiality}
\label{app:triaxial_torque}

In this Appendix, we detail the derivation of the expression of the triaxial torque in the general case, given by \eq{torquetri}. This torque is expressed as a function of the planet spin angle ($\spinangle$) and true anomaly ($\truea$) as \citep[e.g.][Eq.~(2)]{GP1966} 

\begin{equation}
\torquetri \define - \frac{3}{2}  \left( \Binertpla - \Ainertpla\right) \frac{\Ggrav \Mstar}{\rstar^3} \sin \left[ 2 \left( \spinangle - \truea \right) \right],
\end{equation}

\noindent which can be rewritten as $\torquetri = \imag{\torquetriC}$, where

\begin{equation}
\torquetriC \define - \frac{3}{2}  \left( \Binertpla - \Ainertpla\right) \norb^2 \expo{2 \inumber \spinangle} \gHansen{-3}{-2} \left( \ecc , \meana \right).
\label{torquetriC}
\end{equation}

\noindent By using \eq{ghansen1} and the notation $\anglequads = \spinangle -  \left( \ss / 2 \right) \meana$, we thus obtain 

\begin{equation}
\torquetriC = - \frac{3}{2} \left( \Binertpla - \Ainertpla \right) \norb^2 \sum_{\ss = - \infty}^{+ \infty} \Hansen{\ss}{-3}{2} \left( \ecc \right)  \expo{2 \inumber \anglequads}, 
\end{equation}

\noindent and 

\begin{equation}
\torquetri = -\frac{3}{2} \left( \Binertpla - \Ainertpla \right) \norb^2 \sum_{\ss = - \infty}^{+ \infty}   \Hansen{\ss}{-3}{2} \left( \ecc \right)  \sin \left( 2 \anglequads \right).
\end{equation}



\begin{onecolumn}
\section{Nomenclature}
\begin{longtable}{lll}
\caption{\label{Notations} List of the notations used along this work and their designations in order of appearance in the text.}\\
\hline\hline
\textsc{Symbol} & \textsc{Definition} & \textsc{Reference}  \\
\hline
\vspace{-0.3cm}  \\  
\endfirsthead
\caption{continued.}\\
\hline\hline
\textsc{Symbol} & \textsc{Definition} & \textsc{Reference}   \\
\hline
\vspace{-0.3cm}  \\
\endhead
\hline
\endfoot
	$\ecc$		& Planet eccentricity & \sect{sec:intro} \\
	$\eccasync$	& Critical eccentricity for asynchronous rotation & \sect{sec:intro} \\
	$\Mpla$		& Planet mass & \sect{sec:setup} \\
	$\Rpla$		& Planet radius & \sect{sec:setup} \\
	$\Hoc$		& Ocean depth & \sect{sec:setup} \\
	$\rhooc$		& Ocean density & \sect{sec:setup} \\
	$\norb$		& Planet mean motion & \sect{sec:setup} \\
	$\norbvect$	& Orbital angular momentum of the planet& \sect{sec:setup} \\
	 $\framegeo$	& Planeto-centric reference frame & \sect{sec:setup} \\ 
	 $\framerot$	& Reference frame co-rotating with the planet & \sect{sec:setup} \\ 
	$\spinvect$	& Spin vector & \sect{sec:setup} \\
	$\spinrate$	& Rotation rate of the planet  & \sect{sec:setup} \\
	$\time$		& Time & \sect{sec:setup} \\
	$\rr$			& Radial coordinate (spherical coordinates) & \sect{sec:setup} \\
	$\col$		& Colatitude (spherical coordinates) & \sect{sec:setup} \\ 
	$\lon$		& Longitude (spherical coordinates) & \sect{sec:setup} \\ 
	$\left( \er , \etheta, \ephi \right)$		& Spherical unit-vector basis & \sect{sec:setup} \\ 
	$\ggravi$		& Planet surface gravity & \sect{sec:setup} \\ 
	$\rhopla$ 		& Planet mean density & \sect{sec:setup} \\
	$\Moc$		& Mass of the ocean & \sect{sec:setup} \\ 
	$\Mcore$ 		& Mass of the solid part & \sect{sec:setup} \\
	$\Rcore$		& Radius of the solid part & \sect{sec:setup} \\ 
	$\rhocore$	& Mean density of the solid part & \sect{sec:setup} \\ 
	$\fdrag$		& Rayleigh drag frequency &\sect{sec:setup}  \\ 
	$\taudrag$	& Typical timescale of core-ocean coupling by viscous friction & \sect{sec:setup}  \\
	$\rstar$		& Star-planet distance  & \eq{stellar_potential} \\ 
	$\Mstar$		& Mass of the host star & \eq{stellar_potential} \\ 
	$\UstarR$		& Gravitational potential of the host star & \eq{stellar_potential} \\ 
	$\Rvar{~}$	& Symbol used to highlight real quantities & \eq{stellar_potential} \\
	$\Cvar{~}$	& Symbol used to highlight complex quantities & \eq{stellar_potential} \\ 
	$\UtideR$		& Real tidal gravitational potential & \eq{UtideR} \\ 
	$\Re$		& Real part of a complex number & \eq{Utide} \\ 
	$\Im$		& Imaginary part of a complex number & \eq{Utide} \\ 
	$\inumber$	& Imaginary number & \eq{Utide} \\ 
	$\Utide$		& Complex tidal gravitational potential & \eq{Utide} \\ 
	$\llat$ 		& Latitudinal degree (spherical harmonics) & \eq{Utide} \\ 
	$\mm$		& Longitudinal degree (spherical harmonics) & \eq{Utide} \\ 
	$\ss$ 		& Eccentricity degree & \eq{Utide} \\ 
	$\ftide$		& Forcing frequency & \eq{Utide} \\
	$\ftide_{\mm,\ss}$	& Forcing tidal frequency of the mode associated with the doublet $\left( \mm , \ss \right)$ & \eq{Utide} \\ 
	$\Plmnorm$	& Normalised associated Legendre functions & \eq{Utide} \\ 
	$\Ulms$ 		& $\left( \llat , \mm , \ss \right)$-component of the forcing tidal gravitational potential at the planet surface & \eq{Utide} \\ 
	$\smaxis$		& Semi-major axis of the planet & \eq{Utidelms} \\ 
	$\Alms$		& Dimensionless eccentricity functions & \eq{Utidelms} \\ 
	$\kroind{\llat}{\kk}$	& Kronecker symbol & \eq{Alms_ecc} \\
	$\kroinf{\ss}$	& Coefficient equal to 1 for $\ss < 0$, 0 otherwise & \eq{Alms_ecc} \\ 
	$\Plm$		& Unnormalised associated Legendre functions & \eq{Alms_ecc} \\ 
	$\Xslm$		& Hansen coefficients & \eq{Alms_ecc} \\ 
	$\ftidequads$ & Forcing frequency of the degree-$\ss$ eccentricity term & \eq{ftide2s} \\
	$\Utidequad$ & Degree-$\ss$ component of the perturbing tidal potential & \eq{U22s} \\ 
	$\Utidelms{2}{2}{2} $ & Semidiurnal component of the perturbing tidal potential & \eq{U222} \\
	$\Utidelms{2}{2}{1}$	& Degree-1 eccentricity component of the perturbing tidal potential & \eq{U221} \\
	$\sigma_{2,1} $	& Forcing frequency of the degree-1 eccentricity component & \eq{U221} \\
	$\Utidelms{2}{2}{3}$	& Degree-3 eccentricity component of the perturbing tidal potential & \eq{U223} \\ 
	$\sigma_{2,3} $	& Forcing frequency of the degree-3 eccentricity component & \eq{U223} \\
	$\ecctrans$	& Transition eccentricity  & \sect{ssec:perturbing_potential} \\ 
	$\mupla$		& Effective unrelaxed shear modulus of the solid core& \sect{ssec:solid_part}  \\ 
	$\stresstens$	& Stress tensor & \eq{strain_stress} \\ 
	$\straintens$	& Strain tensor & \eq{strain_stress} \\ 
	$\Cmu$		& Complex shear modulus of the solid part & \eq{Hooke} \\ 
	$\GammaF$	& Gamma function & \eq{Andrade_mu} \\ 
	$\alphaA$		& Rheological exponent of the material in the Andrade model & \eq{Andrade_mu} \\ 
	$\tauM$		& Maxwell relaxation time of the material & \eq{Andrade_mu} \\
	$\viscosity$	& Viscosity of the material & \eq{Andrade_mu} \\ 
	$\tauA$		& Andrade time of the material in the Andrade model & \eq{Andrade_mu} \\ 
	$\tautide$		& Tidal period & \sect{ssec:solid_part} \\
	$\betaA$		& Andrade parameter & \sect{ssec:solid_part} \\ 
	$\kl$			& Gravitational Love number of the degree-$\llat$ mode & \eq{love_solid} \\ 
	$\hl$			& Displacement Love number of the degree-$\llat$ mode & \eq{love_solid} \\ 
	$\kloadl$		& Gravitational load Love number of the degree-$\llat$ mode & \eq{love_solid} \\
	$\hloadl$		& Displacement load Love number of the degree-$\llat$ mode & \eq{love_solid} \\ 
	$\Cmul$		& Dimensionless effective rigidity & \eq{Cmul} \\ 
	$\Al$			& Dimensionless rigidity coefficient associated with the degree-$\llat$ component & \eq{Al_Cmu} \\
	$\xibot$		& Vertical displacement of the oceanic floor & \sect{sec:ocean_response} \\ 
	$\Vvect$		& Velocity vector of the horizontal component of tidal flows & \sect{sec:ocean_response} \\ 
	$\Vtheta$		& Latitudinal component of the velocity field & \sect{sec:ocean_response} \\ 
	$\Vphi$		& Longitudinal component of the velocity field & \sect{sec:ocean_response} \\ 
	$\xisurf$		& Vertical displacement of the ocean surface & \sect{sec:ocean_response} \\ 
	$\xilayer$ 		& Variation of the ocean thickness & \sect{sec:ocean_response} \\ 
	$\dd{}{\XX}$	& Partial derivative with respect to $\XX$ & \eq{momentum_eq} \\ 
	$\Ftide$		& Perturbed potential encompassing the gravitational forcing and coupling with the solid part & \eq{momentum_eq} \\ 
	$\gradh$		& Horizontal gradient operator & \eq{gradh} \\
	$\divh$		& Horizontal divergence operator & \eq{divh} \\ 
	$\Ftidems$	& Fourier component of the perturbed potential & \eq{Fourier_var} \\ 
	$\Vvectms$	& Fourier component of the velocity vector & \eq{Fourier_var} \\
	$\xilayerms$	& Fourier component of the vertical variation of the ocean depth & \eq{Fourier_var} \\ 
	$\Ftidelms$	& Degree-$\llat$ component of $\Ftidems$ expanded in series of associated Legendre functions & \eq{legendre_series} \\ 
	$\Vvectlms$	& Degree-$\llat$ component of $\Vvectms$ expanded in series of associated Legendre functions & \eq{legendre_series} \\ 
	$\xilayerlms$	& Degree-$\llat$ component of $\xilayerms$ expanded in series of associated Legendre functions & \eq{legendre_series} \\ 
	$\ftidedrag$	& Complex tidal frequency  & \eq{freq_spinpar} \\
	$\spinpar$		& Complex spin parameter & \eq{freq_spinpar}  \\ 
	$\Laplace$	& Laplace's tidal operator & \eq{Laplace_eq} \\ 
	$\nn$		& Degrees of Hough functions & \eq{Hough_var} \\
	$\Ftiden$		& Degree-$\nn$ component of $\Ftidems$ expanded in series of Hough functions & \eq{Hough_var} \\ 
	$\Thetan$		& Degree-$\nn$ Hough function & \eq{Hough_var} \\ 
	$\Lambdan$	& Eigenvalue associated with the degree-$\nn$ Hough function & \eq{Laplace} \\
	$\heqn$		& Equivalent depth of the degree-$\nn$ Hough mode & \eq{heqn} \\ 
	$\Anl$		& Coefficients of Hough functions expanded in series of the $\Plmnorm$ & \eq{Hough_Plm} \\ 
	$\Bln$		& Coefficients of the $\Plmnorm$ expanded in series of Hough functions & \eq{Plm_Hough} \\
	$\Clnk$		& Overlap coefficients weighting the degree-$\nn$ component of the oceanic tidal torque & \eq{Clnk} \\ 
	$\tiltU$		& Tilt factor associated with the tidal gravitational forcing of the perturber & \eq{tiltU} \\
	$\tiltxi$		& Tilt factor associated with the distortion of the oceanic layer & \eq{tiltxi} \\
	$\sig{\nn}{\kk}$	& Complex characteristic frequencies of the planet & \eq{fnk} \\
	$\khori$		& Horizontal wavenumber of the degree-$\nn$ Hough mode & \eq{fnk} \\ 
	$\Fforcen$	& Component of the force vector inducing the tidal perturbation & \eq{Fforcen} \\
	$\torquepla$	& Tidal torque exerted on the planet & \eq{torque_pla} \\
	$\kp$		& Quadrupolar component of the effective gravitational Love number of the planet & \eq{kpla} \\
	$\torquesol$	& Tidal torque exerted on the solid part in absence of oceanic layer & \eq{torque_sol} \\
	$\kquad$		& Quadrupolar tidal Love number of the solid part in absence of oceanic layer & \eq{k2sol_asymp} \\
	$\Aquad$		& Dimensionless rigidity coefficient associated with the $\llat = 2$ component & \eq{Aquad} \\ 
	$\torqueoc$	&  Tidal torque exerted on the ocean in case of solid part of infinite rigidity & \eq{torque_ocean} \\ 
	$\focnplus$	& Complex eigenfrequency of the degree-$\nn$ Hough mode in the positive-frequency range & \eq{xilayern} \\ 
	$\focnmoins$	& Complex eigenfrequency of the degree-$\nn$ Hough mode in the negative-frequency range & \eq{xilayern} \\ 
	$\focn$		& Characteristic frequency of the degree-$\nn$ surface gravity mode & \eq{focn} \\
	$\fpeakn$		& Frequency at which the resonant peak of a mode reaches a maximum& \eq{fpeakn1} \\ 
	$\left. \imag{\kocquad} \right|_{\ipeak ; \nn}$	&  Maximum of $\imag{\kocquad}$ reached by the resonance peak  & \eq{maxpeak} \\
	$\ftide_{2,\ss}^\blacktriangleup$	& Degree-$\ss$ eccentricity frequency associated with a peak & \sect{ssec:pure_oceanic_response} \\ 
	$\spinrateeqs$	& Rotation rate corresponding to the degree-$\ss$ spin-orbit resonance & \sect{ssec:pure_oceanic_response} \\ 
	$\Porb$		&  Orbital period of the planet & \sect{ssec:frequency_behaviour} \\ 
	$\Msun$		& Mass of the Sun & \sect{ssec:frequency_behaviour} \\ 
	$\spinparquadsemid$	& Spin parameter associated with the semidiurnal tidal component & \eq{Coquadsemid} \\ 
	$\gamCn$		& Parameter controlling the existence of asynchronous states in the case of pure oceanic tide  & \eq{gamCn} \\ 
	$\Xfreq$		& Normalised frequency  & \eq{poly_sol} \\ 
	$\Aconst$		&  Dimensionless constant & \eq{poly_sol} \\
	$\ualpha$		& Maximum of the  polynomial function defined by \eq{poly_sol}  & \eq{ualpha} \\
	$\Prot$		& Rotation period of the planet & \sect{ssec:tidal_locking} \\
	$\Hocs$		& Ocean depth associated with the resonance of the degree-$\ss$ eccentricity component & \sect{ssec:tidal_locking}  \\
	$\Csol$		& Constant factor in the expression of the solid tidal torque in the non-resonant regime  & \eq{Csol} \\
	$\torqueplatid$	& Tidal torque exerted on the planet  & \eq{spin_evolution_eq} \\ 
	$\torquetri$	& Torque due to triaxiality & \eq{spin_evolution_eq} \\
	$\Dt{}$		& Time derivative & \eq{spin_evolution_eq} \\ 
	$\Ainertpla, \Binertpla, \Cinertpla$	& Principal moments of inertia of the planet (in increasing magnitude) & \eq{torquetri} \\
	$\anglequads$	& angle associated with the degree-$\ss$ perturbation component & \eq{torquetri} \\
	$\spinangle$	& Planet rotation angle & \eq{torquetri} \\ 
	$\meana$ 	& Planet mean anomaly & \eq{torquetri} \\
	$\Dtt{}$		& Second order time derivative & \eq{spin_evolution_angle} \\ 
	$\Deltares$	& Width of the resonance & \eq{Deltares} \\
	$\torqueeven$	& Even component of the tidal torque & \eq{torqueeven} \\
	$\torqueodd$	& Odd component of the tidal torque & \eq{torqueodd} \\ 
	$\captureprob$	& Capture probability of the $\ss / 2$ spin-orbit resonance & \eq{captureprob} \\
	$\DeltaProt$	& Normalised width of the $1{:}1$ spin-orbit resonance & \eq{DeltaProt} \\ 
	$\gamma$  	& Logarithm of the ocean depth & \sect{sec:evolution_timescale} \\ 
	$\inertieplanorm$	& Normalised moment of inertia & \sect{sec:evolution_timescale} \\
	$\tauevol$		& Evolution timescale of the planet spin rotation & \eq{tauevol} \\
	$\gHansen{\llat}{\mm}$ & Eccentricity and mean anomaly function & \eq{gHansen} \\
	$\truea$		& True anomaly & \eq{gHansen} \\ 
	$N$			& Order of the fast Fourier transform (FFT) in the calculation of Hansen coefficients & \append{app:Hansen} \\
	$K$			& Truncation index of the series of Hansen coefficients calculated using the FFT method & \append{app:Hansen} \\
	$\torquetriC$	& Complex triaxial torque & \eq{torquetriC} \\
\end{longtable} 
\end{onecolumn}

\end{document}